\documentclass[useAMS,usenatbib]{mn2e}
\newcommand\actaa{\rmfamily{ACTAA}}

\usepackage{epsfig}
\usepackage{amsmath}
\usepackage{amssymb}
\usepackage{ifthen}
\usepackage{txfonts}
\usepackage{rotating}
\usepackage{url}
\usepackage{varioref}
\usepackage{verbatim}
\usepackage{latexsym}
\usepackage{graphicx}
\DeclareSymbolFont{cmletters}{OML}{cmm}{m}{it}

\DeclareMathSymbol{v}{\mathord}{cmletters}{"76}

\def\ergs{{\rm\,erg\,s^{-1}}}

\def\ergs{\rm \,erg\,s^{-1}}
\def\be{\begin{equation}}
\def\ee{\end{equation}}
\newcommand{\kb}{ k_{\rm B} }

\newcommand{\tp}{\tilde{p}}

\newcommand{\msun}{{\rm M_{\odot}}}

\newcommand{\<}{\langle}
\renewcommand{\>}{\rangle}

\newcommand{\alf}{Alfv\'en}
\newcommand{\alfic}{Alfv\'enic}

\newcommand{\erg}{{\rm\,erg}}
\newcommand{\kev}{{\rm\,keV}}

\newcommand{\vfs}{v_{\rm fs}}

\newcommand{\cut}[1]{\hbox{}}

\newcommand\araa{\rmfamily{ARA\&A}}%
\newcommand\apj{\rmfamily{ApJ}}%
\newcommand\apjl{\rmfamily{ApJ}}%
\newcommand\apjs{\rmfamily{ApJS}}%
\newcommand\apss{\rmfamily{Ap\&SS}}%
\newcommand\aap{\rmfamily{A\&A}}%
\newcommand\mnras{\rmfamily{MNRAS}}%
\newcommand\pra{\rmfamily{Phys.~Rev.~A}}%
\newcommand\prd{\rmfamily{Phys.~Rev.~D}}%
\newcommand\pasj{\rmfamily{PASJ}}%
\newcommand\solphys{\rmfamily{Sol.~Phys.}}%
\newcommand\nat{\rmfamily{Nature}}%
\newcommand\grl{\rmfamily{Geophys.~Res.~Lett.}}%
\newcommand\jgr{\rmfamily{J.~Geophys.~Res.}}%
\title[GRB Reconnection Switch]
{A Reconnection Switch to Trigger Gamma-Ray Burst Jet Dissipation}

\author[J.~C. McKinney, \& D.~A. Uzdensky]
   {Jonathan C. McKinney,$^1$$^2$\thanks{\hbox{E-mail: jmckinne@stanford.edu~(JCM);} \hbox{uzdensky@colorado.edu~(DAU);} }
    Dmitri A. Uzdensky$^3$\footnotemark[1] \\
  $^1$Department of Physics and Kavli Institute for Particle Astrophysics and Cosmology, Stanford University, Stanford, CA 94305-4060, USA \\
  $^2$Chandra Fellow \\
  $^3$Center for Integrated Plasma Studies, UCB 390, Department of Physics, University of Colorado, Boulder, CO 80309
  }

\begin{document}
\date{Accepted 2011 August 29. Received 2011 July 11; in original form 2010 November 02}
\pagerange{\pageref{firstpage}--\pageref{lastpage}} \pubyear{2011}
\maketitle

\label{firstpage}
\begin{abstract}

Prompt gamma-ray burst (GRB) emission requires
some mechanism to dissipate an ultrarelativistic jet.
Internal shocks or some form of electromagnetic dissipation are candidate mechanisms.
Any mechanism needs to answer basic questions, such as
what is the origin of variability,
what radius does dissipation occur at,
and how does efficient prompt emission occur.
These mechanisms also need to be consistent with
how ultrarelativistic jets form and stay baryon pure
despite turbulence and electromagnetic reconnection near the compact object
and despite stellar entrainment within the collapsar model.
We use the latest magnetohydrodynamical models of ultrarelativistic jets
to explore some of these questions in the context
of electromagnetic dissipation due to
the slow collisional and fast collisionless reconnection mechanisms,
as often associated with Sweet-Parker and Petschek reconnection, respectively.
For a highly magnetized ultrarelativistic jet and typical collapsar parameters,
we find that significant electromagnetic dissipation
may be avoided until it proceeds catastrophically near the jet photosphere at large radii
($r\sim 10^{13}$--$10^{14}{\rm cm}$),
by which the jet
obtains a high Lorentz factor ($\gamma\sim 100$--$1000$),
has a luminosity of $L_j \sim 10^{50}$--$10^{51}\ergs$,
has observer variability timescales of order $1$s (ranging from $0.001$-$10$s),
achieves $\gamma\theta_j\sim 10$--$20$ (for opening half-angle $\theta_j$) and so is able to produce jet breaks,
and has comparable energy available for both prompt and afterglow emission.
A range of model parameters are investigated and simplified scaling laws are derived.
This reconnection switch mechanism allows
for highly efficient conversion of electromagnetic energy into prompt emission
and associates the observed prompt GRB pulse temporal structure
with dissipation timescales of some number of reconnecting current sheets embedded in the jet.
We hope this work helps motivate the development of
self-consistent radiative compressible relativistic reconnection models.

\end{abstract}


\section{Introduction}
\label{sec_intro}

Gamma-ray bursts (GRBs) are thought to originate from
core-collapse events or compact object mergers
leading to magnetars or accreting black holes capable of launching ultrarelativistic jets.
The prompt emission from standard cosmological long-duration GRBs
has an energy of about $10^{51}\erg$ over a few seconds
that is beamed into a jet with an opening half-angle of a few degrees \citep{Frail:2001:BGR,Bloom:2003:GRB}.
The prompt emission is typically presumed to occur in internal shocks \citep{sp97}.
The internal shock model is reasonable because such shocks are expected in an unsteady outflow,
which then has an observed variability on timescales related to the central engine.
Within the collapsar \citep{woosley_gamma_ray_bursts_1993,pac98,mw99} or other GRB models,
the observed variability may arise
indirectly from activity near the central compact object,
indirectly from entrainment driving propagation instabilities \citep{aloy00b,zhang04,mlb07,waz08,bucc08},
or directly from relativistic turbulence \citep{lyutikov_grbs_2003,narayankumar09,zmw09},

However, the internal shock model has some unresolved problems.
For example,
highly relativistic relative motion between interacting shells is required
in order to efficiently generate photons \citep{kobayashi01,maxhamzhang09} ;
only a small fraction of electrons should be accelerated
in order to obtain consistency with the observed peak energy \citep{shenzhang09} ;
internal shocks produce a steeper spectral slope than observed \citep{ghisellini00,asano09} ;
the afterglow energy would dominate the prompt energy as opposite to observed \citep{willingale07} ;
particle-in-cell (PIC) simulations show that shocks are dominated in energy by Maxwellian electrons
whose emission may be inconsistent with the observed double power-law Band function \citep{gianniosspitkovsky09} ;
and if GRB jets contain strong toroidal fields, then the shocks
should inefficiently dissipate the kinetic energy \citep{kc84,zk05,nkt11}
and inefficiently accelerate particles to high energies \citep{sironi09}.
One alternative to the internal shock model includes
dissipation from nuclear and Coulomb collisions \citep{bel10a}.

In light of these issues with shock models and since GRB jets are expected to be highly magnetized,
it is interesting to explore alternatives for which electromagnetic
dissipation directly leads to acceleration and emission
\citep{rl92,thompson94,usov94,levinson97,lyutikov01,lyutikov06,thompson06}.
Such investigations are motivated by studies of
pulsar winds \citep{kc84,coroniti90,lyubarskykirk01,kirkskj03,nagata08},
field reversals in active galactic nuclei (AGN) jets \citep{lovelace94,lnr97},
and magnetars \citep{td95,lyutikov03,lyutikov06b}.
Electromagnetic dissipation mechanisms are appealing
because they do not require the generation of
highly relativistic relative motion between shells of matter
or highly relativistic turbulent motion.
Dissipation can instead proceed in situ.
Indeed, high-resolution 3D magnetohydrodynamical (MHD) simulations
of jet propagation show that toroidal magnetic fields shield jets against many of the shear instabilities
noticed in purely hydrodynamical simulations \citep{kmbv09,mignone10},
which suggests that it is difficult for a magnetized jet with a strong toroidal field
to produce numerous shells moving at varying relativistic speeds
as required by the internal shock model to generate temporal variability and efficient dissipation.
Further, modern simulations of magnetized GRB jets
suggest that even if relatively relativistic shells were generated,
internal shocks would be unable to explain the high efficiency of prompt GRB emission \citep{nkt11}.
This implies that local electromagnetic dissipation may be the only efficient dissipation mechanism
possible in highly-magnetized jets.

Further, several observations point to a requirement
of strong electromagnetic fields in GRB jets.
The typical absence of a thermal photospheric emission component
requires the jet to contain a significant low-entropy (e.g. magnetized) component,
and the presence of non-pair-producing GeV photons
requires emission to be at larger radii than expected
in the internal shock model \citep{zhangpeer09}.
The typical non-detection by Fermi of a GeV spectrum excess due to
synchrotron self-Comptonization (SSC) in prompt GRBs
can be achieved by a highly magnetized jet since fewer
electrons are required to support a strong electromagnetic field,
as compared to the number of electrons required by the internal shock model
that should generate an SSC component \citep{fan09}.
Possible measurements of highly polarized gamma-rays from GRBs
could require an ordered magnetic field in the emitting region \citep{lpb03}.
The absence of bright optical flashes in the very early afterglow may
require the reverse shock region to be somewhat magnetized \citep{mizuno09,mga09}.

While electromagnetic fields are commonly understood to be dynamically
important in the jets produced by GRB engines \citep{npp92,thompson94,usov94,vla03a,lyutikov06,um06,um07},
only recently have self-consistent magnetohydrodynamic (MHD)
simulations been able to model the formation and
large-scale axisymmetric structure of such jets \citep{kvkb09,lyu10a},
including what happens beyond a stellar envelope
in the collapsar model \citep{tnm09}.
Such large-scale idealized MHD jet simulations are complemented by
small-scale general relativistic MHD (GRMHD) axisymmetric and 3D simulations
that consider the role of magneto-rotational instability (MRI) driven turbulence \citep{bal91}
and magnetic field geometries close to the black hole \citep{mck06jf,bhk08,mb09}
and by realistic GRMHD simulations of the engine in the collapsar model \citep{mizuno04,barkov08,nagataki09}.
Of relevance to GRBs is that a dipolar field appears
required to launch a jet from the black hole \citep{bhk08,mb09},
yet magnetic field advection from a presupernova core collapsing onto a black hole leads to too small
of a magnetic flux to generate the power of a cosmological GRB \citep{kombar09}.
The MRI must then be invoked \citep{Akiyama:2003:MIC}, but that only generates small-scale (not dipolar) field.
One solution is for a magnetic dynamo to generate quasi-periodic large-scale dipolar fields
over tens of dynamical times as occurs for toroidal fields in accretion disks \citep{dsp10},
but this necessarily implies one must understand electromagnetic dissipation at the interface between each
successive flip of the jet's dipolar field.

However, it remains difficult to understand how
the electromagnetic field dissipates in GRB jets.
The most advanced simulations of relativistic jets from collapsing
stars forming black hole accretion flows use the MHD approximation \citep{kombar09},
so they cannot incorporate all the plasma effects observed in PIC simulations
that are required to self-consistently study magnetic reconnection.
Even the most advanced PIC simulations are inapplicable to GRB jets
that require relativistic reconnection in the presence of pair creation and annihilation, neutrinos, photons,
radiative cooling in both optically thick and thin regimes, relativistic compression, etc.
There is still no coherent physical theory of relativistic reconnection
(see, e.g., \citealt{bf94,lu03,lyubarsky05,jh09}; and also section 5.5 of \citealt{um06}).
Therefore, quasi-analytical approaches must be used
to investigate the role of electromagnetic dissipation in GRB jets.

\begin{figure}
  \begin{center}

      \includegraphics[width=2.5in]{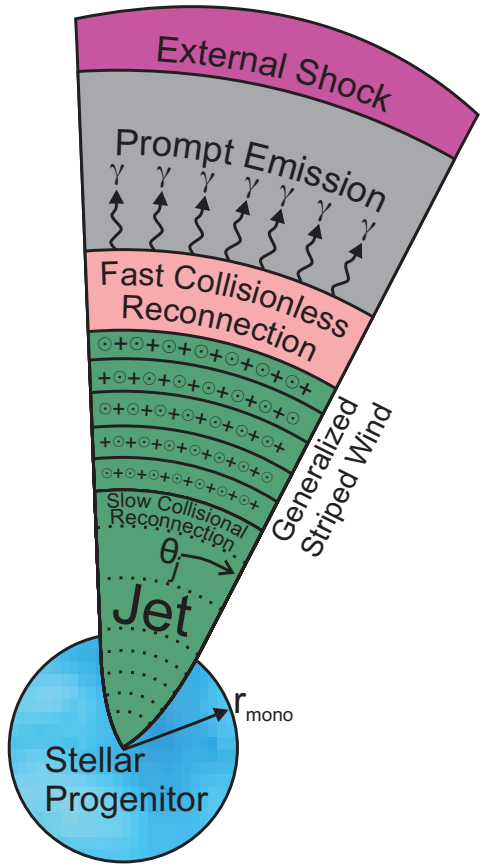}
  \end{center}
  \caption{
  Reconnection switch concept:
  Collapsar model or some other system produces
  a jet (with opening half-angle $\theta_j$) corresponding to
  a generalized stripped wind containing many
  field reversals that develop into dissipative current sheets.
  The jet collimates due to confinement by the stellar envelope
  out to a radius $r_{\rm mono}$, after which the jet (having become ultrarelativistic)
  cannot expand laterally and becomes a nearly radial (monopolar) flow.
  The figure shows toroidal field polarity reversals produced by dynamo processes
  near the central engine or by turbulent entrainment at boundary layers.
  In the reconnection switch model,
  these current sheets avoid significant dissipation while in the collisional regime
  until they reach sufficiently large radii where the plasma becomes collisionless
  and fast collisionless reconnection is triggered.
  The curved dotted lines denote that the striped wind continues down to the central engine.
  This paper focuses on the dissipation processes within current sheets
  that ultimately lead to prompt emission.
  }
  \label{fig_generalpicture}
\end{figure}

In prior quasi-analytical work, electromagnetic dissipation through magnetic reconnection
in GRB jets has been suggested as a possible source of
acceleration and emission at large radii from the central compact object
\citep{thompson94,sdd01,drenkhahn02,lyutikov06,giannios06}.
It was considered problematic for ideal MHD processes
to efficiently accelerate and collimate jets,
so non-ideal electromagnetic dissipation was favored to accelerate GRB jets \citep{thompson94,drenkhahn02}
or be a source of energy for emission in GRB jets \citep{thompson94}.
Prior works also assume that the flow is already relativistic ($\gamma\gg 1$)
and continues to speed-up with radius as based upon one-dimensional jet models.
The dissipation rate was assumed to be~\alfic~or some fraction of it,
which assumes a fast reconnection mode.

However, it is now known that
a relativistic quasi-conical outflow (i.e. split-monopole near compact object and otherwise unconfined)
accelerates efficiently near the polar axis in ideal MHD \citep{tmn09},
and any sufficiently globally collimated outflow is efficiently accelerated \citep{kvkb09,tnm09}.
This dramatically changes the behavior of the Lorentz factor as a function of radius
compared to prior one-dimensional models (for a review, see \citealt{spruit10}.)
In addition, jets following ideal MHD acceleration near the compact object
start with $\gamma\approx 1$ until reaching the~\alf~surface.
So, if dissipation were always nearly~\alfic, then (unabated by time dilation)
fast reconnection could occur near the jet base.
Reconnection near the jet base can severely quench the Blandford-Znajek effect \citep{bhk08,mb09},
which may be required to drive the relativistic electromagnetic jet in the first place.
Also, if reconnection were quite close to~\alfic, then the jet could dissipate
deep inside the photosphere leading to a hot fireball instead of an electromagnetic jet.
This would lead to a dominant thermal spectrum, which is inconsistent with observations.

Prior quasi-analytical works have also investigated
the potential role of the current-driven magnetic kink instability,
which is assumed to operate rather efficiently on~\alfic~timescales \citep{sdd01,giannios06}.
However, modern stability analyses have shown that the kink instability
operates somewhat slower than comoving~\alf~timescales for collimating flows \citep{nlt09},
and much of a relativistic quasi-conical flow can be out of causal contact with itself
to eliminate the instability except within narrow regions right around the polar axis
that contain little electromagnetic energy flux \citep{sdd01,mb09}.

Prior work also considered other aspects of electromagnetic dissipation.
For example, a violation of ideal MHD occurs when the plasma density drops
below the Goldreich-Julian density, and this leads to dissipation \citep{lyutikov01}.
However, for reasonable lower-limits on the baryon-loading \citep{le03},
such a violation of ideal MHD only occurs at $r\gg 10^{19}{\rm cm}$ \citep{sdd01}.
Reconnecting layers in accelerated jets have been found to be potentially unstable
to magnetic interchange instabilities leading to enhanced reconnection \citep{lyu10b},
although they considered a sharp pressure boundary
instead of a distributed radiative photosphere as present in GRB jets.
Turbulence-driven electromagnetic dissipation from magnetic instabilities induced by internal shocks
has also been recently considered \citep{zy10},
although the conditions for turbulence in highly magnetized jets remains undetermined.

One aspect of electromagnetic dissipation not investigated so far is the 
role of critical transitions between collisional and collisionless
reconnection and its association with, respectively,
rather slow resistive-MHD reconnection
(perhaps as slow as Sweet-Parker reconnection; \citealt{sweet58,parker57})
and fast Petschek or Petschek-like reconnection \citep{petschek64}.
(For a review, see \citealt{priestforbes00}.)

Prior simulations and laboratory experiments show
that collisional plasmas avoid the fast Petschek type reconnection
in favor of the slow Sweet-Parker type reconnection.
For collisional plasmas the Petschek reconnection configuration
promptly collapses to a thin Sweet-Parker layer \citep{uk98,uk00}.
No fast (inflow velocity of order the~\alf~velocity) Petschek reconnection
is seen to occur in collisional plasmas \citep{biskamp86,mb96,ji98,uk00, bj03,mlk05,kulsrud01}.

On the other hand, a collisionless plasma allows for
non-ideal MHD effects that can force a plasma into a fast
Petschek-like reconnection regime occurring on the ion skin depth scale
and operating independently from the classical resistivity.
(For a review, see \citealt{nla.cat-vn2113803,kulsrud05,zweibel09}.)
Simulations of pure pair plasmas also show fast reconnection but on the electron skin depth
(instead of the ion skin depth) as perhaps due to the electron pressure effect \citep{besshobhatt07}.
Petschek reconnection plays an important role in the Sun,
where nanoflares remain the most plausible source of coronal heating \citep{parker88,klimchuk09}.
In fact, the solar corona may even exist in a balanced marginally-collisionless state
governed by the density-controlled transitions between these two modes of reconnection \citep{uzdensky2007b,uzdensky07c}.
There is also observational evidence that the transitions between the collisional and collisionless reconnection regimes may also
control coronal heating in black-hole accretion disks \citep{gu08}.
Solar and stellar flares could also be triggered by these critical transitions \citep{cassak06,cassak08}.
Therefore, it is important to determine whether a reconnecting system is collisional or collisionless
in order to determine whether current sheets operate in the fast reconnection mode
\citep{mlk05,cassak05,yamada06,uzdensky2007b,uzdensky07c,besshobhatt07}.

There do exist purely resistive-MHD alternatives to fast collisionless Petschek reconnection,
including due to externally driven MHD turbulence  \citep{lv99,kowal09,loureiro09}
and plasmoid-dominated reconnection \citep{lsc07,kowalreview09,samtaney09, cassak09,huang10,uls10}.
However, relativistic Poynting-flux dominated jets are
strongly electromagnetically-dominated and relativistically expanding flows,
and hence they may only be susceptible to turbulence or similar cascades in regions
that remain in causal contact across the jet -- such as very close to the polar axis
where there is little total electromagnetic energy and so little possible emission.
These alternatives require further consideration in such relativistic regimes.

The present paper builds upon three novel key elements.
The first key element of this work is that
we consider the role of transitions between slow collisional and fast collisionless reconnection.
The fact that the reconnection rate depends upon the collisionality of the plasma
is particularly important for GRB jets that are collisional at the jet base and out to large radii.
We compute the properties of the collisional and collisionless reconnecting layers
to determine which reconnection mode dominates at each radius and angle within the GRB jet.
Once collisionless reconnection is initiated at some ``transition radius,''
reconnection proceeds at a rate much faster than the collisional rate
and can initiate prompt GRB dissipation and emission.

The second key element of this work is that
the collisional layer's properties are determined as due to collisions among species
of ions, photons, electrons, positrons, and neutrinos.
The current layer structure is treated similarly
as done for radiative accretion disks (see, e.g., \citealt{knp05}).
We also make use of relativistic reconnection work by \citet{lyubarsky05},
which shows that the relativistic Sweet-Parker layer
has a reconnection rate as expected from non-relativistic theory.
We incorporate radiative effects into the reconnection switch model by following \citet{um11},
who showed that the current layer compresses in the strong cooling limit
of non-relativistic radiative reconnection.

The third key element of this work is that the large-scale jet
is modelled using the latest ideal MHD models of ultrarelativistic jets
rather than assuming inefficiently accelerated one-dimensional flows.
We also extend the study of \citet{sdd01}
and determine a range of possible ways for field polarity reversals to occur
that lead to current sheets in the substructure of the jet.

A basic version of the overall argument of the paper is provided in \S\ref{sec_argument}.
The GRB jet structure is described in \S\ref{sec_jetstructure},
some possible field substructures are described in \S\ref{sec_substructure},
the reconnection models are described in \S\ref{sec_reconnectionphysics},
results for GRBs and other jet systems are described in \S\ref{sec_jetdissipation},
a discussion is in \S\ref{sec_discussion},
and conclusions are provided in \S\ref{sec_conclusions}.
In Appendix~\ref{sec_fulljetstructure}, the generalized full jet structure solution
that is used throughout the paper is obtained.
In Appendix~\ref{baseeos}, the equation of state for all species
within the radiative current layer is presented.
In Appendix~\ref{sec_collisionalcollisionlessreconnection},
the collisional and collisionless reconnection models are discussed.
We assume a flat space-time in spherical polar coordinates ($r,\theta,\phi$),
an orthonormal basis for all vectors,
and Gaussian-cgs-Kelvin-radian units.

\section{Basic Argument}
\label{sec_argument}

In this section, a basic argument is presented to demonstrate the existence of a reconnection
switch mechanism that leads to dissipation near the GRB jet photosphere.

Figure~\ref{fig_generalpicture} shows a basic picture of the reconnection switch model
for a jet containing multiple current sheets each corresponding
to a layer wherein oppositely-directed magnetic field lines are dissipated.
Near the central engine, dissipation proceeds via slow collisional reconnection
due to the high collisional rate.
This allows the electromagnetic field to avoid significant dissipation despite the presence of current sheets,
and this allows the generation of a baryon-pure
strong electromagnetic field for launching an ultrarelativistic jet.
At large radii, pairs annihilate and the densities go down
leading to infrequent collisions.
Collisionless plasma effects (e.g. due to electron-proton or electron-positron decoupling)
can then initiate the much faster collisionless reconnection mode (operating in a Petschek-like geometry)
that disrupts the slow collisional mode (operating in a Sweet-Parker-like geometry) \citep{cassak05}.
This leads to a {\it reconnection switch} that triggers jet dissipation,
which (as shown below) initiates prompt emission near the GRB jet photosphere.
As mentioned in the introduction,
the concept of a reconnection switch has been applied to many astrophysical phenomena,
and in this paper we simply apply the same concept to GRB jets.

The relevant length scale for fast collisionless reconnection
in baryon-dominated plasmas is the proton skin depth
\begin{equation}
d_p = \frac{c}{\omega_{pp}} ,
\end{equation}
for proton plasma frequency $\omega_{pp}=\sqrt{4\pi n_p e^2/m_p}$,
proton number density $n_p$, charge $e$, proton mass $m_p$, and speed of light $c$.
The relevant scale for pair-dominated plasmas is the pair skin depth
\begin{equation}
d_e = \frac{c}{\omega_{pe}} ,
\end{equation}
with pair plasma frequency $\omega_{pe}=\sqrt{4\pi n_{e,\rm tot} e^2/m_e}$,
electron+pair number density $n_{e,\rm tot} = n_e + n_{\rm pairs}$,
and electron mass $m_e$.

Collisions ensure that resistive MHD applies,
which forces the current layer to avoid the fast Petschek-like geometry
in favor of a Sweet-Parker-like geometry (see., e.g., \citealt{uk00}).
The Sweet-Parker solution requires pressure equilibrium across the current sheet.
Assume, as valid for most of this paper,
that the reconnecting field is not significantly weaker than the guide field.
Then, the Sweet-Parker solution without a guide field
can be used to estimate any quantities to order unity.
The electromagnetic pressure ($p_{\rm EM}=u_{\rm EM}$, where $u_{\rm EM}$ is the electromagnetic energy density)
and thermal gas pressure
(which, as borne out in this paper, is dominated by photon pressure $p_\gamma = u_\gamma/3$,
where $u_\gamma$ is the photon energy density) balance via
\begin{equation}
p_{\rm EM} \sim p_\gamma .
\end{equation}
Next, assume, as also borne out in this paper,
that the collisional resistivity against current-carrying electrons and positrons
is dominated by Compton scattering,
then the corresponding magnetic diffusivity is
\begin{equation}
\eta \approx (4/3)d_e^2 (u_\gamma \sigma_T c)/(m_e c^2) ,
\end{equation}
\citep{gu08}, for Thomson scattering cross section $\sigma_T$.
Then, for electromagnetically-dominated jets with relativistic~\alf~speeds,
the collisional Sweet-Parker current sheet thickness is
\begin{equation}
\delta_{\rm SP}\sim \sqrt{\frac{L_0 \eta}{c}} ,
\end{equation}
where $L_0$ is the comoving length of the current layer
as estimated by the scale for variations in the electromagnetic field.

The fast Petschek-like collisionless reconnection mode takes over
when $d_p\gtrsim \delta_{\rm SP}$ for baryonic-dominated plasmas, as discussed above.
Using the above equations, the condition for fast collisionless reconnection
is then given by
\begin{equation}\label{simplecondition}
1 \lesssim \left(\frac{d_p}{\delta_{\rm SP}}\right)^2 \sim \frac{(1+n_{\rm pairs}/n_e)^2}{8\tau_{\gamma,\rm sca} \tilde{\mu}} ,
\end{equation}
where $\tilde{\mu} = u_{\rm EM}/(\rho_b c^2)$
is the electromagnetic energy per baryon rest-mass energy in the jet,
$\rho_b \approx 2 m_p n_p$ is the baryon rest-mass density for a proton number density $n_p$
assuming protons and neutrons are at equal number densities,
$n_e=n_p$ is the baryonic-associated electron density from charge neutrality,
$\tau_{\gamma,\rm sca} \approx n_{e,\rm tot} \sigma_T L_0$ is the scattering optical depth,
and $L_0$ corresponds to the density scale-height for photons, electrons, and positrons
for an emitting slab of length $L_0$ \citep{popham95}.
For a pair-dominated plasma in the limit that baryons play no role,
the condition for fast reconnection becomes
\begin{equation}
1\lesssim \left(\frac{d_e}{\delta_{\rm SP}}\right)^2 \sim \left(\frac{m_e}{m_p}\right)\frac{(1+n_{\rm pairs}/n_e)}{8\tau_{\gamma,\rm sca} \tilde{\mu}} .
\end{equation}
Future studies can clarify how fast reconnection operates
in a plasma with baryons that typically dominate the rest-mass energy
even if pairs have a non-negligible number density.

Assume $n_{\rm pairs}\lesssim n_e$,
then the condition for transition to fast reconnection simply becomes
\begin{equation}\label{verysimplecondition}
\tau_{\gamma,\rm sca} < (8\tilde{\mu})^{-1} .
\end{equation}
Because MHD jets for GRBs have $\tilde{\mu}\sim 1$ at large radii \citep{tnm09},
{\it fast reconnection in GRB jets is predicted to occur near the photosphere}.
Such a dissipation is required by dissipative photosphere models (see, e.g., \citealt{rm05}).
Further, once fast dissipation starts, the value of $\tilde{\mu}$ drops,
which forces the fast collisionless reconnection condition to be maintained.
Even if fast reconnection were relativistic with speed order $c$,
dissipation is still suspended until large radii so that
a baryon-pure ultrarelativistic jet can form.

The rest of this paper computes
the radius where the reconnection switch occurs,
the photon optical depth at this radius to determine
whether quasi-thermal photospheric emission is possible,
and some other observables.
Realistic MHD GRB jet models are used since the polar jet's ideal MHD acceleration is substantially
different than considered in prior reconnection jet models.
The important pair contribution is non-trivial to compute
because near the photosphere the photons and pairs are only marginally optically thick.
Also, the plasma temperature near the transition to fast reconnection
ends up low enough that pairs are deep in the suppressed regime.
A simple exponential suppression factor would lead to more than an order of magnitude
error in the transition radius, so pairs must be treated more accurately.
Despite pairs being in the suppressed regime, the pair number density can
greatly exceed the baryonic-associated electron number density
and so crucially affect the transition radius as seen from Equation~(\ref{simplecondition}).
We must also consider neutrino cooling near the compact object,
because strong cooling can lead to an effective drop in $\delta_{\rm SP}$ \citep{um11}
and so trigger fast reconnection causing the jet to dissipate before it is launched.
Overall, this requires us to compute the properties of a jet filled
with a complex of multiple reconnecting slabs,
each consisting of baryons with electrons, pairs, and neutrinos at arbitrary optical depths.
This is achieved by treating the current layer structure
as similarly done for radiative accretion disks (see, e.g., \citealt{knp05}).

\section{Large-Scale Jet Structure}
\label{sec_jetstructure}

In this section, the large-scale axisymmetric structure of relativistic
GRB jets from black holes or magnetars is presented.
The large-scale jet structure acts as a key constraint on the reconnection physics
by fixing the radial dependence of magnetic pressure.
During reconnection, the magnetic pressure balances the thermal pressure within current layers,
which determines the contributions of baryons, photons, pairs, and neutrinos.

\begin{figure}
  \begin{center}

      \includegraphics[width=2.5in]{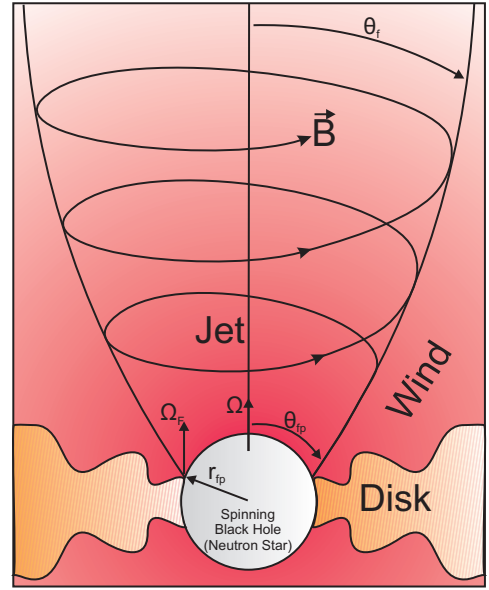}
  \end{center}
  \caption{
  GRB jet structure near the central engine:
  A relativistic MHD jet is driven by a
  compact object with rotation frequency $\Omega$
  that causes the field line at foot point radius $r_{\rm fp}$ and opening half-angle $\theta_{\rm fp}$
  to have a field line rotation frequency $\Omega_{\rm F}$ (offset from the axis for clarity).
  At larger radii, each field line follows a collimating trajectory with opening half-angle $\theta_{\rm f}$,
  while the entire jet has an opening half-angle $\theta_j$ corresponding to the largest $\theta_{\rm fp}$
  allowed by the central engine's accretion disk, corona, or wind.
  }
  \label{fig_jetaxi}
\end{figure}

Figure~\ref{fig_jetaxi} shows the basic elements of the central engine
and the production of an axisymmetric jet.
The jet could be confined laterally (up to a radius $r_{\rm mono}$) by
an accretion disk, disk corona, disk wind, stellar envelope, or some ambient medium.
A generalized jet solution that applies for both small and large radii is obtained
in Appendix~\ref{sec_fulljetstructure},
which collects together results from recent analytical works and simulations \citep{tmn08,tmn09,tnm09}.
However, the full solution can be cumbersome for obtaining simple scaling laws
for how the results depend upon model parameters.

Therefore, this section outlines a simplified jet structure that
approximates the full jet structure model
and helps to highlight the basic elements of the full solution.
The radius is assumed to be much larger than the size of the compact object (i.e. $r\gg r_{\rm fp}$),
at which a field line attaches at a footpoint radius ($r_{\rm fp}$) and at a footpoint angle ($\theta_{\rm fp}$)
across the surface of the compact object (see Figure~\ref{fig_jetaxi}).
The radius is also assumed to be larger than the deconfinement radius (i.e. $r\ge r_{\rm mono}$).
The following expressions apply for $0\le \theta \le \pi/2$ with an assumed equatorial symmetry.

The degree of collimation inside $r_{\rm mono}$ is given by a parameter $\nu$,
such that for angles $\theta\ll 1$ (i.e. cylindrical radius $R\ll r$)
one can show that the field lines obey
\begin{equation}\label{rofth}
\frac{r}{r_{\rm fp}} \approx \theta_{\rm f}^{-2/\nu}\left(2\sin{\frac{\theta_{\rm fp}}{2}}\right)^{2/\nu} ,
\end{equation}
where the field collimates with $2>\nu>0$ out to $r_{\rm mono}$
after which it follows the $\nu=0$ type radial geometry on a particular field line with fixed opening half-angle of $\theta_f$ giving
\begin{equation}\label{thetaofrmono}
\theta_{\rm f} \approx \left(\frac{r_{\rm mono}}{r_{\rm fp}}\right)^{-\nu/2} \left(2\sin{\frac{\theta_{\rm fp}}{2}}\right) ,
\end{equation}
where the total jet opening angle ($\theta_j$) corresponds to the value of $\theta_{\rm f}(\theta_{\rm fp})$
for the largest value of $\theta_{\rm fp}$ that is allowed
by the presence of a confining medium near the compact object.

For $r\ge r_{\rm mono}\gg r_{\rm fp}$,
the radial field strength is
\begin{equation}
B_r \approx B_{r,\rm fp} \left(\frac{r_{\rm mono}}{r_{\rm fp}}\right)^{\nu-2} \left(\frac{r}{r_{\rm mono}}\right)^{-2} .
\end{equation}
The $\theta$-component is small and so can be neglected.
The toroidal $\phi$-component is
\begin{equation}
B_\phi \approx B_{r,\rm fp} \left(\frac{-2r_{\rm fp}\Omega_{\rm F}}{c}\right) \left(\frac{r_{\rm mono}}{r_{\rm fp}}\right)^{\nu-1}  \left(\frac{r}{r_{\rm mono}}\right)^{-1} \tan{(\theta_f/2)} ,
\end{equation}
where for rapidly rotating black holes or neutron stars $r_{\rm fp}\Omega_{\rm F}\lesssim 0.25 c$
and $\Omega_{\rm F}$ is the field rotation frequency one can set at each foot point.
For a black hole angular rotation rate of $\Omega=\Omega_{\rm H} = (j c)/(2 r_{\rm H})$
with horizon radius $r_{\rm H}$ and dimensionless black hole spin $j$,
the value of the field line angular rotation rate is $\Omega_{\rm F}\approx \Omega_H/2$.
For Lorentz factor of $\gamma\gg 1$, one can show that $b^2\approx B_\phi^2/\gamma^2$,
which allows one to obtain the electromagnetic pressure and energy density via $p_{\rm EM}=u_{\rm EM}=b^2/(8\pi)$,
where $|b|$ is the comoving field strength.
The baryonic rest-mass density is
\begin{equation}
\rho_b \sim \rho_{b,\rm fp} \left(\frac{B_r}{B_{r,\rm  fp}}\right) \left(\frac{1}{\gamma}\right) ,
\end{equation}
where $\gamma$ is some estimate of the Lorentz factor for $r\ge r_{\rm mono}$.
We define a magnetization parameter as
\begin{equation}
\zeta\equiv \frac{B^2_{r,\rm fp}}{8\pi\rho_{b,\rm fp}c^2} ,
\end{equation}
which is used to define $\rho_{b,\rm fp}$ from $B_{r,\rm fp}$.
Near the compact object, the value of $\zeta$ is similar to the electromagnetic energy per particle.

Overall, for a given $\zeta$, $B_{r,\rm fp}$, $r_{\rm mono}$, $\theta_{\rm fp}$,
and an estimate of $\gamma$, one can determine the radial and angular dependence of
$b^2$, $\rho_b$, $\theta_f$, and $\theta_j$,
and other quantities defined in Appendix~\ref{sec_fulljetstructure}; such as
the electromagnetic energy per unit rest-mass-energy ($\tilde{\mu}$),
the electromagnetic energy flux per unit rest-mass flux ($\mu$),
the electromagnetic energy flux per unit mass-energy flux ($\sigma$),
and the jet power ($P_j$).
This MHD jet solution only has to be applicable up to the radius where significant dissipation occurs,
because determining the dissipation radius is the primary goal of this work.

For all models, we choose a rapidly rotating black hole
foot point radius of $r_{\rm fp}\approx 4.4{\rm km} (M_{\rm BH}/\msun)$
($M_{\rm BH}$ is black hole mass)
with a rapid rotation rate such that $r_{\rm fp}\Omega_{F,\rm fp}=0.25c$.
The collapsar case with a black hole of mass $M_{\rm BH}=3\msun$ is chosen.
For the collapsar model $r_{\rm mono}\approx 3\times 10^{10}{\rm cm}$ is typically chosen
(i.e. radius of the progenitor star),
while for the short-duration GRB compact object merger model
$r_{\rm mono}\approx 1.2\times 10^7{\rm cm} = 120 {\rm km}$ is chosen
(i.e. extent of newly formed disk-corona-wind that helps collimate the jet ;  \citealt{tmn08,tnm09}).

A black hole is chosen instead of a neutron star
because it is more likely to be a generator of a powerful baryon-pure jet.
First, the black hole cleans magnetic field lines of mass \citep{macdonald_thorne_bh_forcefree_1982,le93},
so the magnetization can be quite high in the black hole case.
The magnetization $\mu$ could be limited by neutron diffusion \citep{le03},
which still leads to quite high magnetizations of $\mu\sim 10^3$--$10^4$ \citep{mckinney2005a}
(corresponding to $\zeta\sim 10^3$--$10^4$).
Second, a black hole can produce a more powerful jet since its Kerr parameter is $j\sim 1$.
For the magnetar case, the outflow only becomes highly magnetized
at late time when the power has significantly diminished,
and neutron stars have a Kerr parameter only up to $j\sim 0.6$ before break-up.
These issues make it potentially more difficult for the magnetar
to operate as both an efficient and powerful engine
of a highly magnetized ultrarelativistic jet \citep{mtq07,bqamt08,mgtbq10}.

One must use Appendix~\ref{sec_fulljetstructure}
to obtain an accurate dependence for all MHD quantities (i.e. including $\gamma$)
within the jet as a function of radius and angle.
Consider typical collapsar model parameters that would lead to an ultrarelativistic jet with $\gamma\sim 1000$.
If $\zeta=10^4$, $B_{r,\rm fp}=3.2\times 10^{15}$G, $r_{\rm mono}=3\times 10^{10}$cm, and $\theta_{\rm fp}=\pi/2$,
then $\mu\approx 5400$,
$\gamma(r=10^{14}{\rm cm})\approx 800$,
$P_j(r=r_{\rm fp})\approx 2.2\times 10^{51}\ergs$,
$[\gamma\theta](r=10^{14}{\rm cm})\approx 18$,
and $\sigma(r=10^{14}{\rm cm})\approx 6$.
At $r=10^{14}{\rm cm}$ there is about $\approx 6$ times less kinetic energy than electromagnetic energy,
which can be tapped for prompt GRB emission if there exists some mechanism to dissipate the energy.

\section{Jet Field Substructure: Generalized Striped Wind}
\label{sec_substructure}

In this section, we consider the process
whereby small-scale field reversals and current sheets become
embedded within the large-scale jet structure.
The comoving length scale, $\Delta_0$, of these jet field substructures
plays a prominent role in later calculations
because the dissipation rate due to {\it collisional} reconnection
is dominated by the smallest value of $\Delta_0$.

\begin{figure}
  \begin{center}

      \includegraphics[width=2.5in]{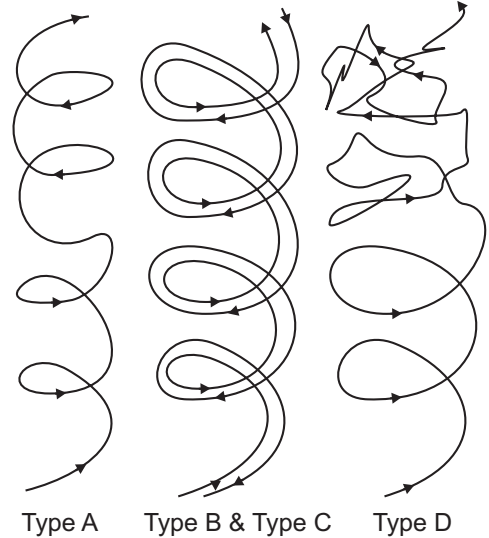}
  \end{center}
  \caption{
  Generation of electromagnetic field reversals:
  Different jet field geometries at the jet base lead to current sheets with different orientations.
  Type A corresponds to a time-dependent polarity for an axisymmetric dipolar field.
  Type B corresponds to a time-independent non-axisymmetric multipolar field.
  Type C corresponds to a time-independent axisymmetric multipolar field.
  Near the jet base, types B and C are similar except the alternating field polarities
  are displaced either in the $\phi$-direction (type B) or the $\theta$-direction (type C).
  Type D corresponds to a dipolar field that is unstable at large radii.
  All types are expected to some degree.
  }
  \label{fig_fieldgeom}
\end{figure}

Figure~\ref{fig_fieldgeom} shows the different types of field geometries that are considered.
All of these substructure types probably coexist to some degree.
The jet structure solution presented in section~\ref{sec_jetstructure}
was formally constructed with a single
polarity for the electromagnetic field where the return field is presumed
to connect back to the region slightly beyond the compact object.
However, the polarity of the field can have arbitrary reversals without changing
the large-scale force balance of the jet as long as the
current sheets are inserted so that only the field direction changes.
For an assumed field perturbation of approximately $\exp{(i(\vec{k}\cdot \vec{R}_j - \Omega_{\rm F} t))}$
with cylindrical radius $R_j$ for the entire jet,
Maxwell's equations give an overall smallest characteristic electromagnetic field length scale
of $\Delta\sim 1/(1/(\pi R_j) + \Omega_{\rm F}/(2\pi c))$.
The comoving length scale ($\Delta_0$) generally depends upon the substructure type.

Substructure type A corresponds to a time-dependent alternating polarity
for a dipolar field near the compact object.
The injection of field reversals occurs at an angular rate of $\sim m \Omega_{\rm F}$ near the compact object
for some spherical harmonic quantum number $m\ge 0$.
Such an injection can occur when
the compact object's rotation axis is misaligned with the magnetic field axis
as in the standard striped wind \citep{coroniti90,thompson94,sdd01,pgll10},
a neuron star's internal large-scale poloidal field dynamo flips the sign of the field,
the compact object is fed alternating polarity from the accretion disk \citep{lnr97},
or a large-scale poloidal field accretion disk dynamo determines the field near the black hole.
Values of $0<m<1$ are allowed since a magnetic dynamo can operate
on timescales much longer than the dynamical timescale.
For example, an accretion disk could have $m=0.1$ \citep{dsp10}.

Substructure type B corresponds to a time-independent non-axisymmetric ($m\ge 1$)
multipolar field threading the compact object or accretion disk \citep{Akiyama:2003:MIC}.
The length scale of the electromagnetic field is determined by reversals
at different foot point angles in the $\phi$-direction near the compact object,
and so type B is also related to the angular rate of $\sim m \Omega_{\rm F}$.

At large distances, substructure types A and B both have field reversals
that are radially stacked such that the lab-frame $\Delta = 2\pi c/(m\Omega_{\rm F})$.
In the lab-frame, the pattern of reversals in $\Delta$
is advected to large radii at roughly the speed of light,
so that the comoving scale is $\Delta_0 \sim \Delta \gamma$ \citep{sdd01}.
There may be many ``stripes'' separated radially by~$\Delta_0$.

\begin{figure}
  \begin{center}

      \includegraphics[width=1.5in]{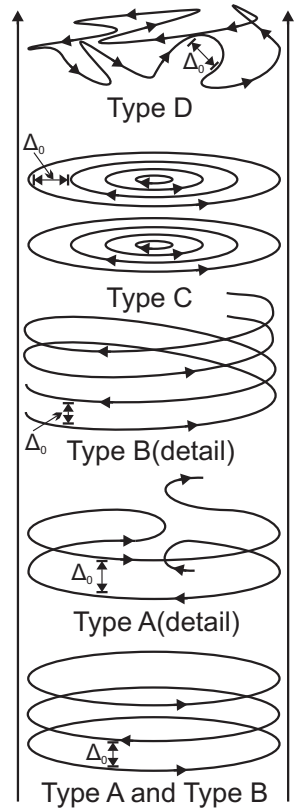}
  \end{center}
  \caption{
  Magnetic field reversals at large radii:
  Shows substructure types A-D at large radii where the toroidal field dominates
  and the flow carries many field reversals harboring current layers separated by length scale $\Delta_0$.
  Type A corresponds to time-dependent reversals with $\Delta_0\gtrsim \gamma (2\pi c/\Omega_{\rm F})$.
  Type B corresponds to a ``barber pole'' of reversals with
  $\Delta_0\sim \gamma  (2\pi c /(m \Omega_{\rm F}))$
  ($m\ge 0$ is the toroidal spherical harmonic quantum number).
  Type C corresponds to reversals in $\theta$ with $\Delta_0\sim (\pi R_j/l)$
  ($l>0$ is the poloidal spherical harmonic quantum number).
  Type D corresponds to reversals due to instabilities or turbulence at boundary layers
  and is some mixture of types A-C.
  In general, types A-D can be described by some particular (more generally a power spectrum)
  of $m\ge 0$ and $l>0$ to obtain $\Delta_0\sim 1/(l/(\pi R_j) + m\Omega_{\rm F}/(\gamma 2\pi c))$.
  }
  \label{fig_subBmany}
\end{figure}

Substructure type C corresponds to a time-independent axisymmetric
(e.g. spherical harmonic $l=1,2,3,\ldots$ corresponding to dipolar, quadrupolar, etc.)
multipolar field threading the compact object or disk.
Notice that $l$ cannot be zero because jet formation requires a non-zero axial field at the jet base.
The scale of the electromagnetic field is determined by reversals
at different foot point angles in the $\theta$-direction near the compact object.
The field reversal scale is $\Delta\sim \pi R_j/l$.
The scale of changes in the field is perpendicular to the flow direction
so that $\Delta=\Delta_0\sim \pi R_j/l$.
If $\gamma\theta\gtrsim 1$, then the jet becomes causally disconnected across in $\theta$
due to finite wave speeds for any finite magnetization.
The causally connected region has a size $\Delta_0\sim \pi R_j/(l\gamma\theta)$,
so physical models effectively have $l\ge \gamma\theta$ when using the original $\Delta_0$ formula.
The model described in figure 2 of \citet{rl92} corresponds to
this substructure type C with $l=2$ and with a prescribed structure in $\phi$.
Another example of type C is a magnetic tower model, where there is
a single large current sheet that separates two regions \citep{lynden96,um06,um07}.
This substructure is qualitatively different than substructure types A and B
because $\Delta_0\propto r$ while the other substructures do not depend upon $r$
once $\gamma$ is roughly constant.

Substructure type D corresponds to an ordered dipolar field at the jet base,
but the jet becomes unstable at large radii.
Such instabilities could be due to intrinsic jet properties
leading to the, e.g., current-driven magnetic kink instability
or due to turbulence at boundary layers such as the jet-wind or jet-envelope boundary.
For example, in the relativistically magnetized limit of MHD jets,
if the comoving toroidal field exceeds the comoving poloidal field by factors of a few to tens,
then the growth rate for the kink instability is $T\sim R_j/c$
generating structure from $\Delta_0\sim 2\pi R_j$ to $\Delta_0\sim r$ \citep{nlt09}.
Such kink modes would be stabilized by jet expansion or finite mass loading
leading to causal disconnection across much of the jet except
within $\theta\lesssim 1/\gamma$ \citep{sdd01}.
In general, the scale of the perturbations could be randomly oriented
across $\sim \pi R_j$ and generated on a time-scale of $2\pi/\Omega_{\rm F}$ in the lab-frame.
Hence, this substructure corresponds to some mixture of substructures A-C
for $m\ge 0$ and $l>0$ within $\Delta_0\sim 1/(l/(\pi R_j) + m\Omega_{\rm F}/(\gamma 2\pi c))$.

Figure~\ref{fig_subBmany} shows the jet substructures at large radii.
Types A and B are similar, but in detail they differ.
For types A-C, at all radii (including near the central compact object),
the comoving poloidal field is at most order unity comparable in magnitude
to the comoving toroidal field.
For types A-C, the toroidal field dominates in the lab and comoving
frames for $r\gg r_{\rm mono}$.
Generally, substructure D might correspond to having a strong poloidal field.
However, magnetic kink driven modes are expected to typically leave the jet's
comoving toroidal and poloidal fields as comparable
(i.e. when the so-called ``safety factor'' is of order unity in comoving frame),
because that is when the instability is initiated.
Overall, one concludes that any guide field (which does not reconnect)
in the current sheet is not much stronger than the reconnecting field.
This means the analysis in section~\ref{sec_argument} and in later
sections that assume a ``weak'' guide field are accurate to order unity or better,
because the Sweet-Parker analysis is only significantly modified
if the guide field is significantly stronger than the reconnecting field.

In summary, the jet can develop various substructures
that have comoving scale $\Delta_0$ satisfying
\begin{equation}
\frac{1}{\Delta_0(l,m)} \sim \frac{l}{\pi R_j} + \frac{m\Omega_{\rm F}}{\gamma 2\pi c} ,
\end{equation}
for some real numbers $l> {\rm max}(0,\gamma\theta)$ and $m\ge 0$.
Only dynamo and jet-medium interaction computations can determine (the power spectrum of) $l,m$,
but many dynamos and instabilities saturate at low $\{l,m\}\sim 1$.
The length scale ($\Delta_0$) over which the comoving electromagnetic field changes
is a natural scale over which any dissipation term enters into the comoving induction equation,
and so $\Delta_0$ plays a central role in determining
the comoving physics of reconnection discussed in the next section.
At large radii, these modes correspond to the jet having current sheets
with a guide field that is not significantly stronger than the reconnecting field,
which allows one to estimate (to order unity) the properties of the collisional layer
using the Sweet-Parker analysis without a guide field.

\section{Reconnection Physics}
\label{sec_reconnectionphysics}

In this section, the conditions for a plasma to undergo
collisional or collisionless reconnection in the comoving frame of the jet are computed.
As described in section~\ref{sec_substructure},
the characteristic length scale of the electromagnetic field (before reconnection occurs)
is given by $\Delta_0$.
As discussed below, such current distributions
can be unstable to collapse to a much thinner size $\delta$.
The jet is assumed to be filled with a complex of multiple current layers each of thickness $\delta\ll \Delta_0$.
Instead of seeking a solution that is in global force balance within the jet,
all fluid quantities ($\rho,u,v_i,B_i$ in the background jet solution)
are assumed to be roughly homogeneous within the current layer at any given radius.

In section~\ref{sec_layerformationandlength},
the formation and length of current sheets is discussed,
because the length enters into models of slow and fast reconnection.
In section~\ref{eos}, the equation of state and emission rates
for the current layer are summarized (see Appendix~\ref{baseeos} for details).
In section~\ref{sec_condition}, the transition
from slow to fast reconnection is discussed
(see Appendix~\ref{sec_collisionalcollisionlessreconnection} for more details).
In section~\ref{sec_jetdiss}, how the electromagnetic field is modified by fast dissipation is computed.
Finally, in section~\ref{sec_jetvar},
the timescales for observed dissipation/emission variability are estimated.

\subsection{Layer Formation and Length}
\label{sec_layerformationandlength}

First, we must establish that current sheet formation is possible on timescales shorter
than the jet propagation timescale to distances relevant for prompt GRB emission.
Current layer formation starts with a distributed current of size $\Delta_0$
that spans a null point in the magnetic field strength.
Such current distributions can collapse to current sheets
on fast~\alf~crossing time scales through a variety of mechanisms:
1) exponential collapse of the distributed current
until the thermalized gas pressure within a thickness $\delta<\Delta_0$
can support the external electromagnetic pressure \citep{kulsrud05} ;
2) collapse as part of the evolution of the global magnetic field
as occurs for the equatorial current sheet in a pulsar wind ;
or 3) intrinsically unstable collapse at an X-point \citep{dung53,dung58,is67,waelbroeck93,loureiro05}.
For example, X-type collapse forms current sheets on a timescale
\begin{equation}
t_r \sim \frac{\Delta_0}{2 v_{\rm A}} \ln S ,
\end{equation}
\citep{pf86},
where $v_{\rm A} = c|b_{\rm in}/\sqrt{4\pi \xi}| \sim c$ is the~\alf~speed outside the newly forming current layer,
$\xi = \rho_{\rm in} c^2 + u_{\rm in} + p_{\rm in} + b^2_{\rm in}/(4\pi)$,
$|b_{\rm in}|$ is the upstream comoving electromagnetic field strength,
$u_{\rm in}$ is the upstream gas internal energy,
and $p_{\rm in}$ is the upstream gas pressure.
The relevant controlling parameter is the global Lundquist number, $S\equiv \Delta_0 v_{\rm A}/\eta$,
describing the ratio of the resistive time ($T_{\rm res}={\Delta_0}^2/\eta$ for resistivity $\eta$)
to the typical electromagnetic advection time ($T_{A} = \Delta_0/v_{\rm A}$).
Typically $S\gg 1$ and so ideal MHD is valid on scales of order~$\Delta_0$.
The above timescale shows that current sheet formation
only depends logarithmically on the resistivity.
For rapidly rotating black holes generating jets with
$\gamma\sim 100$, half-opening angle $\theta_j\sim 2^\circ$,
substructure types A or B with $m=1$, and $S\sim 10^{15}$, current sheet formation occurs
by $r\sim 10^{13}$cm as comparable with the expected radius of prompt GRB emission.
Thus, current sheet formation is a relatively fast process that can occur during jet propagation.

Second, the length of the current layer is needed
because this is what enters the slow and fast reconnection models.
The length of the current layer in the comoving frame ($L_0$)
can be estimated from the characteristic reversal length scale ($\Delta_0$) of the electromagnetic field,
because this is the natural length scale that enters the MHD equations.
The X-type collapse described above can feed off of perturbations introduced initially in the jet,
producing structures with $L_0\sim \Delta_0$ \citep{pf86}.
In addition, the presence of a complex of multiple current layers can itself induce
structures with $L_0\sim \Delta_0$ \citep{ob92,yoml94}.
Ideal MHD type processes such as magnetic Rayleigh-Taylor instability can also produce
structure along the layer with $L_0\sim \Delta_0$ \citep{lyu10a}.
Even in the absence of any current layers (i.e. $m=0$ and $l=1$),
the jet still has a fundamental characteristic scale of $\Delta_0(1,0) = \pi R_j$.

These arguments suggest that the length of the current layer
should be within the range $\Delta_0 \lesssim L_0\lesssim \pi R_j$.
Because $\Delta_0\sim R_j$ is represented by $l\sim 1$ and for most of
the paper $\Delta_0$ never appears explicitly independent from $L_0$,
for simplicity,
the allowed range for $L_0$ is subsumed as an allowed variation in $l$ and $m$.
With the length of the layer determined,
one determine the thickness of the slow collisional reconnection Sweet-Parker layer
in section~\ref{sec_condition}.

\subsection{Current Layer Pressure, Temperature, Density, and Emission Rates}
\label{eos}

The reconnection physics is determined not only
by the externally imposed magnetic field, rest-mass density, internal energy density,
field reversal scale, and length of the current layer as described in the previous sections.
In addition, the reconnection physics is determined
by the form of energy that the magnetic energy is dissipated into.

The solution for the multi-species equation of state (EOS)
and emission rates are described in Appendix~\ref{baseeos},
which follows similar physics used for radiative accretion disks.
In summary, the gas EOS consisting of baryons, photons, non-pair-produced electrons,
pair-produced electrons-positrons, and neutrinos
is given by the total internal energy density
\begin{equation}
u_g = u_b + u_{\gamma} + u_{e} + u_{\rm pairs} + u_\nu ,
\end{equation}
and total thermal gas pressure
\begin{equation}
p_g = p_b + p_{\gamma} + p_{e} + p_{\rm pairs} + p_\nu ,
\end{equation}
where each species is listed respectively
and is a function of temperature ($T$) and baryon density ($\rho_b$).

Given the proton number density ($n_p$), the baryon number density ($n_b=\rho_b/m_b$),
charge neutrality such that the electron number density is $n_e = n_p$,
then the electron fraction is $Y_e\equiv n_p/n_b$.
A value of $Y_e\sim 1/2$ is assumed (see Appendix~\ref{baseeos}).
Also, $n_e\equiv n_{e^-} - n_{e^+}$,
where $n_{e^-}$ and $n_{e^+}$ are the total electron and total positron number densities, respectively.
Let the total electron+positron number density be $n_{e,\rm tot} \equiv n_{e^-} + n_{e^+}$,
then the pair-only number density is $n_{\rm pairs} = n_{e,\rm tot} - n_e$.

The reconnecting layer is treated as a radiating slab,
with a total cooling rate (${\rm erg}~{\rm s}^{-1}~{\rm cm}^{-3}$)
for photons, pairs, and neutrinos of
\begin{equation}
Q_g = Q_{\gamma} + Q_{\rm pairs} + Q_{\nu} ,
\end{equation}
for each species respectively.
The respective optical depths in the jet are given by
$\tau_{\gamma}$, $\tau_{\rm pairs}$, and $\tau_{\nu}$,
which are separated into scattering and absorption optical depths
(e.g., $\tau_{\gamma,\rm sca}$ and $\tau_{\gamma,\rm abs}$, respectively).
A non-negligible $Q_{\rm pairs}$ occurs when $\tau_{\rm pairs}<1$
as allowed when pairs leave the current layer by travelling along field lines.

\subsection{Transition from Slow Collisional to Fast Collisionless Reconnection}\label{sec_condition}

In this section, we discuss how the current layers within the GRB jet
make a transition from collisional to collisionless reconnection,
which triggers the Petschek-like reconnection geometry leading to fast reconnection.
Then, dissipation and emission from fast reconnection
can contribute significantly to prompt GRB emission.

When the collisionless reconnection layer thickness
is larger than the Sweet-Parker thickness \citep{sweet58,parker57},
the layer enters a Petschek-like reconnection geometry \citep{petschek64}
allowing for fast reconnection (see., e.g., \citealt{mb96,uk00,kulsrud01,cassak05,yamada06}).
For current layers where plasma $\beta=p_g/p_b\sim 1$,
the condition that the plasma no longer obeys
the resistive MHD equations on the scale of the current layer
can be shown to be equivalent to
\begin{equation}
\delta_{\rm SP'} < d_{i} = c/\omega_{pi}  ,
\end{equation}
where $\omega_{pi}$ is the ion plasma frequency \citep{kulsrud05}.
Here, $\delta_{\rm SP'} = \delta_{\rm SP} A^{-1/2}$,
where $\delta_{\rm  SP}\sim L_0\, S^{-1/2}$ is the nominal Sweet-Parker
thickness that is determined using a balance of gas pressure ($p_g$)
and electromagnetic pressure ($p_{\rm EM}$) across the layer
by following the Sweet-Parker analysis.
The compression ratio $A\sim {\rm max}(1,Q_g/Q_{\rm SP})$ (where $Q_{\rm SP}$ is the
dissipation rate in the Sweet-Parker layer) accounts for radiative
cooling in the layer \citep{um11}.
Typically, $A\sim 1$ is found in this paper.
The value of the Lundquist number, $S$, is obtained from Spitzer and Compton drag resistivities.
For more details see Appendix~\ref{sec_collisionalcollisionlessreconnection}.

When the above condition is satisfied, the reconnection operates in a fast collisionless mode.
The transition radius to fast collisionless reconnection is then given by
\begin{equation}\label{rtrans}
r_{\rm trans} = r[\delta_{\rm SP'}=d_{i}] .
\end{equation}
This transition radius is used to identify where fast magnetic reconnection starts
and so leads to dissipation and emission.

One interesting caveat is that once collisionless reconnection is initiated,
the higher dissipation rate may lead to different plasma parameters such as the temperature.
The loss of magnetic energy may raise the local temperature in the optically thick limit,
but it can lower the temperature at slightly larger radii if the flow is already optically thin
because a loss of magnetic energy leads to lower gas pressures.
If resistivity increases due to dissipation,
then the Sweet-Parker thickness increases
and may lead to recovery of the collisional mode of reconnection.
In such a situation, the plasma may enter into a ``marginally collisionless'' reconnection mode
in which $\delta_{\rm SP'} = d_{i}$ is maintained
as the plasma dissipates and continues to enter and exit the fast and slow reconnection modes.
It is difficult to estimate the average rate of reconnection
in such a scenario \citep{uzdensky2007a,gu08}.
However, in some studies, the transition back to slow reconnection was shown to be avoided
by a hysteresis effect \citep{cassak05}.

Slow collisional reconnection can be as slow as Sweet-Parker,
giving an inflow velocity of $v_r\sim v_{\rm A}S^{-1/2}\ll v_{\rm A}\sim c$,
or it may be as fast as $v_r\sim 0.01v_{\rm A}$ \citep{uls10}.
Various studies have found that the fast collisionless reconnection rate
is of order $v_r\sim 0.1v_{\rm A}$ or even $v_r\sim v_{\rm A}$
(for details see Appendix~\ref{sec_collisionless}).
For simplicity and definiteness,
the reconnection rate is chosen to be $v_r\sim 0.1c$ in all cases.
The exact number used for $v_r$ only moves the expected dissipation radius
and does not significantly affect the transition radius or any other calculations.
Once significant dissipation has occurred, the value of $v_{\rm A}$ drops below $c$,
but in all cases we will find it only drops to $v_{\rm A}\sim 0.01c$ by the time
the slow collisional mode would recover.  Also, since $v_{\rm A}$ dropping below $c$
indicates most of the electromagnetic energy flux has been dissipated (i.e. the goal of this work),
we simply avoid detailed discussions about the solution at large radii where this occurs.

\subsection{Effect of Magnetic Reconnection on the Jet Structure}
\label{sec_jetdiss}

In this section, we determine how the jet electromagnetic energy density,
$b^2/(8\pi)$, vs. $r$ is modified by fast dissipation
once the transition to fast reconnection has been triggered.
The resulting function for $b^2$ vs. radius modifies the original large-scale jet structure
dependence given in section~\ref{sec_jetstructure} and Appendix~\ref{sec_fulljetstructure}.
The radial extent over which significant dissipation occurs also determines
the photon/pair/neutrino opacity integrals used to determine the equation of state and emission
rates for the radiative current layer in Appendix~\ref{opticaldepthjet}.

The electromagnetic energy dissipated in the jet
is given by the electromagnetic enthalpy flux into both sides of a current
layer where each side harbors a region of size $\Delta_0/2$ from
which magnetic flux can be accumulated for each sheet.
The comoving electromagnetic dissipation energy density rate is then
\begin{equation}\label{Qem}
-Q_{\rm EM} \equiv \frac{d}{dt_{\rm co}}\left(\frac{b^2}{8\pi}\right) = -2\frac{b^2}{4\pi}\frac{v_r}{\Delta_0} ,
\end{equation}
where $v_r$ is the comoving rate of reconnection,
corresponding to the inflow of magnetic flux into the current layer.
Notice that the relevant length scale is $\Delta_0$,
the distance {\rm between} layers instead of the length of the layer,
because magnetic flux must be brought in from the space between layers.

For the jet structure,
the relativistic MHD energy-momentum equations of motion
are dominated by the radial electromagnetic field advection term\footnote{
This calculation can be compared to that in \citet{lyubarskykirk01,ks03},
and also compared with \citet{Drenkhahn:2002:AGO} who made some simplifications
to obtain their equation~31 giving their equation~37.
This simplification was also used to obtain equation~1 in \citet{Drenkhahn:2002:EAR}.
Drenkhahn et al. concluded that no more than $50\%$ of the
electromagnetic energy flux can be converted to optically thin radiation
and that the rest goes into kinetic energy flux due to jet acceleration.
However, no works have yet fully included radiative cooling that (in the optically thin regime)
can remove thermal energy on a timescale comparable to
the timescale for heating and jet acceleration.
For simplicity, we assume dissipation only leads to radiation
(as might occur in a fully radiative calculation) but does not lead to jet acceleration.
This issue does not affect any of our interpretations of results.}
for either an electromagnetically-dominated jet with $\gamma\gg 1$
or for $\gamma$ and $\theta_f$ roughly constant as occurs for $r\gtrsim r_{\rm mono}$
(as applicable to all astrophysically relevant models studied in this paper).
Then, Equation~(\ref{Qem}) is a sink term
to the quasi-steady relativistic MHD energy equation given by
\begin{equation}\label{bsqeq}
0 = -\left(\frac{1}{r^2\sin\theta}\right) \partial_r ( r^2 \sin\theta b^2 \gamma^2 c )  - 4\pi Q_{\rm EM} \gamma .
\end{equation}
For $Q_{\rm EM}\to 0$ (i.e. $v_r\to 0$) corresponding to the ideal MHD jet solution,
the comoving electromagnetic energy density is a power-law dependence with $b^2\propto r^{-2}$.
In general, Equation~(\ref{bsqeq}) can be solved for $b^2(r)$ to give
\begin{equation}
\frac{b^2(r)}{b^2(r_0)} = \exp \left({-\int_{r_0}^r dr \left[ -\frac{2}{r} - \frac{2v_r}{c\gamma\Delta_0} \right] } \right) ,
\end{equation}
which assumes $v_r\propto v_{\rm A}\sim c$ as relevant for electromagnetically-dominated jets.
Once $v_{\rm A}\ll c$, significant dissipation (i.e. the goal of this work) has already occurred,
and then we simply avoid detailed discussions about the overall solution for such radii
rather than seeking the correct $b^2$ vs. $r$ in the weakly magnetized regime.

For $m\neq 0$ modes, the value of $\Delta_0$ is fixed
(i.e. does not scale with $r$ for $r\gtrsim r_{\rm mono}$),
and then the jet solution has an exponential term such that,
\begin{equation}\label{mmodedecay}
\frac{b^2}{b^2[r=r_{\rm trans}]} =  \left(\frac{r}{r_{\rm trans}}\right)^{-2} \exp \left({- \left(r-r_{\rm trans}\right) \left(\frac{2v_r}{c \gamma \Delta_0}\right) }\right) ,
\end{equation}
for a radius of transition to fast reconnection $r_{\rm trans}$ and $r\ge r_{\rm trans}$.

On the other hand, for $l>0$ modes $\Delta_0\propto r$,
and then the dissipation does not lead to exponential decay and instead
leads to a modified power-law given by
\begin{equation}\label{lmodedecay}
\frac{b^2}{b^2[r=r_{\rm trans}]} = \left(\frac{r}{r_{\rm trans}}\right)^{-2 \left(1 + \frac{l v_r}{\pi c\gamma \sin\theta_f} \right)} ,
\end{equation}
for $r\ge r_{\rm trans}$.
In general, one obtains $r_{\rm diss}/r_{\rm trans}\sim f^{-\pi c \gamma\theta_j/(2 l v_r)}$
for a fraction $f$ of magnetic energy dissipated
(e.g. $r_{\rm diss}\sim 10^{9} r_{\rm trans}$ for $l=5$, $v_r/c=0.1$, $\gamma\theta_j=10$, and $f=1/2$).
A simple conclusion would be that negligible change in $b^2$
occurs for large enough $x_r\equiv (\gamma\theta_j/l)(\pi c/v_r)$
as due to a given $l$-mode becoming causally disconnected across the jet.
However, the $l$ modes generate plasmoids with radial size $L_0$
that may remain fixed in size once significant dissipation occurs.
This alternative assumes, perhaps optimistically, that for $l$ modes
the magnetic flux can be accumulated and dissipated
from all directions instead of only in the $\theta$ direction.
So instead of assuming some $l$ modes fail to dissipate,
an alternative is also considered that
the $l$ modes produce plasmoids similar to substructures A and B
except with an effective $\Delta_0\to L_0$ whose value is fixed with radius
as evaluated at the transition radius.

Once fast reconnection starts at $r_{\rm trans}$,
then Equation~(\ref{mmodedecay})
shows that reconnection completes a single e-folding of dissipation/emission by
\begin{equation}\label{rdiss}
r_{\rm diss} \approx r_{\rm trans} + \frac{\gamma c \Delta_0}{2v_r} ,
\end{equation}
assuming $m$ modes or the optimistic case for $l$ modes.
For example, consider a model with
$m=0.1$,
$\gamma\sim 300$,
a rapidly rotating BH such that $\Omega_F\sim 0.25c/r_{\rm H}$,
$\Delta_0\approx 2\pi c\gamma/(m\Omega_F)\approx 3\gamma\times 10^8{\rm cm}$,
and $v_r\sim 0.1c$.
This model gives $r_{\rm diss}\approx r_{\rm trans} + 10^{14}{\rm cm}$,
which results in a reasonable radius for prompt GRB emission for small enough $r_{\rm trans}$.
For $r\gtrsim r_{\rm diss}$, most of the dissipation has completed
and leaves relatively few photons or pairs.
So, this radial extent can be used to determine
the opacity integral for photons, pairs, and neutrinos
within a relativistic jet as done in Appendix~\ref{opticaldepthjet}.

Between the transition and dissipation radii
most of the jet's electromagnetic power, $P_j^{(EM)}$,
is lost as radiation from plasma blobs moving at Lorentz factor $\gamma$
being observed at some angle away from the jet axis.
For a lab-frame radial dissipation range of $dr\sim \gamma c \Delta_0/v_r$,
lab-frame cross-sectional area $\sim \pi R_j^2$,
and lab-frame dissipation rate $\approx \gamma Q_{\rm EM}$,
the total isotropic jet luminosity is
\begin{equation}
L_j \sim (\gamma Q_{\rm EM}) (\gamma c \Delta_0/v_r) (\pi R_j^2) \sim \gamma^2 b^2 c R_j^2 \sim P_j^{(EM)} \approx P_j,
\end{equation}
which shows that the lab-frame loss rate of electromagnetic energy equals
the lab-frame electromagnetic jet power.
This also shows that the reconnection rate, $v_r$,
does not determine the luminosity from the region $dr$.
Instead, a larger number of layers within the range $dr$
dissipate simultaneously for smaller $v_r/c$,
so a fixed jet power leads to a fixed total isotropic luminosity.

\subsection{Variability from Fast Reconnecting Layers}
\label{sec_jetvar}

In this section, we compute various timescales from reconnecting layers
that may be imprinted in observations of the prompt GRB emission.
The jet substructure type determines the temporal structure
of the prompt GRB emission via the number of dissipating layers.
Also, emission timescales can be linked to the total event timescale,
the event timescale per unit number of reconnecting current layers embedded in the jet,
the dissipation timescale of a single current layer,
and the thermal photon radiation emission timescale.

Consider a reconnecting layer (plasma blob) that begins to dissipate at a transition radius $r_{\rm trans}$
with a comoving dissipation timescale of $dt_{\rm co}$.  This is related to the lab-frame
timescale via $dt_{\rm lab} \approx \gamma dt_{\rm co}$.
Photons emitted at the start of fast reconnection travel outward as the plasma blob also travels outward.
The observed photon time difference from beginning to end of current sheet dissipation
is $dt_{\rm obs} \approx dt_{\rm lab}/(2\gamma^2)$,
such that the observed photon timescale accounting for light-travel effects
is $dt_{\rm obs} \approx dt_{\rm co}/(2\gamma)$.

For the $m$ modes and for the optimistic case for $l$ modes,
the comoving dissipation timescale is $dt_{\rm diss,co}\sim \Delta_0/(4v_r)$,
corresponding to an observed timescale of
\begin{equation}\label{dtobsdiss}
dt_{\rm obs, diss} = \frac{\Delta_0}{8 \gamma v_{r}} .
\end{equation}

Each layer present in the jet over the duration
of the event gives one expected emission timescale.
For substructure types A and B the lab-frame cycle time
to produce a single slab is $dt_{\rm lab,m}\sim 2\pi/(m\Omega_{\rm F})$.
Therefore, one expects a variability timescale of
\begin{equation}
dt_{\rm obs, m} \sim \frac{2\pi}{m\Omega_{\rm F}} ,
\end{equation}
associated with each shell that emerges from $r=r_{\rm trans}$.
For substructure type C, there are $l$ simultaneous emitting layers
that in principle could dissipate stochastically in time rather than going off simultaneously.
In that case, for an event duration $T$, one expects a variability timescale of
\begin{equation}
dt_{\rm obs, l} \sim \frac{T}{l} .
\end{equation}

Reconnection is expected to proceed stochastically,
but it will also have a short timescale associated with the transit
time of fluid through a single reconnection layer of exhaust length $L_0$.
The comoving transit time is order $dt_{\rm co, transit} = L_0/v_{\rm A}$,
such that each reconnection event has an observed time scale of
\begin{equation}
dt_{\rm obs, tra} = \frac{L_0}{2\gamma v_{\rm A}} .
\end{equation}

The thermal photon radiation emission timescale is
\begin{equation}
dt_{\rm obs,\gamma} \sim \frac{u_\gamma}{2\gamma Q_\gamma} ,
\end{equation}
corresponding to the timescale over which the thermal radiation contributes
fully to its component of force balance against the magnetic pressure in the current layer complex.

For scale-height $H$ and mean free path of $\lambda=1/(d\tau/ds)$,
the photon diffusion timescale is
\begin{equation}
dt_{\rm obs,d} = \frac{1}{2\gamma} {\rm max}\left[ \frac{3H^2}{\lambda c} , \frac{H}{c}  \right] ,
\end{equation}
where for $H=H_{\rm abs}$ one has $H/\lambda \equiv \tau$, the optical depth.
The observer absorption diffusion timescale is denoted $dt_{\rm obs,ad}$  when setting $H=H_{\rm abs}$,
while the observer scattering diffusion timescale is denoted $dt_{\rm obs,sd}$ when setting $H=H_{\rm sca}$.

\section{Results: Jet Dissipation}
\label{sec_jetdissipation}

This section computes several results obtained with the procedure:
\begin{itemize}
\item Given the independent variables $\zeta$, $B^r_{\rm fp}$, $\nu$, $r_{\rm mono}$, and $\theta_{\rm fp}$
and the jet structure from section~\ref{sec_jetstructure} (specifically, using Appendix~\ref{sec_fulljetstructure}),
one obtains $\rho_b(r,\theta)$, $b^2(r,\theta)$, $\gamma(r,\theta)$, $\Omega_{\rm F}(r,\theta)$, and $\theta_f(r)$.
\item Given $l$, and $m$ and the jet substructure from section~\ref{sec_substructure},
one obtains the single current layer spacing of $\Delta_0(l,m)$ and layer length of $L_0(l,m)$ at any point in the jet.
\item The multiple current layer complex equations determined in section~\ref{sec_reconnectionphysics},
with the non-ideal MHD jet solution for $b^2(r)$
(equations~\ref{mmodedecay} and~\ref{lmodedecay}) used after $r=r_{\rm trans}$,
are iteratively solved as a set of simultaneous equations
for the temperature $T$, pair number density $n_{\rm pairs}$, and the compression ratio $A$.
A single (unique) solution is always found.
\item Given $T$, $n_{\rm pairs}$, and $A$, remaining quantities are iteratively computed.
\end{itemize}
The solution at each radius is obtained independently\footnote{Each
solution takes $1$--$24$ hours to compute on a modern Xeon core,
where over all models about $100,000$ CPU-hours are required.}.
The transit time through the current layer is roughly the equilibrium timescale,
which is sufficiently small compared to the jet flow time for this to be a valid approximation
except for radii much larger than the transition or dissipation radii.

In what follows,
a fiducial model is discussed in detail in section~\ref{sec_fulljet},
a model parameter exploration is presented in section~\ref{sec_fulljetdissipation},
the simplified jet structure given in section~\ref{sec_jetstructure}
is used to write down scaling laws for results as functions of independent variables,
a comparison is made to prior work in section~\ref{sec_compare},
and finally in section~\ref{sec_otherjetsystems},
a discussion is provided about how the reconnection switch model
applies to other (non-GRB) jet systems in order to identify
whether GRB jets are special.

\subsection{Results: Fiducial Model with Full Jet Structure}
\label{sec_fulljet}

\begin{figure*}
  \begin{center}

      \includegraphics[width=5.5in]{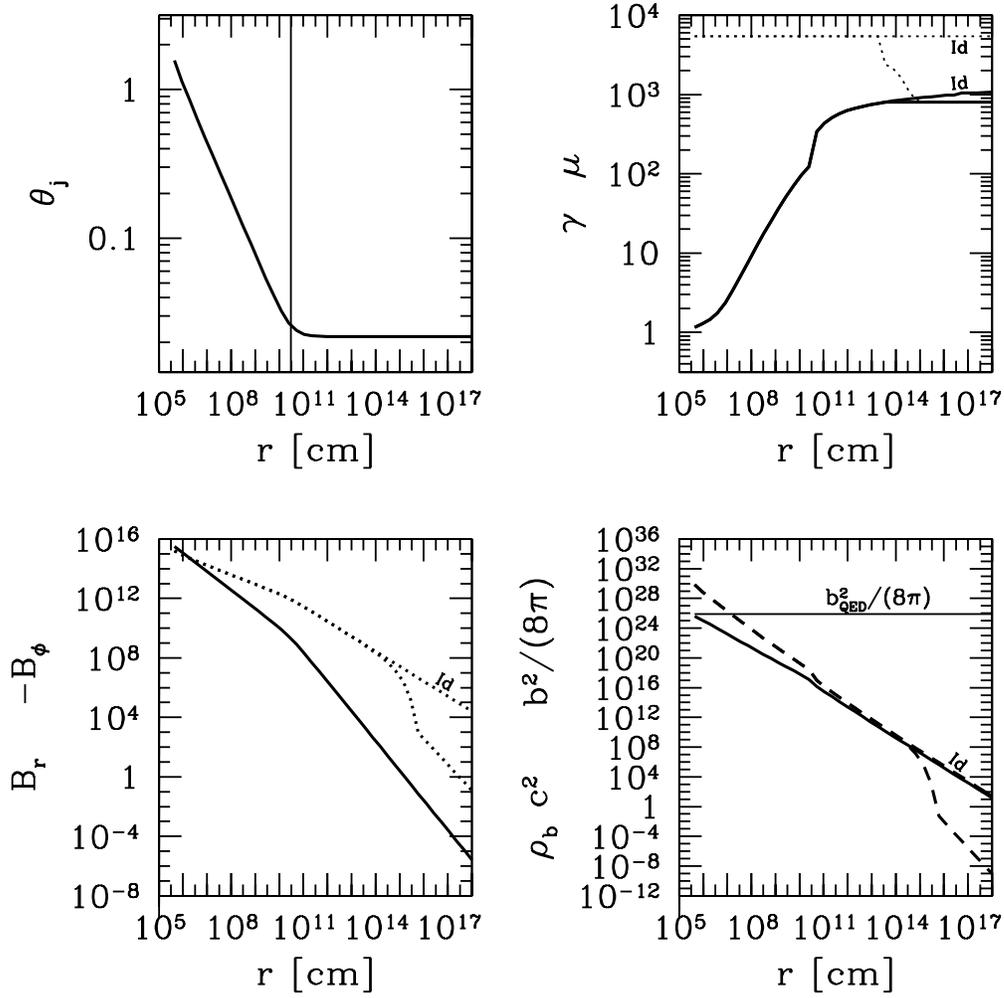}
  \end{center}
  \caption{
  Jet structure for fiducial rapidly rotating black hole model
  with $\zeta=10^4$, $B_{r,\rm fp}=3.2\times 10^{15}$G, $\nu=3/4$, $r_{\rm mono}=3\times 10^{10}$cm, $\theta_{\rm fp}=\pi/2$,
  and modes $m=0.1,l=1$,
  for quantities along the outer-most field line.
  Top-left panel:
  Opening half-angle ($\theta_j$) as thick solid line.
  The location of $r=r_{\rm mono}$ is shown by the thin vertical line.
  Top-right panel:
  Lorentz factor ($\gamma$) as thick solid line.
  Total energy flux per unit rest-mass flux ($\mu\ge \gamma$) as mostly horizontal dotted line.
  In this and other panels, the label ``Id'' marks the same quantity for that line type,
  but corresponding to the ideal MHD jet solution without electromagnetic dissipation.
  The ideal MHD jet solution has constant $\mu$ and $\gamma$ continues to rise beyond the dissipation radius.
  Bottom-left panel:
  Magnetic field components ($B_r$ as solid line and $-B_\phi$ as dotted line).
  Ideal MHD jet solution at large radii follows roughly $-B_\phi \propto 1/r$.
  Bottom-right panel:
  Baryon rest-mass energy density ($\rho_b c^2$) as solid line
  and comoving electromagnetic energy density ($b^2/(8\pi)$) as short-dashed line.
  The thin horizontal solid line shows $|b|=b_{\rm QED}$, the QED critical field.
  As with the toroidal field strength, the comoving electromagnetic energy is dissipated away
  in the non-ideal MHD jet solution beyond the transition radius.
  This is in contrast to the ideal MHD jet solution that would continue with roughly $b^2\propto r^{-2}$,
  which is a similar power-law scaling as rest-mass density at large radii.
  Overall, by $r\approx 10^{14}$cm the jet obtains an opening half-angle of $1^\circ$,
  a Lorentz factor of $\gamma\approx 800$ with $\gamma\theta_j\approx 18$,
  and a jet power of $P_j\approx 8\times 10^{51}\erg{\rm s}^{-1}$.
  }
  \label{jetfull}
\end{figure*}

Consider a fiducial model of a rapidly rotating black hole with
a magnetization $\zeta=10^4$ that leads to $100\lesssim \gamma \lesssim 1000$ ;
a radial magnetic field at the compact object of $B_{r,\rm fp}=3.2\times 10^{15}$G that leads to $P_j\sim 10^{50}$--$10^{52}\ergs$ ;
a collimating field geometry with $\nu=3/4$
as roughly consistent with a shocked stellar envelope (see discussion in \citealt{tmn08,tnm09}) ;
a deconfinement radius of $r_{\rm mono}=3\times 10^{10}$cm
as roughly the radius of a Wolf-Rayet star in the presupernova phase ;
and a foot point opening half-angle of $\theta_{\rm fp}\approx \pi/2$
as applicable to a neutrino-dominated accretion disk that takes on a quite thin
geometry near a rotating black hole where the jet is launched (see, e.g., \citealt{knp05,cb07}
and for simulations of such magnetized thin disks see, e.g., \citealt{shafee08,pmntsm10}).
The chosen $\zeta=10^4$ corresponds to an electromagnetic energy flux per unit rest-mass flux of $\mu\approx 5400$
and an electromagnetic energy per baryon rest energy
at the field line foot point of $\tilde{\mu}\equiv b^2/(8\pi\rho_b c^2)\approx 17000$.
For simplicity $r_{\rm fp}\Omega_{F,\rm fp}=0.25c$ is assumed,
as applicable for a rapidly rotating neutron star or black hole.

Given the ambiguity in whether small $l$ (with $m=0$)
modes actually efficiently dissipate (see end of section~\ref{sec_jetdiss}),
let the fiducial model correspond to $l=1$
(dipolar jet with no separated current sheets in the $\theta$-direction),
and let $m=0.1$ (low order non-axisymmetric mode).
The choice of $m=0.1$ is motivated by recent studies of accretion disks showing
non-axisymmetric modes generated on tens of dynamical times \citep{dsp10}.

This fixes the values of all the free model parameters,
defining our fiducial model explored in this section.
The figures show $40$ positions in radius along a field line within the jet.
The effects of varying the model parameters on the solution are explored in the next section.

Figure~\ref{jetfull} shows the structure of the jet along a single field line
from the compact object to large radii.
At a radius of $r\sim r_{\rm mono}$,
the opening angle rapidly becomes constant and the Lorentz factor jumps up
due to a rarefaction wave that forces the jet towards the monopolar solution.
The jet is cold in the sense that
$p_{\rm EM} \gtrsim p_g$ and $\gamma\gg u_g/(\rho_b c^2)$ up to the dissipation radius,
so the cold ideal MHD jet structure solution
is roughly valid at large radii even when current sheets are dissipating.

For MHD jets, the Lorentz factor
is limited such that strictly $\gamma\le \mu$ at all radii.
The Lorentz factor is dominated by radial motion for our jet solutions.
As discussed later, the loss of electromagnetic energy occurs near the jet photosphere,
so no dominate thermal fireball is generated.
This loss of electromagnetic energy causes the total energy flux per unit rest-mass flux ($\mu$)
to drop down to the bulk Lorentz factor.

Ideal MHD jets are characterized by starting with a poloidal field stronger
than the toroidal field at the foot point on the compact object,
while at larger radii the toroidal field dominates.
In the present model, the field near the compact object is in the super-critical QED regime.
The comoving field at large radii is $|b|\sim |B_\phi|/\gamma$,
$|b|\gg |B_r|$ before significant dissipation occurs
because $B_r\propto r^{-2}$ in the monopole regime beyond $r=r_{\rm mono}$,
and $\gamma$ remains roughly constant.
In the $\theta$-direction the electromagnetic pressure is constant as required by force balance.

Figure~\ref{jetfull} also shows how the ideal MHD jet solution
is modified by the current sheet dissipation.
Beyond the transition radius at $r_{\rm trans}\approx 3\times 10^{13}$cm,
dissipation proceeds due to fast collisionless reconnection.
This causes $|b|$ (and so $|B_\phi|$) to drop exponentially for several e-foldings.
Fast dissipation finally ceases once the transition back to slow collisional
reconnection occurs due to low temperatures leading to a dominant Spitzer resistivity.
Despite the significant drop in the comoving toroidal field,
only by $r\approx 6\times 10^{15}$cm does it become smaller than the comoving poloidal field.
Beyond this radius, any current sheet analysis requires consideration of a guide field,
but this is beyond the dissipation radius so that all of our work remains valid.

\begin{figure*}
  \begin{center}

      \includegraphics[width=5.5in]{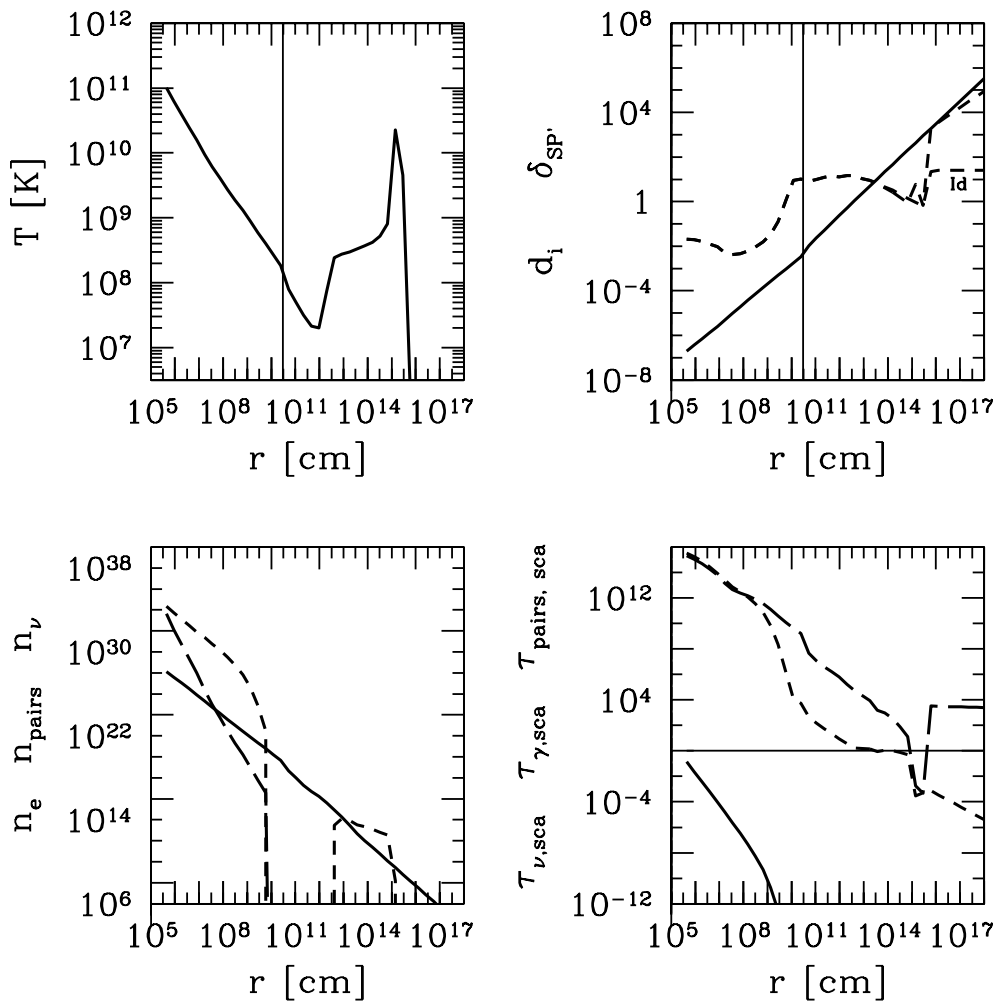}
  \end{center}
  \caption{
  Reconnection layer solution using the jet model shown in Figure~\ref{jetfull}.
  Top-left panel:
  Current layer temperature ($T$).
  The vertical line shows the location of $r_{\rm mono}$, where the jet becomes deconfined.
  The jumps in temperature are resolved by several points and explained in the text.
  Top-right panel:
  Fast collisionless (Petschek-like) ion skin depth ($d_{i}$) as solid line
  and slow collisional (Sweet-Parker-like) layer thickness ($\delta_{\rm SP'}$) as short-dashed line.
  Notice that $d_{i}=\delta_{\rm SP'}$ at $r\approx 3\times 10^{13}$cm,
  where fast collisionless reconnection is triggered.
  As in Figure~\ref{jetfull}, the label ``Id'' marks the case when the ideal MHD jet solution
  is used to determine the properties of the collisional layer.
  Bottom-left panel:
  Electron number density ($n_e$) as solid line,
  pair number density ($n_{\rm pairs}$) as short-dashed line,
  and neutrino number density ($n_{\nu}$) as long-dashed line.
  Jet expansion leads to cooling and a loss of pairs at $r\sim 10^{10}$cm
  until they reemerge with $n_{\rm pairs}\approx n_e$ at $r\approx 8\times 10^{12}$cm.
  Bottom-right panel:
  Neutrino scattering optical depth ($\tau_{\nu,\rm sca}$) as solid line,
  photon scattering optical depth ($\tau_{\gamma,\rm sca}$) as short-dashed line,
  and pair scattering optical depth ($\tau_{\rm pairs,\rm sca}$) as long-dashed line.
  The thin horizontal line shows an optical depth of unity,
  where $\tau_{\gamma,\rm sca}\approx 1$ at $r\approx 5\times 10^{13}$cm.
  Overall, the key result is that the fast collisionless reconnection mode
  is triggered at $r\approx 3\times 10^{13}$cm where $\tau_{\gamma,\rm sca}\approx 1$,
  so the transition roughly coincides with where the jet becomes optically thin for photons.
  }
  \label{jetfullrec1}
\end{figure*}

Figure~\ref{jetfullrec1} shows the temperature,
collisional and collisionless current layer thicknesses,
number densities, and optical depths for neutrinos, photons, and pairs as functions of~$r$.
The sequence of events from small to large radii is as follows:
1) The optically thick current layer's temperature decreases until pairs drop out $r \approx 10^{10}$cm ;
2) Photons dominate the gas pressure.
But after the photon absorption photosphere at $r \approx 2\times 10^{11}$cm,
the photon densities decrease faster ;
3) The loss of photons causes the temperature to begin to increase at $r \approx 2\times 10^{12}$cm,
which eventually leads to a reemergence of pairs at $r\approx 4\times 10^{12}$cm
(with $n_{\rm pairs}=n_e$ at $r\approx 8\times 10^{12}$cm)
and then to a slight quenching of the rise in temperature
because the pairs suspend the decrease in the photon scattering opacity ;
4) The transition to fast collisionless reconnection occurs at $r_{\rm trans}\approx 3\times 10^{13}$cm,
where $n_{\rm pairs}\approx 7 n_e$ ;
5) The transition radius is nearly coincident with where photons begin to free-stream
leading to $\tau_{\gamma,\rm sca}\sim 1$ at $r\approx 5\times 10^{13}$ ;
6) Eventually the pairs drop out again at $r \approx 10^{15}$cm
due to the pair absorption and scattering optical depths approaching unity,
after which the temperature rises rapidly again ;
and 7) An e-folding of electromagnetic dissipation occurs by $r_{\rm diss} \approx 2\times 10^{15}$cm,
leading to a decrease in temperature due to a loss of electromagnetic and hence thermal pressures.

The key point from Figure~\ref{jetfullrec1}
is that $d_{i}=\delta_{\rm SP'}$ at $r_{\rm trans}\approx 3\times 10^{13}$cm,
marking the transition radius to fast collisionless reconnection.
Beyond $r_{\rm trans}$, the collisional resistivity is no longer effective
and dissipation proceeds due to collisionless processes that
operate on the larger scale of the ion skin depth.
This disrupts the current layer into a Petschek-like geometry,
which allows a relatively fast reconnection rate of $v_r\sim 0.1$--$1.0c$
rather than a slower Sweet-Parker-like rate.
This is why we identify this as a catastrophic {\it reconnection switch}.

Note that the radial range where electrons are degenerate is far inside the transition radius,
so degeneracy does not affect the resistivity nor the value of $r_{\rm trans}$.
Also, note that even with the loss of pairs for $r\sim 10^{10}$cm--$4\times 10^{12}$cm,
the proton-electron plasma remains highly collisional,
which forces the transition to fast collisionless reconnection to be at large radii.

The transition to fast collisionless reconnection might lead to a marginally collisionless
state that forces $d_{i}\sim \delta_{\rm SP'}$.
Using the full non-ideal MHD jet solution shows that
the collisional layer thickness rises once dissipation occurs,
and fast dissipation ceases when the two thicknesses become roughly equal
because the collisional mode recovers.
However, by this radius, significant electromagnetic dissipation has already occurred --
and in fact this is what allows the collisional mode to recover.
In addition, the possibility for a return to collisional reconnection
was tested by using different $v_r$ up to $v_r\sim c$,
which leads to a different radial dependence of $b^2$.
We find that marginal collisionality was unable to be established for any $v_r$
until there has already been significant electromagnetic dissipation.
Also, notice that the dissipation included in the model does not
itself control the transition to fast reconnection.
This was tested by using the ideal MHD jet solution for $b^2$ vs. $r$,
which shows that the thickness for the collisional layer
remains much smaller than for the collisionless layer
once fast collisionless reconnection would have started.

The condition for fast reconnection in electron-positron plasmas
is the pair-dominated switch condition $d_e=\delta_{\rm SP'}$.
This condition occurs at $r\approx 6\times 10^{15}$cm in the ideal MHD jet solution
that assumes no dissipation up to this radius.
If the baryon-dominated switch condition applies and dissipation did occur,
then the pair-dominated condition is reached at $r\approx 2\times 10^{15}$cm
after significant dissipation has already occurred.
However, the pair-dominated switch condition
is not expected to be relevant for the fiducial model
or for a broad range of variations in model parameters.
Pairs only reemerge after the photon scattering opacity becomes of order unity,
and the transition to fast collisionless reconnection has occurred to order unity.
Also, the total mass-energy of the plasma is found to be carried by protons
by the radius where $d_e=\delta_{\rm SP'}$ or $d_i=\delta_{\rm SP'}$.
So, the condition for fast reconnection may still be dominated by the protons,
although future studies should consider how fast reconnection proceeds
when pairs dominate in number density but ions dominate in energy density.

Figure~\ref{jetfullrec1} also shows the scattering optical depth for neutrinos, photons, and pairs.
The neutrinos are never optically thick in this model.
The quantity $\tau_{\gamma}$ corresponds to the opacity towards the observer
accounting for locally generated pairs and downstream baryonic electrons
as computed via Equation~(\ref{tauparorig}).
The photons become optically thin to scattering at $r\sim r_{\rm trans}$,
which is consistent with our basic argument presented in section~\ref{sec_argument}.
This means that the reconnection switch mechanism occurs
once non-thermal and quasi-thermal photons can be produced.
Beyond $r\sim r_{\rm trans}$, the non-ideal MHD jet solution has a photon scattering opacity
that deviates only slightly from that of the ideal MHD jet solution.
So dissipation is not crucial for lowering the photon scattering optical depth at larger radii.
The pair scattering opacity is dominated by Coulomb interactions when the temperature
is low, but then the pair density is negligible
(e.g. beyond $r\sim r_{\rm diss}$ leading to a jump in the pair opacity).
For the ideal MHD jet solution, the pair optical depth roughly follows that of the photons
and continues to drop at all larger radii.

Perhaps the most distinct non-intuitive feature of the solution
is the two-step rise and final drop in temperature.
Let us consider a simplified calculation to explain
why the temperature is forced to rise once the jet becomes optically thin to scattering.
Cyclo-synchrotron radiation dominates over free-free emission at all radii,
so the optically thin energy density loss rate is
\begin{equation}
Q \approx n_{e,\rm tot} \frac{4}{3} \sigma_T c (\gamma_e^2-1) u_{\rm EM} ,
\end{equation}
where $\gamma_e \approx 1 + \Theta_e/(\Gamma-1)$ (recall $\Theta_e\equiv \kb T/(m_e c^2)$)
is the electron thermal Lorentz factor.
Then, $\tau_{\gamma,\rm abs} \approx Q \Delta_0/(c u_{0,\gamma})$
is the absorption optical depth from Kirchhoff's law,
and the scattering optical depth is $\tau_{\gamma,\rm sca} \sim \sigma_T n_{\rm pairs} \Delta_0$
since pairs (when present) typically dominate the photon scattering opacity.
The energy density of photons is obtained from the two-stream approximation using Equation~(\ref{gtau}),
which in the limit $\tau_{\gamma,\rm abs}\ll 1$ and $\tau_{\gamma,\rm sca}\sim 1$ gives
\begin{equation}\label{ugammasimple}
u_\gamma \approx 3.2\tau_{\gamma,\rm abs} u_{0,\gamma} ,
\end{equation}
where $u_{0,\gamma} = a_{\rm rad} T^4$ is the optically thick radiation energy density,
and $p_{\gamma}=u_{\gamma}/3$ is the radiation pressure that dominates over other pressures.
Then, the pressure equilibrium condition across the current layer
of $p_{\rm EM}=u_{\rm EM}=p_\gamma$ determines the temperature implicitly via
\begin{equation}\label{tauscaT}
\tau_{\gamma,\rm sca}^{-1} \approx 2.9\left(\frac{\Theta_e}{\Gamma-1}\right) + 1.4 \left(\frac{\Theta_e}{\Gamma-1}\right)^2
\end{equation}
when $\tau_{\gamma,\rm abs}<1$ (otherwise the temperature is set by $p_{0,\gamma}=p_{\rm EM}$).
So a drop in the scattering optical depth leads to a rise in the temperature in this limit.
One also obtains
\begin{equation}
\tau_{\gamma,\rm abs} \approx \frac{u_{\rm EM}}{u_{0,\gamma}} ,
\end{equation}
which indicates that the absorption opacity drops significantly when the
electromagnetic energy density is not high enough to produce optically thick photons.
Equation~(\ref{tauscaT}) shows that once the flow becomes
optically thin, the temperature is forced to rise up to $T\sim 10^9$K
regardless of the temperature in the optically thick regime.
The full solution has a slightly lower temperature of $4\times 10^8$K
because the cyclo-synchrotron emission is slightly dominated
by higher energy electrons than those with the mean temperature.
This gives a solution closer to using $\Gamma=4/3$ than $\Gamma=5/3$ in the above simplified equations.
Overall, the behavior of the temperature vs. radius can be explained as follows:
1) The temperature decreases with radius due to adiabatic expansion in the optically thick regime ;
2) The temperature rises when the photons become optically thin
due to the need to maintain pressure equilibrium with electromagnetic field ;
3) The slight flattening in the temperature during its rise (two-step rise)
is because pairs reemerge and increase the scattering opacity, which suppresses the rapid temperature rise ;
4) Eventually the pair opacity also decreases and the pairs drop out,
which leads to a continued rise in the temperature as the scattering opacity decreases ;
and
5) Finally, the temperature decreases independently of the photon scattering opacity
once a significant loss of electromagnetic energy has occurred.

The scattering and absorption photospheres lead to thermal emission
and might correspond to the observed peak energy.
In addition, low energy electrons' cyclo-synchrotron emission is self-absorbed
and then Comptonization can dominate their emission leading to a Comptonized thermal component
as the peak energy corresponding to $E_{\rm peak}\sim 10\kev$--$2$MeV \citep{rm05,tmr07}.
Electromagnetic dissipation at the transition radius
(coincident with the photon scattering photosphere)
leads to a thermal peak of $E_{\rm peak} \approx \gamma T \sim 20{\rm MeV}$
for the fiducial model value of $\gamma\approx 800$.
The thermal peak obtained at the photon absorption photosphere is smaller at $T\approx 10^7$K,
giving $E_{\rm peak}\sim 600{\rm keV}$ at this relatively high $\gamma\approx 800$.
This peak energy is roughly consistent with the observed peak energy for cosmological long-duration GRBs.

\begin{figure*}
  \begin{center}

      \includegraphics[width=5.5in]{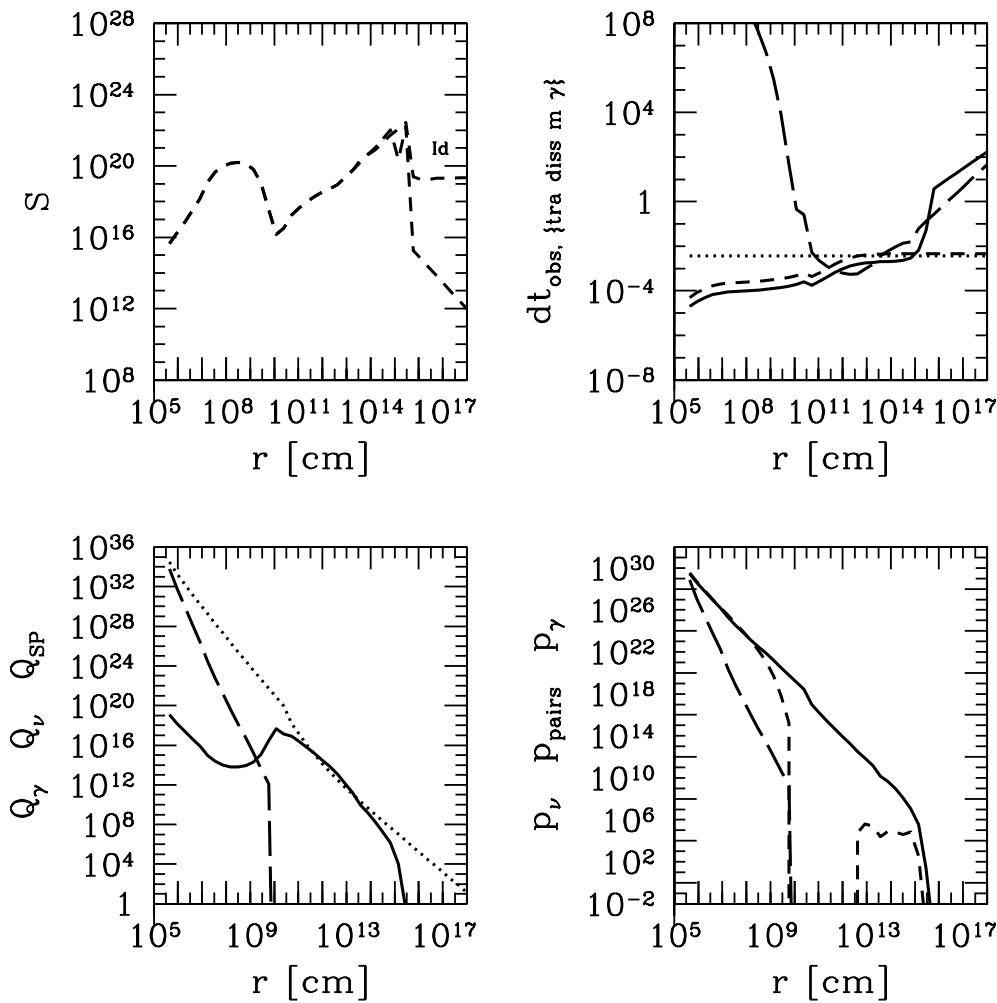}
  \end{center}
  \caption{
  Continued reconnection layer solution (otherwise identical to Figure~\ref{jetfullrec1}).
  Top-left panel:
  Lundquist number ($S$) as short-dashed line.
  As in Figure~\ref{jetfull}, the label ``Id'' marks the ideal MHD jet solution.
  Using the ideal MHD jet solution for the current layer shows that the Lundquist number
  stays roughly constant, while the full non-ideal MHD jet solution shows the Lundquist number
  drops as associated with a high Spitzer resistivity that dominates over the Compton drag resistivity.
  Top-right panel:
  Observer time for transit through the current layer ($dt_{\rm obs,tra}$) as solid line,
  observer time for single e-folding for fast reconnection dissipation ($dt_{\rm obs,diss}$) as short-dashed line,
  observer time for $m=0.1$ radial structure induced variability ($dt_{\rm obs, m}$) as dotted line,
  and observer time for photon emission from the current layer ($dt_{\rm obs,\gamma}$) as long-dashed line.
  Overall, the observer timescales during significant electromagnetic dissipation
  are order $1$s and span $0.001$s to $10$s.
  Bottom-left panel:
  Photon energy density cooling rate ($Q_{\gamma}$) as solid line,
  neutrino energy density cooling rate ($Q_{\nu}$) as long-dashed line,
  and upper limit for cooling rate in strong-cooling regime ($Q_{\rm SP}$) as dotted line.
  The layer is not in the strong-cooling regime because $Q_g\le Q_{\rm SP}$,
  which means the layer thickness can be estimated by the standard Sweet-Parker type solution.
  Bottom-right panel:
  Photon pressure ($p_{\gamma}$) as solid line,
  pair pressure ($p_{\rm pairs}$) as short-dashed line,
  and neutrino pressure ($p_\nu$) as long-dashed line.
  Baryonic-associated pressures are always negligible until significant dissipation has already occurred.
  The photon pressure is comparable or dominates all other sources of thermal pressure.
  }
  \label{jetfullrec2}
\end{figure*}

Figure~\ref{jetfullrec2} shows the Lundquist number,
observer timescales, energy density loss rates, and pressures
within the collisional current layer at each radius.

For the ideal MHD jet solution, one would estimate that
the resistivity is dominated by Compton drag at all radii,
with the Spitzer (due to proton-electron Coulomb collisions)
becoming nearly as important at $r\approx 7\times 10^{8}$cm and at $r\approx 2\times 10^{15}$cm.
For the full dissipative MHD jet solution, Spitzer resistivity begins to play a role
only after significant electromagnetic dissipation occurs once
there is a significant decrease in the temperature within the layer at $r\approx 6\times 10^{15}$cm.
The associated Lundquist number is order $S\sim 10^{20}$
and eventually drops at large radii due to MHD dissipation.

Figure~\ref{jetfullrec2} also shows various observed timescales for photon emission.
Once fast collisionless reconnection kicks in at $r_{\rm trans}$,
these timescales give an estimate for observed variability.
The rise in the timescales at large radii is due to the jet being optically
thin and so the diffusion timescale reaches the light crossing time that increases with radius.
The transit time for fluid to pass through
the current layer would lead to variable emission directly due to reconnection,
while some of the other timescales give variability due to photon diffusion and emission rates.
The $m$ mode produces many pulses due to new reconnection layers passing beyond $r_{\rm trans}$.
Emission would decay over the timescale for completing dissipation.
The observer time for photon absorption diffusion from the current layer photosphere ($dt_{\rm obs,ad}$)
and the observer time for photon scattering diffusion from the current layer photosphere ($dt_{\rm obs,sd}$)
roughly follow $dt_{\rm obs,\gamma}$.
Between $r_{\rm trans}$ and $r_{\rm diss}$, these timescales are
$dt_{\rm obs,tra}\sim 0.002$s--$0.007$s,
$dt_{\rm obs,m}\sim 0.004$s,
$dt_{\rm obs,ad}\sim 0.001$s--$0.04$s,
$dt_{\rm obs,sd}\sim 0.002$s--$0.04$s,
$dt_{\rm obs,\gamma}\sim 0.004$s--$0.06$s,
and $dt_{\rm obs,diss}\sim 0.005$s.
After a few e-foldings of dissipation, the timescales are within the range of $1$s--$10$s.
These timescales are roughly consistent the observed GRB pulse durations
that typically are $\sim 0.5$s, but range from $0.01$s to $10$s \citep{norris96}.
Because the timescale is always associated with a fixed radius,
pulse timescales do not evolve over the event duration as happens in the internal shock model.

Figure~\ref{jetfullrec2} shows that neutrinos dominate
the energy density loss rate up to $r\sim 10^{10}$cm, after which photons dominate.
Near the jet base, neutrino energy density loss rates do not exceed $Q_{\rm SP}$,
which means that the current layer does not compress due to radiative cooling
and the non-radiative Sweet-Parker analysis is approximately valid.
The gas energy cooling rate $Q_g\sim Q_{\rm SP}$
within the dissipation region after fast collisionless reconnection is triggered,
so radiative cooling has become marginally dynamically important
to the current layer structure.

Figure~\ref{jetfullrec2} also shows that
the photon pressure is comparable or dominates other pressures at all radii
including in the optically thin regime.
Electron and baryon pressures are negligible at all radii.

Lastly, consider some checks on the composition of the jet.
The electron and positrons are found to be non-degenerate at all radii except for a
region between $r_{\rm mono}$ and about $100 r_{\rm mono}$.
Nucleons are computed to everywhere be non-degenerate
and $\beta$-equilibrium implies $Y_e=1/2$ everywhere as consistent with this work's assumptions.
However, the nuclear statistical equilibrium (NSE) timescale
can be compared to the jet flow time such that complete
NSE occurs at $r_{\rm NSE}\sim c \gamma dt_{\rm NSE}$.
For the fiducial model, the value of $r_{\rm NSE}$ is always found to be much greater than the local radius.
NSE only completes to about a tenth of a single e-folding NSE timescale.
So, the composition of the jet would primarily be determined by the accretion disk
(either directly by advection and reconnection or by neutron diffusion)
rather than by NSE or $\beta$-equilibrium.

\subsection{Results: Parameter Dependence for Full Jet Structure}
\label{sec_fulljetdissipation}

\begin{figure}
  \begin{center}

      \includegraphics[width=3.0in]{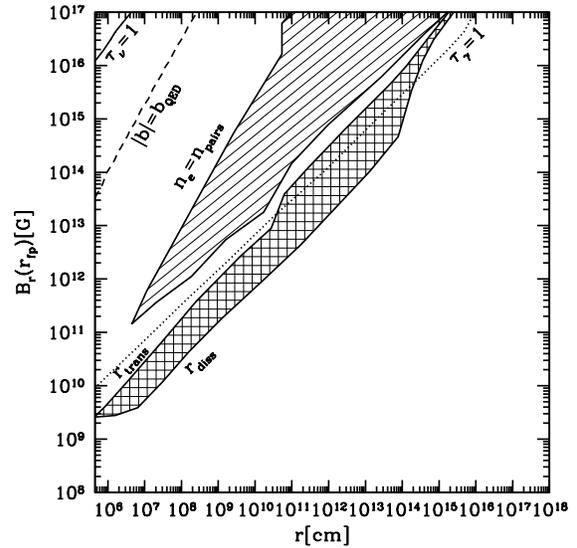}
  \end{center}
  \caption{
  Contour plot for radial magnetic field strength at the field line foot point ($B_r(r_{\rm fp})$) vs. $r$
  showing the
  location of transition radius for fast collisionless reconnection ($r_{\rm trans}$ as left-most diagonal solid line bounding cross-hatched region),
  location of a single e-folding of fast collisionless dissipation ($r_{\rm diss}$ as right-most diagonal solid line bounding cross-hatched region),
  the location of the photosphere defined by the optical depth to radial infinity $\tau_{\gamma}=1$ (diagonal dotted line roughly following the transition radius),
  regions where electrons dominate pairs in number density (region with shade lines going from bottom-left to top-right corner),
  the region where neutrinos are optically thick (region in upper-left corner),
  and the region where the comoving electromagnetic field is QED super-critical (left of the short-dashed line).
  For this and similar figures,
  note that the pairs drop out again within the dissipation region due to significant dissipation and drop in temperature.
  However, for clarity this is not shown here because
  it always starts within the cross-hatched region.
  Overall, the jet becomes optically thin towards the observer
  near the transition radius to fast collisionless reconnection and typically long before dissipation completes,
  which allows significant dissipation to produce non-thermal or quasi-thermal emission.
  }
  \label{fig_brzeta1e4}
\end{figure}

In this section, several model parameters are explored to determine how they affect the results.
All fiducial parameters are held fixed except a single parameter that is allowed to vary.
For each model parameter and across all radii,
a total of $40\times 40$ solutions (i.e. radius $\times$ model parameter) are sought.

Figure~\ref{fig_brzeta1e4} shows results
for $B_r(r_{\rm fp})$ vs. $r$ at $\zeta=10^4$ for which $\mu\approx 5400$.
We find that a stronger electromagnetic field near the compact object
leads to a transition to fast reconnection at larger radii.
For expected collapsar parameters with $B_{r,\rm fp}\sim 10^{15}$G,
the transition to fast collisionless reconnection occurs at $r\sim 10^{13}$--$10^{14}$cm.
In all cases, at $r\gtrsim 10^{14}$cm the Lorentz factor is $\gamma\sim 800$ for the ideal MHD jet solution.
The electromagnetic energy flux per unit mass-energy flux is $\sigma\sim 6$,
indicating a large reservoir of energy exists to be electromagnetically dissipated.
This solution has $\gamma\theta_j\sim 18$ for all field strengths
at a radius of $r\sim 10^{14}$cm, while asymptotically $\gamma\theta_j\sim 20$.
Jet breaks can occur because $\gamma\theta_j\gtrsim 1$.
For all of parameter space, $A\sim 1$ so the current layer is not in the strong-cooling limit.
The break in the behavior of $r_{\rm diss}$ at $B_{r,\rm fp}\sim 10^{14}$G
is due to the Lorentz factor saturating at large radii.

When $\tau_{\nu}\gtrsim 1$, nuclear statistical equilibrium is found to hold
and free nucleons are generated in the jet independently of the accretion disk composition.
Above $B_{r,\rm fp}\sim 10^{15}$G, near the compact object nuclear statistical equilibrium partially holds
and the jet contains a non-negligible fraction of free nucleons that typically freeze-out
before they might convert to $\alpha$-particles at large radii.
Below $B_{r,\rm fp}\sim 10^{13}$G, the jet is found to never be in nuclear statistical equilibrium on jet
flow timescales, and the composition would be set by the accretion disk that feeds the jet with baryons.
The $\beta$-equilibrium value of $Y_e\sim 1/2$ holds for all field strengths.
However, NSE timescales are typically long compared to the jet flow time,
so reaction rates need not be considered.
Thus, for the field strengths required to explain typical cosmological GRBs,
the composition in the accretion disk would determine the jet composition
rather than equilibrium conditions or reaction rates.

Plots similar to Figure~\ref{fig_brzeta1e4}, but with different $\zeta$, show similar results.
For all $\zeta\gtrsim 10^2$,
the flow is optically thin when fast collisionless reconnection initiates.
Consider the case with $\zeta=10^2$ for which $\mu=55$.
The transition radius to fast collisionless reconnection
is also roughly linear in log-log and varies
from $r_{\rm trans}=10^{5.7}{\rm cm}\sim r_{\rm H}$ to $r_{\rm trans}=10^{17}$cm
for a variation of $B_{r,\rm fp}=10^9$G to $B_{r,\rm fp}=10^{17}$G, respectively.
Also, $r_{\rm diss}\sim 10 r_{\rm trans}$ for $B_{r,\rm fp}\lesssim 10^{13}$G,
while otherwise $r_{\rm diss}\sim r_{\rm trans}$.
In addition, the region where pairs drop out shifts down along with this $r_{\rm trans}$ line.
Asymptotically, $\gamma\sim 50$ and $\sigma\sim 1$.
However, the solution always has $\gamma\theta_j\lesssim 2$,
which implies jet breaks may not be discernible for such solutions
even if they produce efficient non-thermal or quasi-thermal emission.
For collapsar parameters and this value of $\zeta\sim 10^2$,
the transition to fast collisionless reconnection occurs at $r\sim 10^{15}$cm.

\begin{figure}
  \begin{center}

      \includegraphics[width=3.0in]{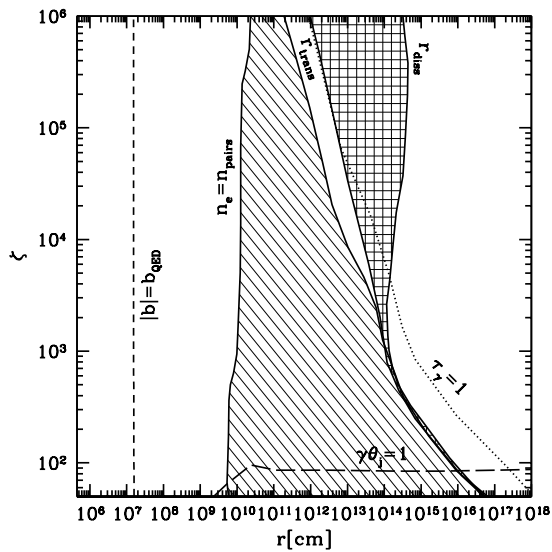}
  \end{center}
  \caption{
  Contour plot for the magnetization ($\zeta$) vs. $r$
  showing
  $\gamma\theta_j=1$ (lower mostly horizontal dashed line) above which jet breaks can occur,
  and otherwise like in Figure~\ref{fig_brzeta1e4}.  Neutrinos are never optically thick for this case.
  By $r\gtrsim 10^{14}$cm and for $\zeta\sim 10^4$, the Lorentz factor is $\gamma\sim \mu/7 \sim \zeta/13$.
  Also, $\sigma\sim 1$ for $\zeta=10^2$ and increases for increasing $\zeta$,
  while $\gamma\theta\sim 1$ for $\zeta=10^2$ and increases for increasing $\zeta$.
  This implies jet breaks are predicted for a broad range in $\zeta$.
  For $\zeta\gtrsim 5000$, the flow becomes optically thin towards the observer
  near the transition radius to fast collisionless reconnection.
  This shows for collapsar parameters with $B_r\sim 10^{15}$G
  that the transition to fast collisionless reconnection occurs at $r\sim 10^{14}$cm
  if $\zeta\sim 10^4$ corresponding to $\mu\sim 5400$ and $\gamma\sim 800$.
  }
  \label{fig_zeta}
\end{figure}

Figure~\ref{fig_zeta} shows how changing the effective magnetization parameter, $\zeta$,
determines the solution for otherwise typical collapsar parameters
that we adopted for our fiducial model in section~\ref{sec_fulljet}.
This shows that dissipation starts at $r\sim 10^{14}$cm if $\zeta\sim 10^4$,
which gives $\gamma\sim 800$ with $\sigma\sim 6$ and $\gamma\theta\sim 18$.
We consider this as a fiducial model that is fairly insensitive to other free parameters.
This type of progenitor system
generates a jet fast enough to avoid the compactness problem,
harbors about $6$ times more energy available for prompt emission than for afterglow emission
(assuming the energy is mostly dissipated and radiated away instead of accelerating the jet),
and can produce jet breaks.
For $\zeta<5000$, the dissipation completes within the region where photons scattering opacity is unity,
which forces a non-negligible fraction of photons to thermalize.
Such events should have a thermal emission component in their spectra.
In this parameter space for the ideal MHD jet solution,
the comoving toroidal field has increased
to a few times the comoving poloidal field,
triggering the kink instability \citep{nlt09},
at $r\sim 10^{13.3}$cm for $\zeta>10^{3.5}$ and at $r\sim 10^{10.8}$cm for $\zeta=50$.
However, for all $\zeta$, the transition to vigorous kink instability still
occurs beyond $r_{\rm mono}$ where $\gamma\theta_j\gg 1$,
so dissipation takes place only near the rotational axis within $\theta<1/\gamma$
due larger angles being causally disconnected.
This angle is typically small compared to the jet opening angle,
and so this region contains very little electromagnetic energy flux
due to both the electromagnetic power dependence on angle
and the efficient ideal MHD acceleration at small angles \citep{tmn09,tnm09}.

All other free model parameters were similarly explored
for otherwise fixed fiducial parameters.
The free parameter $\nu$ controls the field geometry ranging from monopolar ($\nu=0$)
to parabolic ($\nu=1$) with the most likely value for the collapsar model being $\nu\sim 3/4$.
The solution dependence on $\nu=\{10^{-4},2\}$ is weak,
where $r_{\rm trans}$ varies from $10^{13}$cm to $10^{14}$cm for $\nu=10^{-4}$ to $\nu=2$
and
where $r_{\rm diss}$ varies from $10^{13.1}$cm to $10^{15}$cm for $\nu=10^{-4}$ to $\nu=2$.

The free parameter $r_{\rm mono}$ controls when the flow transitions from a collimating solution
(e.g. as caused by a stellar envelope or accretion disk corona/wind)
to a monopolar solution as occurs when the jet becomes unconfined (i.e. no collimating agent is present).
We find that for $r_{\rm mono}=\{GM/c^2,10^{18}{\rm cm}\}$,
the transition radius to fast collisionless reconnection
increases slowly/monotonically from $\{r_{\rm trans},r_{\rm mono}\}$ of $\{10^{13},10^{5.7}\}$cm
to $\{10^{15},10^{18}\}$cm. For $r_{\rm mono}>10^{15.3}$cm,
jet breaks are not expected because $\gamma\theta_j<1$ once $r\sim 10^{15.4}$cm.

Consider short-duration GRBs formed by BH-NS (black hole - neutron star) or NS-NS mergers
that involve only an accretion disk instead of both an accretion disk and a stellar envelope.
Then, $r_{\rm mono}\sim 10^7$cm is applicable,
because this corresponds to the cylindrical radial extent of the accretion disk that collimates the jet.
Such a system is found to have a jet opening half-angle of order $\theta_j\sim 0.4\approx 23^\circ$,
and the transition radius to fast collisionless reconnection is at $r_{\rm trans}\approx 10^{13}$cm.

The dependence on $\theta_{\rm fp}=\{0.16,\pi/2\}$, over which much of the jet power resides, is very weak.
The transition radius to fast collisionless reconnection
only varies from $\{10^{13.7}{\rm cm}$ to $\{10^{13.8}{\rm cm}$.
There is little electromagnetic energy flux for field lines with $\theta_{\rm fp}\lesssim 0.1$,
so dissipation there is not important compared to the total jet dissipation.

\begin{figure}
  \begin{center}

      \includegraphics[width=3.0in]{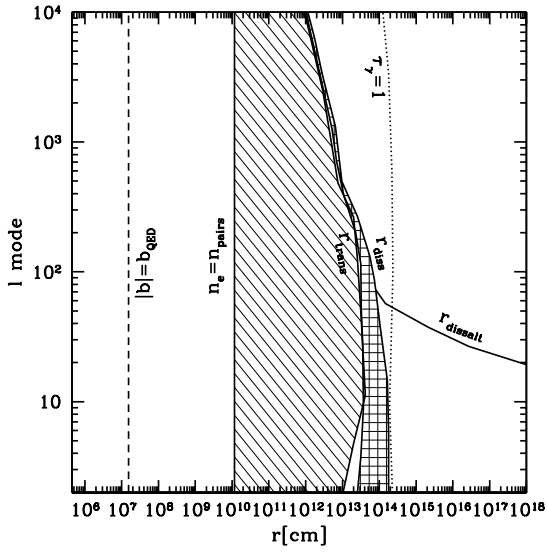}
  \end{center}
  \caption{
  Contour plot for the substructure $l$ mode vs. $r$
  and otherwise like in Figure~\ref{fig_brzeta1e4}.
  Unlike small $l$, large $l$ causes the transition radius to fast collisionless reconnection
  to occur within the optically thick region.
  The line labeled by $r_{\rm dissalt}$ corresponds to Equation~(\ref{lmodedecay}) for
  $f=0.5$ corresponding to dissipating half of the electromagnetic energy flux.
  This is a stronger constraint on the dissipation radius,
  which shows that for $l\lesssim 100$ the dissipation is very slow
  even for fast reconnection due to the (possible) continued radial and lateral expansion
  of the jet on dissipating structures.
  }
  \label{fig_lmode}
\end{figure}

\begin{figure}
  \begin{center}

      \includegraphics[width=3.0in]{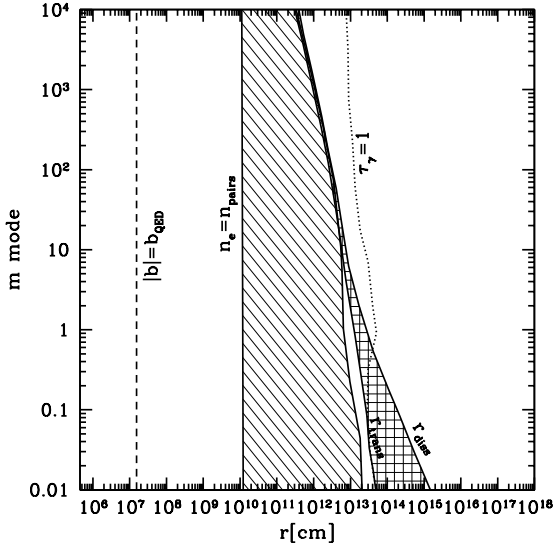}
  \end{center}
  \caption{
  Contour plot for the substructure $m$ mode vs. $r$
  and otherwise like in Figure~\ref{fig_brzeta1e4}.
  Unlike small $m$, large $m\gtrsim 0.8$ causes the transition radius to fast collisionless reconnection
  to occur within the optically thick region.
  }
  \label{fig_mmode}
\end{figure}

Figure~\ref{fig_lmode} and Figure~\ref{fig_mmode}
show that the variations in $l$ and $m$ lead to significant changes in the behavior of the solution.
A significant portion of the dissipated electromagnetic energy is thermalized
and produces a hot magnetic fireball when $l\gtrsim 10^2$
and $m\gtrsim 0.8$ rather than direct non-thermal emission.
This effect is caused by the larger $l,m$ leading to
a smaller current sheet length and hence a smaller Sweet-Parker scale $\delta_{SP}$,
which then becomes equal to the ion skin depth at smaller radii.
One may generally have a spectrum of $l,m$ modes,
and then dissipated electromagnetic energy contributes
both thermal and non-thermal components to the photon spectrum.
A stronger constraint on the $l$ modes is also shown
using Equation~(\ref{lmodedecay}) with
$f=0.5$ corresponding to dissipating half of the electromagnetic energy flux.
This would be the dissipation radius if the collisional structures
(that formed before or at the transition radius) continue to expand with radius
as naively predicted by $\Delta_0$.
It seems more likely that once the collisionless mode is triggered,
then the dissipation decouples from any collisionally-induced structures,
in which case the weaker constraint can be used.

Overall, for collapsar type parameters,
the parameter space study shows that the transition to fast reconnection
often occurs beyond or near the photosphere
unless the magnetization is too low ($\zeta\lesssim 10^3$).
The transition occurs near, but below, the photosphere
when the $l$ mode is too high ($l\gtrsim 10^2$)
or the $m$ mode is too high ($m\gtrsim 1$).
For collapsar type parameters, the transition radius is roughly linear in a log-log
plot of the foot point field strength ($B_r(r_{\rm fp})$[G])
vs. radius, such that, by coincidence, the transition radius happens
to have a value in centimeters
roughly the same as the value of the foot point field in Gaussian units.
Interestingly, the natural $m\sim 1$ spiral mode \citep{mb09}
demarcates a boundary where the transition to fast reconnection occurs
at the jet photosphere.  So small parameter variations
can lead to events with thermal or non-thermal
components in the prompt GRB spectrum.

\subsection{Results: Parameter Dependence for Simplified Jet Structure and Current Layer Physics}
\label{sec_simplejetdissipation}

\newcommand{\rfpnew}{{\tilde{r}_{\rm fp}}}
\newcommand{\rmononew}{{\tilde{r}_{\rm mono}}}
\newcommand{\Brfpnew}{{\tilde{B}_{r,\rm fp}}}
\newcommand{\gammanew}{{\tilde{\gamma}}}
\newcommand{\thetajnew}{{\tilde{\theta}_j}}
\newcommand{\zetanew}{{\tilde{\zeta}}}
\newcommand{\fnew}{{\tilde{f}}}
\newcommand{\thetafpnew}{{\tilde{\theta}_{\rm fp}}}
\newcommand{\npairsnew}{{\tilde{n}_{\rm pairs}}}
\newcommand{\lnewl}{{\tilde{L}_{0,l}}}
\newcommand{\lnewm}{{\tilde{L}_{0,m}}}
\newcommand{\rnewl}{{\tilde{r}_l}}
\newcommand{\rnewm}{{\tilde{r}_m}}

In this section, the simplified jet structure presented in section~\ref{sec_jetstructure}
is used to obtain formulae that describe the essence of the above results.
First, one must obtain an estimate of the bulk jet Lorentz factor, $\gamma$,
such as can be obtained from the full jet structure solution in Appendix~\ref{sec_fulljetstructure}.
Second, one must approximate the current layer physics,
which is possible for a some portions of the relevant parameter space.
The results are not generally accurate to within an order of magnitude,
but the calculation at least highlights the basic ideas and principles.

For typical collapsar parameters
and at the transition radius to fast collisionless reconnection,
the photon pressure dominates,
and the absorption opacity (dominated by cyclo-synchrotron process)
is very small while the scattering opacity is not small.
Cyclo-synchrotron radiation can be approximated
with non-relativistic electrons having a delta function distribution at energy $3 \kb T/2$.
Radiative cooling within the layer is often weak, so that $A=1$ can be chosen.
The Coulomb logarithm is order $30$.
Photon drag resistivity generally dominates over the Spitzer resistivity
up to and far beyond the transition radius.
A non-relativistic temperature approximation is accurate,
although pairs do provide non-negligible cyclo-synchrotron radiation and opacity.
The pair and photon opacities are such that
$g[\tau_{\rm pairs}]$ and $g[\tau_\gamma]$ (see Appendix~\ref{gtau}) are roughly comparable.
These approximations greatly simplify the force balance condition given by Equation~(\ref{forcebalance}),
leaving a couple of equations for $T$ and $n_{\rm pairs}$ to be solved numerically.

Even with these simplifications,
a difficulty in obtaining a closed-form approximation
is that the temperature suppression of pairs is transcendental.
However, one can leave $n_{\rm pairs}$ as a free parameter,
which can be estimated from the full solution
or the approximate solution in the preceding paragraph.

We find that the value of $\zeta$ is implied from the value of $\gamma$,
because at large radii $\gamma\approx \zeta/14$ for models that end up with $\sigma$ order several.
For models near the collapsar type parameters and $\zeta < 10^6$,
the expressions in the rest of this section tend to be accurate to a couple orders of magnitude.
For very large $\zeta > 10^6$, the dependence of $\gamma$ on free parameters
can become difficult to estimate without the full jet structure,
and so the below expressions should be used with caution when arbitrarily varying $\gamma$.

For compactness of the below formulae,
we rescale the free model parameters by collapsar values:
$\rfpnew = r_{\rm fp}/r_g$ with $r_g=3G\msun/c^2$,
$\rmononew = r_{\rm mono}/(3\times 10^{10}{\rm cm})$,
$\Brfpnew = B_{r,\rm fp}/(3.2\times 10^{15}{\rm G})$,
$\gammanew = \gamma/700$,
$\thetafpnew = \sin{[\theta_{\rm fp}/2]}/\sin{[\pi/4]}$,
and $\zetanew = \zeta/10^4$.
Both $l$ and $m$ modes are considered but in general have different properties.
For the case considered, however, $\npairsnew = n_{\rm pairs}/(6 n_e)$ is reasonable
for both mode types.
Such a choice for $n_{\rm pairs}$ is a reasonable estimate for only a limited range of model parameters
such as chosen for the fiducial model near the transition radius as estimated by this simplified model.
Generally, both $\gamma$ and $n_{\rm pairs}$ need to be chosen based upon the results from the full solution.
Otherwise, for $l$ modes, free model parameters are rescaled as follows:
$\rnewl = r/(1\times 10^{14}{\rm cm})$,
and $\lnewl = (L_0)/(\pi R_{\rm jet}/l)$ such that $l=\gamma\theta_j$,
where $\lnewl$ assumes that typically $L_0$ scales with $R_{\rm jet}$ as when $m=0$.
For $m$ modes, free model parameters are rescaled as follows:
$\rnewm = r/(1\times 10^{14}{\rm cm})$,
and $\lnewm = (L_0)/(\gamma c /(m\Omega_{\rm F}))$ with $m=0.1$,
which assumes that $L_0$ does not scale with $R_{\rm jet}$ as for $l=1$ and $m\neq 0$.
As applicable to most of parameter space,
we assume the optical depth integrals use $L_0\sim \Delta_0\lesssim r/\gamma$ for $m$ modes
and $L_0\sim \Delta_0 \gtrsim r/\gamma$ for $l$ modes.

Then, the transition radius is
\begin{equation}\label{rtransscalel}
r_{\rm trans,l} \sim 10^{14}{\rm cm}~\Brfpnew^{4/3}~\lnewl^{1/3}~\rfpnew^{4/3-2\nu/3}~\rmononew^{2\nu/3}~\thetafpnew^{2/3}~\gammanew^{-4/3}~\npairsnew^{-1/3}~\zeta^{-1/3} ,
\end{equation}
for $l$ modes, and
\begin{equation}\label{rtransscalem}
r_{\rm trans,m} \sim 10^{14}{\rm cm}~\Brfpnew~\lnewm^{1/4}~\rfpnew^{5/4-\nu/2}~\rmononew^{\nu/2}~\thetafpnew^{1/2}~\gammanew^{-1/2}~\npairsnew^{-1/4}~\zeta^{-1/4} ,
\end{equation}
for $m$ modes.
This transition radius is consistent with our full solution from the previous sections.
These estimates are valid for both non-relativistic or relativistic temperatures,
whereas other scaling laws obtained later tend to
require a non-relativistic approximation for reliable scalings.
Notice that the transition radius is only weakly sensitive to $n_{\rm pairs}$.
In the case where $n_{\rm pairs}\lesssim n_e$, one can roughly take $n_{\rm pairs}\to n_e$ in the above expressions.
For remaining estimates, $n_{\rm pairs}\gg n_e$ within the current layer is assumed,
although one can readily solve the system of equations that emerge in the other limit.

The dissipation radius is
\begin{equation}\label{rdissl}
r_{\rm diss,l} \sim r_{\rm trans,l} + 10^{14}~{\rm cm}~(\Brfpnew~\lnewl)^{4/3}~\rfpnew^{4/3-2\nu/3}~\rmononew^{2\nu/3}~\thetafpnew^{2/3}~(\npairsnew~\zetanew)^{-1/3}~\gammanew^{-4/3} ,
\end{equation}
for $l$ modes, and
\begin{equation}\label{rdissm}
r_{\rm diss,m} \sim r_{\rm trans,m} + 10^{14}~{\rm cm}~\lnewm~\rfpnew~\gammanew^2 ,
\end{equation}
for $m$ modes.

The optical depth towards the observer for thermal photons is
\begin{equation}
\tau_{\gamma,l} \sim 2~\lnewl~\rnewl~\npairsnew~\gammanew^{-1} + 0.04~\Brfpnew^2~\rfpnew^{2-\nu}~\rmononew^{\nu}~\gammanew^{-2}~\rnewl^{-1}~\zetanew^{-1} ,
\end{equation}
for $l$ modes, and
\begin{equation}
\tau_{\gamma,m} \sim 1~\lnewm~\rfpnew~\npairsnew + 0.1~\Brfpnew^2~\rfpnew^{2-\nu}~\rmononew^{\nu}~\gammanew^{-2}~\rnewm^{-1}~\zetanew^{-1} ,
\end{equation}
for $m$ modes.
The left term corresponds to the optical depth due to the current layer,
and the right term corresponds to the optical depth
due to electrons associated with baryons downstream in the jet.
The local pair contribution tends to dominate the downstream electron contribution to $\tau_\gamma$.
However, as $L_0$ decreases, the right term dominates.

Optically thin synchrotron could be the source of the observed high-energy photons
near the peak of the energy spectrum for prompt GRB emission.
Internal shocks are a fast heating process that could
accelerate particles to high non-thermal energies.
However, reconnection is a slow heating process
such that heating and optically thin synchrotron cooling must be in balance,
which leads to low-energy particles and too low synchrotron photon energies
to explain the observed peak energies
(unless, e.g., high-energy electrons cool downstream where field is much weaker) \citep{gc99}.

Quasi-thermal emission could instead explain the observed prompt spectrum \citep{rm05,tmr07,giannios08c}.
There is necessarily incomplete photon thermalization because $\tau_\gamma\lesssim 1$ for $r\gtrsim r_{\rm trans}$.
Equation~(\ref{tauscaT}) already established that at a unity scattering optical depth
there is a stable temperature of $T\approx 4\times 10^8$K independent of model parameters.
Because the transition to fast reconnection occurs once $\tau_{\gamma,\rm sca}=1$,
this provides a naturally stable temperature.
The temperature is high and would only explain the observed peak energy for $\gamma\lesssim 50$.
However, several scatters may be required to thermalize the photons,
and incomplete thermalization might lead to lower temperatures allowing
up to $\gamma\sim 1000$ while still obtaining consistency with the observed peak energy
\citep{thompson94,gianniosspruit05,giannios06b,gianniosspruit07,giannios08c}.

Incomplete thermalization at the scattering photosphere
may lead to a temperature closer to that at the absorption photosphere.
Let $\tilde{\Theta}_e\equiv \Theta_e/0.0004$,
and assume $n_{\rm pairs}\ll n_e$ as applicable for many model parameters,
then the temperature at $\tau_{\gamma,\rm abs}=1$ satisfies
\begin{equation}
\tilde{\Theta}^{-1}_e \approx \left(\frac{\Brfpnew\rfpnew}{\zetanew}\right)^{4/3} \tilde{\Theta}_e^2 + 0.05\left(\frac{\Brfpnew\rfpnew}{\zetanew}\right)^{2/3} \tilde{\Theta}_e ,
\end{equation}
for both $l$ and $m$ modes,
such that for fiducial parameters $\tilde{\Theta}_e=1$ or $\Theta_e=0.0004$.
Typically, the left-hand term dominates,
then $\tilde{\Theta}_e \approx (\zetanew/(\Brfpnew\rfpnew))^{4/9}$.
Then the observed thermal peak energy is at
\begin{equation}
E_{\rm peak, obs} \sim~\gammanew~\tilde{\Theta}_e~140{\rm keV} ,
\end{equation}
which is close to the observed peak energy.
However, a more detailed study is required to determine whether $E_{\rm peak}\propto \gamma^{13/9}$
is too much variation to be consistent with, e.g., the Ghirlanda relation \citep{ggl04}.

One of the shortest timescales for observed prompt variability
can be estimated from the transit time (as seen by an observer)
for plasma to pass through the reconnection layer as given by
\begin{equation}
dt_{\rm obs,tra,l} \sim 0.02~{\rm s}~\lnewl~\rnewl~\gammanew^{-2} ,
\end{equation}
for $l$ modes, and
\begin{equation}
dt_{\rm obs,tra,m} \sim 0.004~{\rm s}~\lnewm~\rfpnew ,
\end{equation}
for $m$ modes.
This and the other timescales in section~\ref{sec_jetvar}
give a range of values from $0.001$s up to about $10$s for both $m$ and $l$ modes,
which is roughly consistent with observations.
For field substructures types A, B, and C,
this suggests that the prompt GRB emission probes
the turbulent field generated in the accretion disk near the black hole.
For field substructure type D,
the prompt emission probes the turbulent layer that develops between the jet
and stellar envelope or accretion disk corona/wind.
While one expects reconnection to proceed stochastically,
the saturated dissipation of a collection of current sheets
would lead to a delay in the dissipation of current sheets just passing the fast transition radius.
So one expects a correlation between pulse widths and intervals,
and one expects no spreading of pulses at later times because the transition radius is roughly fixed.
More modeling is required to validate these suggestive agreements with observations \citep{piran2004}.

\subsection{Comparison with Other Works}
\label{sec_compare}

Pioneering studies of reconnection in GRB jets by \citet{sdd01}
determined the radius (their equation~61) at which dissipation proceeds.
In this present paper's language, they assumed substructure type A (and type B with $m=1$) to obtain
a dissipation radius of $r_d \approx (\pi c)/(\epsilon \Omega_{\rm F}) \gamma^2$,
where $\epsilon\sim 0.1$ controls the reconnection velocity normalized to~\alf~velocity.
They assumed that fast reconnection occurs over the entire length scale $\Delta_0$,
giving a lab-frame timescale for dissipation of $T\sim \gamma (\Delta_0/v_r)$.
They conclude that $r_d\sim 2\times 10^{12}{\rm cm}$
for $\Omega_{\rm F}\approx 10^4{\rm s}^{-1}$ and a {\it fixed} value of $\gamma\approx 300$.

One should check that this condition holds at smaller radii where $\gamma\sim 1$
in order to ensure the jet can form in the first place without dissipation disrupting
the jet formation process.
At small radii, the jet has
$\gamma\approx 1 + R \Omega_{\rm F}/c$ and $\theta\approx (r/r_{\rm fp})^{-\nu/2} (2\sin{(\theta_{\rm fp}/2)})$.
For any $\nu$, one can solve for the dissipation radius
accounting for the changes in the Lorentz factor.
Their conclusions appear accurate as long as the jet contains substructure type A (or B with $m\sim 1$)
and dissipation occurs over $\Delta_0$.
Some problems may arise in their model because the dissipation radius decreases inversely with $m$.

More serious is that if the jet contains field reversals in $\theta$ as in substructure type C,
then their equation~(57) shows that dissipation would occur near the jet base
because $\gamma\theta_j\sim 1$ at small radii for a magnetized jet\footnote{A
hot MHD jet can be pushed to slightly larger $\gamma\theta_j$ at small radii.}.
GRMHD simulations of accreting black holes show that substructure type C,
which is generated by MHD turbulence, is quite common.
This suggests that only by having a slow reconnection rate at small radii
and a fast reconnection rate at large radii can one ensure magnetized jet formation in the first place.
The reconnection switch model presented in this paper naturally provides such a mechanism.

\subsection{Applications to Other Jet Systems}
\label{sec_otherjetsystems}

AGN and x-ray binaries may also exhibit a transition to fast collisionless reconnection.
Here we consider only a couple example systems.
The ranges $r_g\lesssim r_{\rm mono}\lesssim 10^{10}r_g$ and
$1\lesssim \zeta\lesssim 10^6$ are investigated.
One expects $\zeta\sim \gamma\lesssim 100$ \citep{fender2003}.

First, consider the case of the jet in M87 (similar arguments apply to Blazars),
which has a black hole with mass $M\approx 6\times 10^9\msun$
accreting at $\dot{M}\sim 10^{-2}\msun/{\rm yr}$ with a bolometric
luminosity of $L_{\rm bol}\sim 10^{42}\erg{\rm s}^{-1}$ \citep{gt09},
and so operates at a radiative efficiency of order $\eta\lesssim 0.01$
corresponding to a low-luminosity radiatively inefficient accretion flow.
Equipartition arguments combined with GRMHD simulations of accretion disks
and the Blandford-Znajek funnel region \citep{mckinney2004}
then imply that the magnetic field strength at the base of the jet is of order $B_{r,\rm fp}\sim 10^3$G.

Our solutions show that the M87 jet is already in the fast collisionless regime
at the base of the jet (see also \citealt{gub10}).
This suggests that an electromagnetically-dominated jet from M87 would not survive
unless an ordered dipolar field is present at the jet base \citep{mb09}.
Because the magnetic field should be well-ordered near the black hole to produce a jet in the first place,
this suggests that pairs may dominate the mass-energy
because baryons would only enter the jet base via reconnection or turbulent diffusion.
This may have important implications for polarized disk and jet emission and composition
(see., e.g., \citealt{dafm10,bm10,spm10}).

Second, consider the case of the jet in the BH x-ray binary GRS1915+105,
which has a black hole with mass $M\sim 14\msun$ \citep{greiner01}
accreting at near Eddington rates during outburst
and production of a fast transient jet \citep{fender99}.
The system has $\dot{M}\sim 2.4\times 10^{-8}\msun/{\rm yr}$ \citep{kording06}
and a bolometric luminosity of $L_{\rm bol}\sim 10^{38}\erg{\rm s}^{-1}$,
so operates at a radiative efficiency of order $\eta\sim 0.1$
corresponding to a radiatively efficient accretion flow.
Equipartition arguments combined with GRMHD simulations of accretion disks
and the Blandford-Znajek funnel region \citep{mckinney2004}
then imply that the magnetic field at the base of the polar jet is $B_{r,\rm fp}\sim 10^9$G.

Interestingly, for GRS1915+105 with $\zeta\lesssim 10^4$
and only within $20$ gravitational radii,
the jet is in the slow reconnection regime (see also \citealt{gu08}).
Such a confined region where reconnection is essentially
marginally collisionless could promote the production of the fast transient
jet as triggered by a transition from slow to fast collisionless reconnection.
More modelling is required to investigate this effect.

\section{Discussion}
\label{sec_discussion}

For GRB jets, why must reconnection be slow enough near the central engine
and yet fast at large radii?

Fast relativistic reconnection at speeds up to order $v_r\sim c$ might occur \citep{lu03,lyubarsky05}.
If this operated near the central engine and the jet contained current sheets,
then dissipation of the jet would occur in situ
and no jet might emerge or would be heavily baryon-loaded \citep{bhk08,mb09}.
If somehow a rate of $v_r\sim c$ was suspended until $\gamma\gg 1$, at which point relativistic time dilation would stall dissipation, then still the dissipation radius would be inside the photon's photosphere
(i.e. $r_{\rm diss}<r_{\rm trans}$, where $\tau_{\gamma}\approx 1$ at $r=r_{\rm trans}$).
For example, for $m=0.1$ and otherwise fiducial parameters,
having $v_r\sim c$ leads to $r_{\rm diss}\sim 10^{13}$cm while $\tau_{\gamma}=1$ at $r=10^{14}$cm.
All dissipation would complete inside the photosphere and non-thermal emission would be unlikely.
Hence, in general, fast reconnection should be delayed somehow.

On the other hand, slow collisional reconnection leads to a delayed
dissipation rate relative to fast reconnection by a large factor.
For example, in the most extreme case of Sweet--Parker reconnection,
the slowdown is by a factor of $\sim\sqrt{S}$,
where $S\gg 1$ is the Lundquist number (see Equation~\ref{dtobsdiss}).
At the transition radius, fast reconnection has
an observed timescale on the order of a second for the fiducial model.
A typical Lundquist number at the transition radius is $S\sim 10^{20}$.
This indicates that a time delay of about $10^{10}$s (i.e. hundreds of years)
would occur before the slow reconnection can dissipate a sizable fraction of the available energy,
which is an unacceptable delay.
Thus, slow collisional Sweet--Parker reconnection (with any jet substructure type)
cannot be responsible for powering the prompt GRB emission.
Therefore, fast reconnection is required, but under general considerations
it must be delayed until large radii to avoid complete thermalization of the photons.
The reconnection switch mechanism satisfies both these requirements.

An alternative to fast collisionless reconnection is fast collisional reconnection.
Collisional reconnection may proceed faster than Sweet-Parker
due to secondary tearing instability (plasmoid dominated reconnection) or MHD turbulence.
Preliminary simulations show that, for otherwise fixed parameters,
non-relativistic Petschek may only be $10$ times faster
than non-relativistic collisional reconnection
\citep{lsc07,kowalreview09, samtaney09, cassak09, loureiro09, huang10,uls10}.
While fast collisionless reconnection would dominate this collisional mechanism
once the plasma becomes collisionless, one should consider the effect at smaller radii.
The jet starts with velocity $v\sim c$ on scales of order the black hole horizon
and the Lorentz factor grows rapidly with $\gamma\propto R_j$.
Unless the reconnection rate were $v_r\to c$, the relativistic time dilation
stalls reconnection until $\gamma$ flattens-out for $r>r_{\rm mono}$.
Assuming fast collisional (e.g. plasmoid-dominated) reconnection is $10$ times slower
than the typical speeds of fast collisionless reconnection,
then the dissipation radius is at $10$ times the distance
given by Equation~(\ref{rdissl}) and Equation~(\ref{rdissm}).
So, fast collisionless dissipation already completes before
fast collisional reconnection even becomes important.
Also, for the fast collisional plasmoid-dominated mode,
it remains unclear whether causal disconnection across
the jet would allow for plasmoid chains to grow.
As with suppression of the magnetic kink instability,
turbulence may be avoided except within the narrow region
of $\theta<1/\gamma$ where the jet is still causally connected.
Because this region contains very little power,
this would lead to negligible dissipation compared to the total jet dissipation.

Interestingly, GRB systems are quite unique compared to AGN and x-ray binaries.
GRBs exhibit a transition from slow collisional to fast collisionless reconnection at
roughly billions of gravitational radii from the central compact object.
On the other hand, X-ray binary systems may exhibit a transition to fast reconnection
within tens of gravitational radii,
while AGN jets tend to always be in the fast collisionless regime.
This means that while GRB systems might
sustain a disordered field at the jet base,
it appears unlikely that a disordered field
can be sustained at the jet base in most AGN and some x-ray binary systems.
Significant dissipation and emission can then occur when the jet harbors current sheets.
This suggests that ultrarelativistic jets may be more difficult
to obtain from AGN and x-ray binaries unless the magnetic field geometry
consists of an organized dipolar field \citep{mb09}.
For AGN, such a requirement is tolerable
if the observed coherence length of magnetic fields in the ISM,
which is high enough to trap a significant flux near a supermassive BH,
is assumed to be typical of galaxies with AGN \citep{nia03}.

This suggests that current sheets within jets in AGN and x-ray binary systems
may limit how efficiently electromagnetic energy flux
can be converted into kinetic energy flux through ideal MHD acceleration.
Normally the electromagnetic energy flux per unit mass flux
($\mu$, measuring the degree of mass-loading)
is understood to limit the terminal Lorentz factor.
However, dissipation of electromagnetic energy in current sheets
can lead to significant decreases in $\mu$.
This may explain why GRB jets tend to have higher Lorentz factors
than jets from AGN and x-ray binary systems.

Another interesting result is that the current-driven kink mode
is an unlikely source of current sheets.
As discussed in section~\ref{sec_substructure},
the $|m|=1$ kink instability operates when the magnitude of
the comoving toroidal field ($|b_\phi|$) is about $3-10$ times
the magnitude of the comoving poloidal field ($|b_p|$) \citep{nlt09}.
The comoving toroidal field's relative growth only occurs for $r\gg r_{\rm mono}$.
In the fiducial model for the ideal MHD jet solution with $\theta_{\rm fp}=\pi/2$,
these comoving fields become comparable by $r\sim 10^{13}$cm
and the comoving toroidal field is about $10$ times larger by $r\sim 8\times 10^{13}$cm.
Across all models, $|b_\phi|$ only grows to $3-10$ times $|b_p|$ for $r\gg r_{\rm mono}$.
So, only after this radius would one expect vigorous comoving kink instabilities.
However, beyond $r=r_{\rm mono}$, the jet loses significant causal contact across in angle
due to the typical value of $\gamma\theta_j\sim 10$--$20$.
So, one only expects the region within $\theta<1/\gamma$ around the rotational axis to undergo
the instability.  For typical parameters of $\gamma\theta_j\sim 15$,
this region contains negligible electromagnetic power.
Further, the structure of the jet has $|b_\phi|\ll |b_p|$ at such small angles
due to the even more efficient acceleration that occurs closer to the rotational axis \citep{tmn09,tnm09}.
Overall, kink instabilities appear ineffective at leading to
significant dissipation of electromagnetic energy.

In addition, the interaction between the jet against a disk wind or stellar envelope
is also an unlikely source of current sheets
(or numerous shells moving at varying relativistic speeds)
due to toroidal-field-dominated relativistic jets tending to
suppress boundary layer instabilities as discovered via
3D relativistic MHD simulations of AGN jets \citep{kmbv09,mignone10},
although specific studies for GRB jets should be performed.

\section{Conclusions}
\label{sec_conclusions}

We set out to explore a jet dissipation mechanism,
denoted a ``reconnection switch,''
that relies on the growing evidence that current sheets dissipate
at different rates in collisional and collisionless plasmas.
GRB jets naturally transition from being collisional to collisionless
at large radii where dissipation is initiated near the jet photosphere
due to a transition to fast collisionless reconnection.

Our picture corresponds to a highly magnetized jet
that is presumed to be launched by a rotating neutron star or black hole.
Current sheets are assumed to be generated by being injected into the jet
by some dynamo action near the jet base
or by instabilities at large radii.
These reversals form a complex of numerous current sheets that
dissipate slowly at small radii when the plasma is collisional
and rapidly at large radii when the plasma becomes collisionless.
The fast Petschek-like collisionless mode occurs once the ion skin depth
is larger than the collisional current layer thickness,
which allows the disruption of the thinner collisional layer geometry
and enables the fast Petschek-like reconnection geometry.

For typical long-duration collapsar GRB parameters,
reconnection stays collisional until
the jet transitions to fast collisionless reconnection at $r\sim 10^{13}$--$10^{14}{\rm cm}$.
Even if the reconnection switch mechanism fails to be valid
and dissipation simply proceeds at the somewhat fast reconnection rate of $v_r\lesssim 0.01c$,
then, regardless, all of our calculations remain intact because
the dissipation radius (where significant electromagnetic dissipation occurs)
is beyond the expected transition radius for many models.
However, if $v_r\gtrsim 0.01c$, then the reconnection switch is required
to avoid complete photon thermalization in many models.

Between the transition and dissipation radii,
the  Lorentz factor is $\gamma\sim 100$--$1000$
and the electromagnetic energy flux in the jet exceeds the kinetic energy flux
by factors of $5$--$10$.
Assuming this energy is mostly dissipated and radiated away instead of accelerating the jet,
then this allows the electromagnetic dissipation to produce a prompt GRB luminosity
that is equal to (or even exceeds) the associated afterglow emission,
which could be what is observed but is difficult to obtain in the internal shock model \citep{willingale07}.
Because $\gamma\theta_j\sim 10$--$20$, the afterglow can exhibit jet breaks.
The jet is electromagnetically-dominated up to the transition radius.
However, a reverse shock can be present because dissipation decreases the
electromagnetic energy flux to roughly the kinetic energy flux
by the radius where an external shock forms.

In our reconnection model, the number of prompt GRB pulses is suggested to be related to
the number of current sheets embedded in the jet by the time the jet enters the dissipation regime
and also related to the timescale for reconnection and radiative emission to occur.
For typical collapsar parameters, pulse durations are of order $1$s and range from $0.001$s--$10$s.
Pulses have a rise time associated with the transit time through a current layer as associated with heating,
while the decay time is the timescale for dissipation of magnetic flux in the jet.
These timescales are associated with a fixed transition/dissipation radius
during the entire event so that pulses do not spread in time for fixed engine parameters.

Significant electromagnetic dissipation only occurs once the
flow has become optically thin for photons that reach the observer,
and hence the compactness problem is naturally avoided.
However, some portion of the dissipated energy could lead
to some photospheric emission for certain model parameters.
This may help identify why some fraction of events seem to have thermal peaks \citep{ryde05}.
More work is required to obtain quasi-thermal emission spectra (see., e.g., \citealt{giannios08c}).

One goal of this work was to motivate future studies of reconnection in the
presence of (non-traditional) effects, including: relativistic flow, pairs, radiation,
super-critical electromagnetic fields, a complex of multiple current sheets, etc.
This paper identifies important new directions in future reconnection research,
and motivates them by identifying an new area of application
of such studies to real astrophysical systems.

\section*{Acknowledgments}

JCM thanks
Alexander Tchekhovskoy and Ramesh Narayan for discussions regarding the jet structure ;
Lukasz Stawarz for discussions regarding jet beaming effects ;
Dmitrios Giannios, Chris Thompson, Bing Zhang, Maxim Lyutikov,
Tsvi Piran, and Gabriele Ghisellini for useful comments ;
and the anonymous referee for helpful suggestions.
This work was supported by NASA Chandra Fellowship PF7-80048 (JCM),
the National Science Foundation (NSF) Grant PHY-0903851 (DAU),
and the NSF through TeraGrid resources provided by
NCSA (Abe), LONI (QueenBee), NICS (Kraken)
under grant number TG-AST080025N.

\appendix

\section{Full Jet Structure}
\label{sec_fulljetstructure}

In this section, a generalized axisymmetric steady-state jet solution
is presented as based upon approximate analytical solutions
that have been confirmed numerically \citep{tmn08,tmn09,tnm09}
using the HARM code \citep{gam03,nob06,mm07,tch_wham07}.

In essence, highly magnetized jets are accelerated by the toroidal field gradient.
The key to efficient acceleration is that field lines around the jet axis
must redistribute so that an electromagnetic ``exhaust nozzle'' forms.
At very small angles near the polar axis, the redistribution is highly non-self-similar,
while slightly offset from the polar axis (where most of the energy flux is)
the redistribution is approximately self-similar.
In this approximately self-similar region,
the toroidal field rapidly unwinds due to a drop in $R^2 B_p$ along field lines \citep{tmn09,tnm09},
where $B_p$ is the poloidal field strength.
One can obtain quite accurate yet simple analytical models
of such jet solutions that apply to most of the jet undergoing acceleration
containing most of the electromagnetic energy flux.
The narrow non-self-similar region very close to the polar axis
contains only a small fraction of the total energy flux,
so only its effects on the rest of the mostly self-similar flow
(not its own internal structure) need to be considered.

Aspects of magnetized jets are well-modelled by a force-free\footnote{See, e.g., \citet{mck06ffcode,mck06pulff}.}
(infinitely magnetized limit of ideal MHD) solution.
The vector potential is roughly independent of rotation and given by
\begin{equation}\label{aphi}
P_{\rm coll} \equiv R A_{\phi, \rm coll} \propto \left(\frac{r}{r_{\rm fp}}\right)^\nu (1-\cos\theta) ,
\end{equation}
where $A_{\phi,\rm coll}$ is the vector potential for a collimating jet\footnote{This
is accurate to 10\% for force-free models with $0\le \nu<1.25$ \citep{tmn08}.
For ideal MHD models, it is accurate for the energetically dominant part of the jet \citep{tmn09,tnm09}.},
$R=r\sin\theta$ is cylindrical radius,
$r_{\rm fp}$ is the foot point radius,
$0\le \nu\lesssim 1$ is a free parameter,
and $\theta$ is the position angle in the domain.
$P_{\rm coll}$ is normalized by assuming that
the radial field has the value $B_r(r_{\rm fp})$ at $\theta\approx \pi/2$.
The MHD invariants are constant on field lines (constant $P_{\rm coll}$),
such that, e.g., the invariant field line rotation frequency can be expressed
as $\Omega_{\rm F}(r,\theta) = \Omega_{\rm F}(r_{\rm fp},\theta_{\rm fp})$,
where $\theta_{\rm fp}$ is the angle
of the field line at the foot point attached to the compact object or disk.
For rapidly rotating black holes or neutron stars
$r_{\rm fp}\Omega_{\rm F}\lesssim 0.25 c$.
Equation~(\ref{aphi}) gives that the poloidal magnetic field obeys
\begin{equation}\label{brequation}
\frac{B_r}{B_r(r_{\rm fp})} \approx \left(\frac{r}{r_{\rm fp}}\right)^{\nu-2} ,
\end{equation}
\begin{equation}
\frac{B_\theta}{B_r(r_{\rm fp})} \approx -\nu \left(\frac{r}{r_{\rm fp}}\right)^{\nu-2}\tan(\theta/2) ,
\end{equation}
where the poloidal field strength is $B_p = \sqrt{B_r^2+B_\theta^2}$.
Force balance considerations shows that the toroidal field is given by
\begin{equation}\label{bphioriginal}
\frac{B_\phi}{B_r(r_{\rm fp})} \approx \left(\frac{-2 r_{\rm fp}\Omega_{\rm F}}{c}\right) \left(\frac{r}{r_{\rm fp}}\right)^{\nu-1} \tan(\theta/2) ,
\end{equation}
which is consistent with the minimal torque condition that captures
the effect of the~\alf~critical surface \citep{nar07}.

The force-free Lorentz factor
($\gamma_{\rm ff}$, such that $\gamma\to \gamma_{\rm ff}$ for a pure force-free jet solution)
follows the so-called first and second asymptotic regimes and is defined by
\begin{equation}
\frac{1}{\gamma_{\rm ff}^2} \approx \frac{1}{\gamma_1^2} + \frac{1}{\gamma_2^2},
\end{equation}
where in the first asymptotic regime
\begin{equation}\label{gamma1}
\gamma_1 \approx \sqrt{\gamma_0^2 + (\Omega_{\rm F} R/c)^2 - (\Omega_{F,\rm fp} R_{\rm fp}/c)^2 } ,
\end{equation}
where $\gamma_0$ is a free parameter to be determined later,
and $R_{\rm fp}\equiv R(r_{\rm fp},\theta_{\rm fp})$
with similar notation for other foot point quantities.
The first asymptotic acceleration regime is dominated by the winding of the toroidal field.
The second asymptotic regime has
\begin{equation}
\gamma_2 \approx \sqrt{C \left(\frac{R_c}{R}\right)}  ,
\end{equation}
where $R_c$ is the local poloidal radius of curvature of field lines and $C\approx 3$
(to order unity $C$ depends upon the details of the solution).
For large $r$,
\begin{equation}
\gamma_2 \approx \frac{2}{\theta} \sqrt{\frac{C}{(2-\nu)\nu}} .
\end{equation}
The second asymptotic acceleration regime is dominated by the poloidal field curvature.
The force-free Lorentz factor can be used together with MHD
energy conservation constraints to obtain an accurate full MHD Lorentz factor.
Given the total energy flux per unit mass flux
($\mu$, which is the upper limit of $\gamma$)
and the electromagnetic energy flux per unit rest-mass flux ($\sigma$),
then one can show that
$\mu = \gamma - \Phi R\Omega_{\rm F} B_\phi/(c^2\sqrt{4\pi}) = \gamma(1+\sigma)$
with the conserved quantity
$\sqrt{4\pi}\Phi \equiv B_p/(\rho_b u_p) = B_r/(\rho_b u_r) = B_\theta/(\rho_b u_\theta) = B_\phi/(\rho_b(u_\phi-\gamma R\Omega_{\rm F}))$.
With the general estimate that $\gamma_{\rm ff} \approx \mu/\sigma$ with $\sigma\to 0$
as $\gamma\to \mu$, then one has that
\begin{equation}
\frac{1}{\gamma_{\rm coll}} \approx \frac{1}{\gamma_{\rm ff}} + \frac{1}{\mu},
\end{equation}
which accurately predicts the full Lorentz factor (i.e. $\gamma=\gamma_{\rm coll}$) for
collimating field geometries for MHD jets with $\nu>0$ \citep{tmn08,tmn09}.
The value of $\gamma_0$ is chosen such that $\gamma=\gamma_{\rm fp}$ at $r=r_{\rm fp}$.

If the MHD jet was never confined or becomes unconfined,
then the jet has $\nu\sim 0$ and behaves qualitatively differently than a collimating MHD jet \citep{tnm09}.
The first asymptotic remains the same,
while the monopolar solution has a modified second asymptotic solution
given by the cubic equation
\begin{equation}\label{gamma2m}
\gamma_{\rm 2,m} = C_1 \left[ \frac{(\mu - \gamma_{\rm 2,m})}{\sin^2\theta_c} \ln{\left(1+C_2 \frac{r\Omega_{\rm F} \sin\theta_c}{c \gamma_c}\right)}\right]^{1/3} ,
\end{equation}
where $C_1\approx 2$, $C_2\approx 0.4$,
$\theta_c\sim \theta$ is the angle at which the jet passes the causality surface
and $\gamma_c \approx (\mu/\sin^2\theta_c)^{1/3}$ is the Lorentz factor at the
causality surface\footnote{As $\mu\to \gamma_{\rm fp}\sim 1$
these approximations for the causality surface introduce order unity errors.
Generally $\mu\gg 1$ is assumed.} \citep{tmn09}.
The maximum value of $\gamma_{\rm 2,m}\to \mu$ occurs as $\theta\to 0$,
which leads to a maximally efficient conversion of electromagnetic energy to kinetic energy.
This behavior of the Lorentz factor near $\nu\sim 0$
captures the crucial effects of the fast critical surface and the related causality surface.
A sharp transition\footnote{As shown in \citet{tnm09,kvk09},
this sharp transition produces a rarefaction wave that generates the change
in solution towards a monopolar one.  A boundary layer at the outer angular edges of the jet
undergoes a non-self-similar expansion, but this region contains only a small fraction of the total energy flux.}
from the collimating to the monopole solution is assumed to occur
at radius $r=r_{\rm mono}$, which has been shown to be a quite accurate treatment
of the time-dependent ideal MHD solution \citep{tnm09}.
The final Lorentz factor is then $\gamma_{\rm coll}$ inside $r_{\rm mono}$
and otherwise follows the $\nu=0$ monopole solution with the second asymptotic
given by $\gamma_{\rm 2,m}$.
At very small $\theta_{\rm fp}\lesssim 10^{-3}$, these approximations can break down giving $\gamma<1$,
in which case $\gamma=1$ is enforced because it is accurate for such regions.

Now that the full Lorentz factor and magnetic field structure are defined,
the ideal MHD invariants can be used to constrain the rest of the jet structure.
The $\phi$-component of the 4-velocity can be obtained from
the conserved angular momentum flux per unit rest-mass flux
($\lambda = \lambda_{\rm fp} = (R u_\phi - \Phi R B_\phi)/(c\sqrt{4\pi})$)
to obtain the conserved quantity
$\psi = \psi_{\rm fp} = \lambda - \Omega_{\rm F} \mu = \gamma - R \Omega_{\rm F} u_\phi/c^2$,
where $u_\phi$ is the $\phi$-component of the 4-velocity thus given by
\begin{equation}
u_\phi = c\left(\frac{\gamma - \psi}{R\Omega_{\rm F}}\right) .
\end{equation}
$\Phi$ is used to obtain $u_{\phi,\rm fp} = \Omega_{F,\rm fp} R_{\rm fp} + B_{\phi,\rm fp} v_{p,\rm fp}/B_{p,\rm fp}$.
Foot point velocities are used to obtain $\gamma_0(r_{\rm fp},\theta_{\rm fp})$.

The poloidal field-aligned component of the 4-velocity
($u_p\equiv\sqrt{u_r^2 + u_\theta^2}$)
is given by the definition of the Lorentz factor
using $\gamma^2 = 1/(1-v^2/c^2) = 1 + (u_p^2 + u_\phi^2)/c^2$
where $v^2 =v_p^2 + v_\phi^2$ and $u_i = \gamma v_i$
such that $u_p^2 = c^2 (\gamma^2-1) - u_\phi^2$.

The rest-mass density can be obtained from the conservation of mass and magnetic flux such that
\begin{equation}\label{rhoequation}
\rho_b = \rho_{b,\rm fp} \left(\frac{B_p}{B_{p,\rm fp}}\right)\left(\frac{u_{p,\rm fp}}{u_p}\right) .
\end{equation}
A useful measure of the magnetization is the parameter
\begin{equation}
\zeta\equiv \frac{B^2_{r,\rm fp}}{8\pi\rho_{b,\rm fp}c^2} .
\end{equation}
From the definition of $\mu$ one can show that
$\mu \approx \gamma_{\rm fp} + 4 \zeta c(\Omega_{F,\rm fp} r_{\rm fp}/c)^2 \sin(\theta_{\rm fp})\tan(\theta_{\rm fp}/2)/(\gamma_{\rm fp} v_{p,\rm fp})$.
For example, with $\zeta\gg \gamma_{\rm fp}\sim 1$ one has that
$\zeta\approx 2\mu$ when $\theta_{\rm fp}=\pi/2$, $\Omega_{F,\rm fp}=0.25c/r_{\rm fp}$,
and $\gamma_{\rm fp}\approx 1.15$.
In general, a foot point launching velocity of $v_{p,\rm fp}\approx c/2$ is chosen \citep{tmn08},
but it is unimportant as long as $1\lesssim \gamma_{\rm fp}\ll \mu$.

To enforce consistency with the ideal MHD invariants,
$B_\phi$ is recovered from the definition of $\mu$ given the computed $\gamma$ and $u_\phi$.
The true solution's deviation of $B_\phi$ from Equation~(\ref{bphioriginal})
is typically less than $30\%$.
Finally, the consistency with force-balance is checked by computing $b^2$,
which must be constant in $\theta$ at each radius for a cold MHD jet to be in equilibrium.
Generally, the region within $\pi/20$ is found to be in approximately force balance.
To obtain force balance across the entire jet,
one iterates using the constancy of $b^2(\theta)$ in the equation
for $\mu$ to obtain $P_{\rm coll}$ (and so $B_r(\theta)$) and $\gamma(\theta)$.
Most of calculations in the paper focus
on the most powerful part of the jet
at large angles that is already in approximate force balance without this correction.
This final solution satisfies all ideal MHD invariants exactly except the constancy of
$B_\phi/(\rho_b(u_\phi-\gamma R\Omega_{\rm F}))$ once $r\gtrsim 10^{9}r_{\rm fp}$
where this quantity has a relative error of less than $10\%$.

The comoving electromagnetic pressure and energy density are $p_{\rm EM}=u_{\rm EM}=b^2/(8\pi)$,
where the comoving field is
$b_\mu = (B_\mu + (u\cdot B/c) (u_\mu/c))/\gamma$ ($\mu$ index varies from $t$ to $\phi$).
Using the fact that $u\cdot u = -c^2$ and $B^t=0$
gives the comoving electromagnetic field squared of $b^2 = (B^2 + (u\cdot B/c)^2)/\gamma^2$.
For $\gamma\gg 1$ and beyond the first asymptotic regime,
this gives $b^2\approx B_\phi^2/\gamma^2$ and $b_\theta\ll b_p\approx b_r \ll b_\phi$.
So the field is dominated by the toroidal component in both the lab-frame and the comoving frame.

The total power output of the jet is given by the integration over foot points,
such that the single polar power output is
\begin{equation}\label{pjtheta}
P_j(\theta) = 2\pi \int_0^\theta r^2 \sin\theta' d\theta' [\mu \rho_b c^2 u_p] ,
\end{equation}
with $\theta$ allowed up to $\theta(\theta_{\rm fp}=\pi/2)$,
such that varying $B_{r,\rm fp}$ for $\zeta \gg 1$
leads to $P_j\propto B^2_{r,\rm fp}$ for fixed other parameters.
However, because $P_j$ and $\mu$ are non-trivial
(potentially non-monotonic) functions of free parameters,
all plots will be shown with respect to the simple free parameters
$B_{r,\rm fp}$ and $\zeta$.
At large distances beyond the fist asymptotic, the electromagnetic jet power is
$P_j^{(EM)}=2\pi \int_0^\theta r^2 d\theta' \sin\theta' [b^2 \gamma^2 v_p] = 2\pi \int_0^\theta r^2 d\theta' \sin\theta' [B_\phi^2 v_p]$.

We consider variations in $\zeta$, $B_{r,\rm fp}$, $\nu$, $r_{\rm mono}$, and $\theta_{\rm fp}$.
The solution at large radii is insensitive to other parameters (e.g. $v_{p,\rm fp}$).

These jet structure equations do not account for thermal energy of GR effects.
First, if the thermal energy is subdominant to the electromagnetic and rest-mass energy,
then the above jet structure equations are valid and remain accurate
because only weak shocks occur in a highly magnetized flow.
Second, GR effects accumulate magnetic flux towards the black hole spin axis
leading to a non-constant $B_{r,\rm fp}(\theta)$ \citep{mck05,km07,mck07a,mck07b,tnm09b}.
The spin enhancement of the magnetic field would also change
how one would estimate $B_{r,\rm fp}$ from the mass accretion rate \citep{gammie_bh_spin_evolution_2004,mck05}.
However, these effects occur only within a few gravitational radii,
and beyond this radius the flux deconcentrates
without much difference to the non-rotating black hole case.
This suggests a smaller true $\theta_{\rm fp}$ for a rapidly rotating black hole
with an accretion disk of height-to-radius ratio $H/R\sim 0.1$ (as for a neutrino-dominated accretion disk)
should give comparable results as generally choosing $\theta_{\rm fp}\sim \pi/2$.

\section{Densities, Pressures, and  Emission Rates}
\label{baseeos}

In this section, the equation of state and emission rates
for baryonic and radiative species are presented.
For radiative species, the optically thick densities/pressures and optically
thin energy/number loss rates are used with the two-stream radiative transfer
slab approximation presented in section~\ref{twostream}.
This approximation determines the pressure and energy rates for general optical depths.
These are then used in energy and force balance conditions that set
the reconnection layer's physical geometry and reconnection rate in
section~\ref{sec_collisionalrec}.

The distribution function for all species is assumed to be thermal
except as modified by a radiative transport approximation to account
for the range of optically thin and thick behaviors for radiative
species. As tested using Equation~(\ref{ctherm}),
the thermal assumption is often valid for the collisional layer.

\subsection{Baryons}\label{baryon}

A mixture of free nucleons (protons+neutrons) and $\alpha$-particles
can exist at the densities and temperatures of interest.
Baryons are assumed to be at a single temperature~$T$ in thermal equilibrium.
The baryons have a number density of $n_b=\rho_b/m_b$,
particle mass $m_b$,
and are assumed to be non-degenerate within the jet.
Assuming nuclear statistical equilibrium (NSE),
the fraction of free nucleons is
\begin{equation}
X_{\rm nuc} \approx {\rm min}\left[1,296 \rho_{10}^{-3/4} T_{11}^{9/8} {\rm e}^{\left(\frac{-0.8209}{T_{11}}\right)}\right] ,
\end{equation}
\citep{wb92,knp05}.
The comoving timescale for establishing nuclear statistical equilibrium is
$dt_{\rm NSE}\sim \rho_b^{0.2}\exp(1.8\times 10^{11}/T-39){\rm s}$
in the electron degenerate regime and $dt_{\rm NSE}\sim [T/(1.35\times 10^{9})]^{-5}{\rm s}$
in the hot non-degenerate regime \citep{Khokhlov:1989:SDW,qw96}.
These timescales are compared to the jet flow time in section~\ref{sec_fulljet}.

Given $X_{\rm nuc}$, the baryon internal energy density is
\begin{equation}
u_{\rm b} = \rho_b \frac{3\kb T}{8m_N}\left(\frac{(3-E_{\rm ratio})X_{\rm nuc} + 1}{1-E_{\rm ratio}}\right) ,
\end{equation}
where $E_{\rm ratio}\equiv E_{\rm bin}/(4m_N c^2)$,
the $\alpha$-particle binding energy is $E_{\rm bin}=28.3$MeV,
$m_Nc^2\approx (m_n+m_p)c^2/2\approx 938.919$MeV,
and $\kb$ is Boltzmann's constant.
This directly accounts for the nuclear binding energy of $\alpha$-particles
and the effect of photodisintegration directly within the equation of state.
The baryonic pressure is then
\begin{equation}
p_{\rm b} = (\Gamma-1)u_{\rm b} ,
\end{equation}
with $\Gamma=5/3$, corresponding to a sum over all non-degenerate and non-relativistic baryons.
The non-degeneracy assumption for baryons
is ensuring by checking that $n_b\ll (m_b \kb T/2\pi \hbar^2)^{3/2}$ \citep{km02}.
Tests for the assumption that $Y_e=n_p/n_b\sim 1/2$ are discussed in section~\ref{electronicpressure}.

\subsection{Two-Stream Radiation Approximation}\label{twostream}

The energy densities and cooling rates for radiative species are obtained via
the two-stream approximation limit of the Boltzmann equation \citep{rl79,hubeny90,popham95,sawyer03}.
The theory of radiative reconnecting current layers is quite analogous
to the radiative transfer problem in accretion disk theory~\citep{um11},
for which this approximation has been used extensively
for both photons \citep{popham95} and neutrinos \citep{pwf99,dimatteo02,kawanaka07}.
In this one-zone approximation,
radiation is either lost from the slab or remains and is not redistributed within the slab.

The two-stream approximation
requires optical depths for absorption ($\tau_{\rm abs}$), scattering ($\tau_{\rm sca}$), and their total
\begin{equation}
\tau_{\rm tot} = \tau_{\rm abs} + \tau_{\rm sca} ,
\end{equation}

For scattering species $A$
interacting with a bath of scatterers species $B$ with number density~$n_B$,
scatterer density scale-height $H_B$,
and scatter cross section $\sigma_{A-B,\rm sca}$,
the scattering optical depth is
\begin{equation}
\tau_{A, \rm sca} = \sum_B n_{A-B,\rm eff} H_B \sigma_{A-B,\rm sca} .
\end{equation}

Order unity changes in the momentum of species $A$
are required to achieve an effective collision
corresponding to large-angle scattering and an effective exchange of energy.
The actual number density of scatterers $B$ given by $n_B$
is effectively reduced by diffusive (rather than direct) scattering, such that
\begin{equation}\label{nabeff}
n_{A-B,\rm eff} = n_B X[p_A] ,
\end{equation}
and
\begin{equation}\label{Xpa}
X[p_A]\sim {\rm min}\left(\frac{\Delta p_A}{p_A},\left(\frac{\Delta p_A}{p_A}\right)^2\right) ,
\end{equation}
where $p_A$ is the momentum of species $A$.
The center-of-momentum frame change in momenta is
approximated as $\Delta p_A\sim {\rm min}(p_A,p_B)$,
where $p_B$ are the momentum of species $B$.
The ${\rm min}$ conditional in $X[p_A]$ accounts for the fact
that effective collisions occur when $(\Delta p_A)/p_A\gtrsim 1$, while
the momentum undergoes a random walk for $(\Delta p_A)/p_A\lesssim 1$.
This correction is not applied to Coulomb scattering as computed
later because this effect is already included directly in the cross section.

Thermal equilibrium between a pair of particles
is achieved if the transit time of plasma through the current layer
is longer than their collisional energy-exchange time.
This condition is written as
\begin{equation}\label{ctherm}
C_{\rm thermalization,A-B} \sim \frac{L_0}{v_{\rm A}} \nu_{ec,A-B} \gg 1 ,
\end{equation}
where the effective collision energy-exchange rate
due to order unity changes in the momentum of species $A$ is
\begin{equation}
\nu_{ec,A-B}\approx \nu_{ac,A-B} X[p_A] ,
\end{equation}
where $X[p_A]$ is given by Equation~(\ref{Xpa})
and $\nu_{ac,A-B}$ is the actual collisional frequency
for particle type A to collide with particle type B.
In the non-relativistic limit,
$\nu_{ec,e-e}\sim n_e T_e^{-3/2} m_e^{-1/2} \sim \nu_{ec,e-i} \sim (m_i/m_e)^{1/2} \nu_{ec,i-i}\sim (m_i/m_e) \nu_{ec,i-e}$,
where ions are scattered diffusively by electrons.
The value of $C_{\rm thermalization,A-B}$ is checked for all models
in order to ensure sufficient thermalization,
although even in the collisionless regime one expects a dominant thermal
component \citep{gianniosspitkovsky09}.

The absorption opacities are obtained using Kirchhoff's law,
such that for each absorber with number density $n_B$
and some absorption scale-height $H_B$ one obtains
an energy and number density loss rates
based absorption optical depth of
\begin{align}
\tau_{u,A,\rm abs} &\equiv \sum_B n_B H_B \sigma_{A-B,u,\rm abs} = \sum_B \frac{Q_{B-A,0} H_B}{{\vfs}_A u_{0,A}} , \\
\tau_{n,A,\rm abs} &\equiv \sum_B n_B H_B \sigma_{A-B,n,\rm abs} = \sum_B \frac{R_{B-A,0} H_B}{{\vfs}_A n_{0,A}} ,
\end{align}
respectively,
where $u_{0},n_{0}$ are the optically thick
(i.e., those corresponding to the thermodynamic equilibrium at a given temperature)
energy and number densities, respectively,
and $\vfs$ is the free-stream velocity (e.g. $\vfs=c$ for photons and neutrinos).
(Notice that the subscript of $_{0}$ only indicates the quantity is some known quantity
used to construct new quantities that are valid at general optical depths.)
The cross section $\sigma_{A-B,\rm abs}$ is for the inverse reaction
to that operating at the optically thin rate $Q_{B-A,0},R_{B-A,0}$.

Now that the absorption and scattering optical depths have been obtained,
the two-stream approximation can be used to obtain densities and rates
valid for general optical depths.
The internal energy and number density are given by
\begin{align}\label{untot}
u &= u_{0} g[\tau_u] , \\
n &= n_{0} g[\tau_n] .
\end{align}
with
\begin{equation}\label{gtau}
g[\tau] = \frac{\tau_{\rm tot}/2 + 1/\sqrt{3}}{\tau_{\rm tot}/2 + 1/\sqrt{3} + 1/(3\tau_{\rm abs})} ,
\end{equation}
and energy and number surface fluxes lost from the slab of
\begin{align}\label{ftot}
F_u &= \vfs u_{0} h[\tau_u] , \\
F_n &= \vfs n_{0} h[\tau_n] ,
\end{align}
with
\begin{equation}\label{htau}
h[\tau] =  \frac{1/3}{\tau_{\rm tot}/2 + 1/\sqrt{3} + 1/(3\tau_{\rm abs})} .
\end{equation}
The volumetric energy and number density rates are then given by
\begin{align}\label{qtot}
Q &= \frac{F_u}{H_{u,\rm abs}} , \\
R &= \frac{F_n}{H_{n,\rm abs}} ,
\end{align}
respectively, where the absorption depth is
\begin{equation}
H_{\rm abs}[\tau_{\rm abs}] \equiv \frac{\tau_{\rm abs}}{d\tau_{\rm abs}/ds} ,
\end{equation}
where from Kirchhoff's law one obtains
$d\tau_{u,\rm abs}/ds = \sum_B n_B \sigma_{B,u,\rm abs} =  \sum_B Q_{B,0}/(\vfs u_0) = Q_0/(\vfs u_0)$
and
$d\tau_{n,\rm abs}/ds = \sum_B n_B \sigma_{B,n,\rm abs} =  \sum_B R_{B,0}/(\vfs u_0) = R_0/(\vfs n_0)$.
Notice that $Q,R$ become the optically thin energy and density rates $Q_0,R_0$ when $\tau\ll 1$.

\subsection{Optical Depth in a Relativistic Jet}\label{opticaldepthjet}

In order to compute the optical depth in the prior section,
one requires some estimate of the density scale height for scatters, given by $H$,
that is applicable for a relativistic jet.

In the lab-frame the optical depth is
\begin{equation}\label{taugen}
\tau = \int c dt' [\sigma n]  = \int dr [ \sigma n \gamma (1-\beta\cos\theta)] ,
\end{equation}
where $c dt'=dr'$ is the comoving time difference,
$dr'=\gamma (1-\beta\cos\theta) dr$ is the lab-frame radial difference,
$n$ is the comoving number density of particles,
$\gamma$ is the Lorentz factor, $c\beta$ is the 3-velocity,
and $\theta$ is the angle in the lab-frame.
An angle-averaged invariant cross section ($\sigma$) is assumed to be used.
For a fluid with $\gamma\gg 1$ in the lab frame,
a typical photon emitted isotropically in the comoving frame
is emitted parallel to the jet in the lab-frame giving
\begin{equation}\label{tauparorig}
\tau \approx \int_{r}^{r + \Delta r} dr [n \sigma/(2\gamma)] ,
\end{equation}
where $\Delta r$ is the lab-frame distance the photon traverses through the medium of density $n$.
For photons to escape a radiating slab of comoving size $L_0$,
$\Delta r\sim 2\gamma L_0$ per layer would be chosen.

In our case, the dissipating complex of current sheets consists
of numerous narrow dissipation regions covered in a
photon-pair-neutrino photosphere (when a species is optically thick).
As shown in section~\ref{sec_jetdiss},
the range over which the electromagnetic energy is dissipated into these species
is given by $\Delta r\sim c\gamma\Delta_0/(2v_r)$ for all $m$ modes and some $l$ modes.
Assume that radiation densities drop as roughly $r^{-2}$ (accurate for photons up to relevant radii),
and assume that the Lorentz factor and $\sigma(\rho,T)$
vary as some power of $r$ (including constant with radius).
The opacity integral then gives
\begin{equation}\label{Hrad}
\left(\frac{\tau_{\rm rad}}{\sigma n_{\rm rad}}\right)^{-1} \sim \frac{\gamma\sin(\theta_j)}{R_j} + \frac{2v_r}{c \Delta_0} \equiv (\Delta'_0)^{-1} ,
\end{equation}
where $n_{\rm rad}$ is the number density of radiation at radius $r$.
For downstream baryonic-associated electrons,
a similar calculation gives
\begin{equation}\label{Hbaryonic}
\left(\frac{\tau_{\rm baryonic}}{\sigma n_{\rm baryonic}}\right)^{-1} \sim \frac{\gamma\sin(\theta_j)}{R_j} \equiv (R'_j)^{-1} ,
\end{equation}
where baryons are assumed to extend to large radii such that $\Delta r\to \infty$.

So we have determined that $H\sim \Delta'_0$
for scatterers of radiative species generating by the current layers,
and $H\sim R'_j$ for scatterers consisting of baryonic-associated electrons downstream in the jet.
A more detailed relativistic radiative transfer calculation
is left for future work (see, e.g., \citealt{mes92,bel10b}).

The two-stream approximation can now be applied to photons, pairs, and
neutrinos in order to determine their behavior in the optically thick
and marginally optically thin regime when the single temperature
approximation is still good to order unity.  This two-stream
approximation determines the densities ($u$), pressures ($p$), energy rates ($Q$),
and number rates ($R$) for photons, pairs, and neutrinos as obtained from the base
optically thin rates ($Q_0,R_0$), base optically thick densities ($u_0,n_0$),
and scattering cross sections ($\sigma_{\rm sca}$) written down for each species
in the next section.

\subsection{Photons}\label{photons}

Photons are created within the dissipative current layer by processes such as
bremsstrahlung (e.g. free-free emission, free-bound emission),
cyclo-synchrotron,
pair annihilation,
radiative pair annihilation,
and double Compton scattering.
Most photons are created inside the dissipative current layers,
and photons can travel across field lines and interact with other layers.
For photons, the two-stream approximation requires knowing
the scattering and absorption opacities as computed below,
and one also needs the optically thick limit for the energy and number densities.
The optically thick photon internal energy density is
\begin{equation}
u_{0,\gamma} = a_{\gamma} T^4 ,
\end{equation}
where $a_{\gamma} = \pi^2 \kb^4/(15 (\hbar c)^3)$.
The photon pressure is then given by $p_{0,\gamma} = u_{0,\gamma}/3$
as for a $\Gamma=4/3$ ideal gas for adiabatic constant $\Gamma$.
The optically thick number density of photons is
\begin{equation}
n_{0,\gamma} = (2\zeta[3]/\pi^2)(a_{\gamma}/\kb) T^3 \approx (1/4) (u_{0,\gamma}/(\kb T)) .
\end{equation}

\subsubsection{Photon Scattering}\label{scattering}

The scattering opacity for photons is
\begin{equation}\label{tauradsca}
\tau_{\gamma,\rm sca} \sim \sigma_{\rm es} ( n_{\gamma-\rm pairs,\rm eff} \Delta'_0 + n_{\gamma-e,\rm eff} R'_j ) ,
\end{equation}
for an electron scattering opacity $\sigma_{\rm es}$,
where $n_{\gamma-\rm pairs,\rm eff}$ is the effective number density of pairs
and $n_{\gamma-e,\rm eff}$ is the effective number density of electrons
as computed from Equation~(\ref{nabeff}),
and the scale-heights for scatterers ($\Delta'_0$ and $R'_j$)
are given by Equation~(\ref{Hrad}) and Equation~(\ref{Hbaryonic}).
Radiative effects discussed later can lead to an enhancement of the baryon and associated electron density,
but the opacity is an integral of density that is roughly fixed for a conserved amount of mass across the jet.

In the sub-critical QED regime,
the thermal spectrally-averaged cross section for the Klein-Nishina effect is roughly
\begin{equation}\label{sigmaknthermal}
\left(\frac{\sigma_{\rm KN}}{\sigma_{\rm T}}\right)^{-1} \sim 1 +  \left(\frac{3 m_e c^2}{8 \kb T}\right)^{-1} ,
\end{equation}
where $\sigma_{\rm T}$ is the Thomson cross section.

In the super-critical QED regime,
the extraordinary mode (E-mode, electric polarization perpendicular to the magnetic field),
electron scattering cross section
depends upon whether the particle energy is above/below the rest-mass of electrons
and above/below the energy of the first Landau level
given by $\hbar \omega_{Be}(1) = ( (m_e c^2)^2 + 2\hbar c q|b|)^{1/2} - m_e c^2$ \citep{si80,mes92,td95,lairev01}.
For the E-mode, the thermal spectrally-averaged cross section is
\begin{equation}
\left(\frac{\sigma_{\rm B}}{\sigma_{\rm T}}\right)^{-1} \sim 1 +  \left(\frac{\sqrt{5} \pi m_e c^2 \kb T}{\hbar c q b}\right)^{-2} ,
\end{equation}
which only applies for super-critical field strengths.
The high-field suppression of the scattering cross section for the E-mode
and the efficient conversion of the ordinary mode (O-mode) to E-mode \citep{mes92,td95}
means that radiative emission is dominated by the E-mode with this scattering cross section
when above the critical field strength.
Photon splitting and merging are assumed to be
in detailed balance in the optically thick regime \citep{td95},
and often the thermal photons are optically thick for super-critical fields.
Defining an interpolation parameter ${\rm e}_{eB} = \exp{((-m_e c^2)/(\hbar\omega_{Be}(1)))}$,
the total electron scattering cross section for any $T$ and $b$ is
\begin{equation}
\sigma_{\rm es} \sim \sigma_{\rm B} {\rm e}_{eB} + \sigma_{\rm KN} (1-{\rm e}_{eB}) .
\end{equation}

\subsubsection{Photon Absorption and Emission}\label{absorption}

The absorption opacity is determined from the optically thin emission rate and Kirchhoff's law.
Because the true number density of photons is not required to be accurate in this work,
$\tau_{\gamma,\rm abs}\approx \tau_{n,\gamma,\rm abs}\approx \tau_{u,\gamma,\rm abs}$
is set and only energy density loss rates are considered.
Thermal free-free, thermal synchrotron, and thermal pair annihilation are considered.
The total photon absorption optical depth is
\begin{equation}\label{tauradabs}
\tau_{\gamma,\rm abs} \sim (\sigma_{\rm ff} + \sigma_{\rm synch}) ( n_{\rm pairs} \Delta'_0 + n_{e} R'_j ) + \sigma_{e^+ e^- \to \gamma\gamma} n_{\rm pairs} \Delta'_0 ,
\end{equation}
as due to, respectively,
free-free, synchrotron, and pair annihilation given below.
While the non-pair-producing electrons (with number density $n_{e}$)
outside the current layer's absorption photosphere may be initially cold in the jet,
those electrons are assumed to be heated by a sufficient number of photons
bringing that portion of the jet into thermal equilibrium if the region is optically thick.
The local differential absorption optical depth
is $d\tau_{\gamma,\rm abs}/ds = (\sigma_{\rm ff} + \sigma_{\rm synch}) ( n_{\rm pairs} + n_{e} ) + \sigma_{e^+ e^- \to \gamma\gamma} n_{\rm pairs}$.

\subsubsection{Free-Free}\label{freefreeemission}

The optically thin free-free emission rate that includes relativistic effects,
electron/positron-ion collisions, and electron/positron-electron/positron collisions is
\begin{equation}
Q_{1,\rm ff} \sim 1.4\times 10^{-27} T^{1/2} \left(n_{e,\rm tot} \left(1 + \frac{KE_e}{m_e c^2} \right) \right)^2 ,
\end{equation}
where ${\rm KE}_e \approx 3\kb T$
is the average kinetic energy of relativistic electrons/pairs,
$n_e\sim n_p\sim n_b$ is assumed,
and protons are assumed to be non-relativistic \citep{bps99}.
Kirchhoff's law gives
\begin{equation}
\sigma_{0,\rm ff} = \frac{Q_{1,\rm ff}}{n_{e,\rm tot} c u_{0,\gamma}} ,
\end{equation}
for the cross section per emitting-absorbing particle.
In the super-critical QED regime,
the thermal spectrally-averaged cross section is suppressed such that
\begin{equation}
\left(\frac{\sigma_{\rm ff}}{\sigma_{0,\rm ff}}\right)^{-1} \sim 1 +  \left(\frac{\sqrt{5} \pi m_e c^2 \kb T}{\hbar c q b}\right)^{-2} ,
\end{equation}
\citep{td95,lairev01}.
Note that one can define the QED optically thin emission rate, $Q_{0,\rm ff}$,
from Kirchhoff's law: $\sigma_{\rm ff} = Q_{0,\rm ff}/(n_{e,\rm tot} c u_{0,\gamma})$.
These equations give, as required, a free-free contribution to
$Q_g$ of $Q_{0,\rm ff}$ in the optically thin limit.

\subsubsection{Cyclo-Synchrotron}\label{synchrotronemission}

The synchrotron (and approximate cyclotron) energy density loss rate integrated over all angles, frequencies,
and over an isotropic distribution of thermal particles with Lorentz factor $\gamma_e$
and pitch angle $\theta$ is
\begin{equation}
Q_{1,\rm synch} \approx n_{e,\rm tot} \int_{\theta=0}^{\pi} \sin\theta \int_{\gamma_e=1}^{\infty} d\gamma_e [f(\gamma_e) j(\gamma_e,\theta)] ,
\end{equation}
where the frequency-integrated emissivity (erg/s) for each pitch angle is
\begin{equation}
j(\gamma_e,\theta) = \frac{2q^4 b^2 (\gamma_e^2-1)\sin^2\theta}{3m_e^2 c^3} ,
\end{equation}
and for simplicity the non-degenerate relativistic electron thermal distribution,
\begin{equation}
f(\gamma_e) = \frac{1}{\Theta {\rm K}_2(1/\Theta)}\gamma_e^2 \beta_e \exp{(-\gamma_e/\Theta)} ,
\end{equation}
is used, where $\Theta = \kb T/(m_e c^2)$,
$\beta_e=\sqrt{1-1/\gamma_e^2}$,
and $K$ is the BesselK function \citep{rl79,mny96,opn00}.
In cases when the gas is degenerate, synchrotron is found not to be crucial to this study.
The electron and photon azimuthal angles have both already been integrated over.
Kirchhoff's law gives
\begin{equation}
\sigma_{0,\rm synch} = \frac{Q_{1,\rm synch}}{n_{e,\rm tot} c u_{0,\gamma}} .
\end{equation}

QED effects are important for $|b|/b_{\rm QED}\gtrsim 0.01$ where $b_{\rm QED} = (m_e c^2)^2/(c q\hbar)$
and always lead to a suppression of $\sigma_{0,\rm synch}$.
For $\gamma_e-1\gtrsim |b|/b_{\rm QED}$,
well-defined expressions exist for computing a QED version of $\sigma_{\rm synch}$
(e.g. consider the complete integral of equation~(31) in \citealt{baring88b}).
However, the full integrals are computationally expensive.
So instead of directly using the full QED expression,
we numerically derived a suppression factor fitting function.
The suppressed cross section is
\begin{equation}
\left(\frac{\sigma_{\rm synch}}{\sigma_{0,\rm synch}}\right)^{-1/2} \sim 1 + \left(\frac{(m_e c^2)^3}{\sqrt{3}\pi \hbar c q b \kb T}\right)^{-2/3} ,
\end{equation}
which is found to be accurate to order unity for $|b|/b_{\rm QED}\gtrsim 0.01$ and $\Theta\gtrsim 0.01$
as is sufficient for this study.
The asymptotic suppression factor is $\propto (T |b|)^{-4/3}$,
such that both high field strengths and high temperatures induce a suppression effect.
Note that one can define the QED optically thin emission rate, $Q_{0,\rm synch}$,
from Kirchhoff's law: $\sigma_{\rm synch} = Q_{0,\rm synch}/(n_{e,\rm tot} c u_{0,\gamma})$.
These equations give a contribution to $Q_g$ of $Q_{0,\rm synch}$ in the optically thin limit.

The total inverse Compton power is roughly a factor
$(u_{\gamma}/u_{\rm EM})(\sigma_{\rm es}/\sigma_{\rm T})$ of the synchrotron power.
In this work, $u_{\gamma}\lesssim u_{\rm EM}$ and $\sigma_{\rm es}\le \sigma_{\rm T}$,
so that the total inverse Compton power is typically weaker than the total synchrotron power.
In the transition from optically thick to optically thin radiation,
synchrotron at low frequencies can be self-absorbed leading
to a dominant Comptonized emission \citep{giannios08c,lb10}.
This effect, a calculation of (Comptonized) spectra,
and related QED effects \citep{hardinglai06} are left for future work.

\subsubsection{Pair Annihilation into Photons}\label{pairannphotons}

Electron-positron pair annihilation into photons
for thermal pairs has a number density rate of
\begin{equation}
R_{0,e^+ e^-\to \gamma \gamma} \approx \frac{3}{8}\sigma_T c n_+ n_- \left(1 + \frac{2\Theta^2}{\ln(1.12\Theta + 1.3)}\right)^{-1} ,
\end{equation}
for each thermal pair unit (1 electron and 1 positron),
where $\Theta=\kb T/(m_e c^2)$,
$n_+ = (n_{e,\rm tot} - n_e)/2 = n_{\rm pairs}/2$,
and $n_- = (n_{e,\rm tot} + n_e)/2 = n_{\rm pairs}/2 + n_e$ \citep{svensson82,sz86}.
The corresponding energy density loss rate is
\begin{equation}
Q_{0,e^+ e^-\to \gamma \gamma} \approx E_{\rm pairs} R_{0,e^+ e^-\to \gamma \gamma} ,
\end{equation}
where $E_{\rm pairs} \approx 2(m_e c^2 + u_e/n_e)$,
and $u_e$ is the electron/positron internal energy density given later.
By Kirchhoff's law,
\begin{equation}
\sigma_{e^+ e^- \to \gamma\gamma} = \frac{Q_{0,e^+ e^-\to \gamma \gamma}}{n_{\rm pairs} c u_{0,\gamma}} .
\end{equation}
QED effects on the optically thin pair annihilation rates are neglected,
but only tend to be important at very low baryon loading of the jet.

\subsection{Electrons and Positrons}\label{electronicpressure}

Electron-positron pairs can be created by processes such as
photon annihilation, one-photon pair production in super-critical fields,
ion-electron collisions, electron-electron collisions,
ion-photon collisions, and electron-photon collisions.
The pairs do not readily cross field lines and are stuck within the dissipation region
until they can annihilate into photons and spread throughout the photon photosphere
or travel along field lines within the layer.

In thermal equilibrium, the asymmetry between electrons and positrons is
represented by the electron chemical potential $\mu_e \equiv \mu_{e^-} = -\mu_{e^+}$,
which is determined by the condition of charge neutrality among protons, electrons and
positrons given by $n_{e} \equiv n_{e^-} - n_{e^+} = n_p$,
where $n_p$ is the total number density of protons (free or bound).
The degeneracy parameter of electrons is given by $\eta_e = {\mu_e}/{\kb T}$,
where $\eta_{e}\gtrsim 1$ implies electrons are degenerate.
A value of $Y_e\equiv n_e/n_b = 1/2$ is assumed,
but the $\beta$-equilibrium value of $Y_e$ is computed from
$1/(1/Y_e - 1) = n_p/n_n = \exp(Q-\eta_e)$ with $Q=(m_n-m_p)c^2/(\kb T)$,
which assumes only free nucleons are present \citep{km02}.
This gives a test of whether the assumption of $Y_e=1/2$ is violated.

The Fermi-Dirac distribution function of particles in thermal equilibrium is
\begin{equation}\label{fddist}
f_x(E) = \frac{1}{{\rm e}^{(E/(\kb T) - \eta_x)} + 1} .
\end{equation}
The number densities of electrons and positrons are
\begin{equation}\label{nepm}
n_{e^\pm} = \frac{1}{\pi^2(\hbar c)^3}\int^{\infty}_{0}d\tp
\left[ \tp^2 \left(f_{e^\pm}(E_e)\right)\right] ,
\end{equation}
where $\tilde{p}\equiv pc$ and $E_e=\sqrt{\tp^2 + (m_e c^2)^2}$.
For the total number density of (free and bound) protons ($n_p=Y_e \rho_b/m_b$)
and $T$, the value of $\mu_e$ is iteratively solved for from the definition: $n_e(\mu_e)/n_b = Y_e$.
Then $n_{e^-}$ and $n_{e^+}$ and any other quantities where $\mu_e$ appears can be computed.
The electron-positron pressure is
\begin{equation}\label{pepm}
p_{e^{\pm}}= \frac{1}{3\pi^{2}(\hbar c)^3}\int^{{\infty}}_{0} d\tp
\left[ \frac{\tp^4}{E_e} \left(f_{e^\pm}(E_e)\right)\right] .
\end{equation}
The electron rest-mass plus internal energy gives an internal energy density of
\begin{equation}
u_{e^{\pm}}= \frac{1}{\pi^{2}(\hbar c)^3}\int^{{\infty}}_{0} d\tp
\left[ \tp^2 E_e \left(f_{e^\pm}(E_e)\right)\right] .
\end{equation}

The non-degenerate limit gives
\begin{equation}
p_{e} = n_e \kb T ,
\end{equation}
and
\begin{equation}
u_{e} \approx \frac{p_{e}}{\Gamma-1} ,
\end{equation}
where the parameter $\Theta\equiv (\kb T)/(m_e c^2)$ is used
to linearly interpolate from $\Gamma=5/3$ to $\Gamma=4/3$
such that at and beyond a value of unity this gives $\Gamma=4/3$.

Then, the pair pressure, internal energy density, and number density are
\begin{align}\label{npupairs}
p_{0,\rm pairs} &= p_{e^{-}} + p_{e^{+}} - p_{e}, \\
u_{0,\rm pairs} &= u_{e^{-}} + u_{e^{+}} - u_{e}, \\
n_{0,\rm pairs} &= n_{e^+} + n_{e^-} - n_{e}  = 2n_{e^+} .
\end{align}
For more details, see \citet{knp05}.
Note that the sum of, e.g., $p_{e} + p_{\rm pairs}$ that enters
the total gas pressure is unaffected by this decomposition.

\subsubsection{Electrons and Positrons EOS with QED Corrections}\label{electronicpressureqed}

In the super-critical field regime,
the pair internal energy density depends upon whether
the particle energy is above/below the rest-mass of electrons
and above/below the energy of the first Landau level.
Let ${\rm e}_1 = \exp{((-m_e c^2)/(\kb T))}$
and introduce an interpolation factor
${\rm e}_2 \equiv \exp{((-\kb T)/(\hbar\omega_{Be}(1)))}$ \citep{mes92,td95}.
Then, the following approximate QED expressions are used if $e_2>0.1$,
and otherwise the more accurate non-QED expressions given previously are used.

When $\hbar\omega_{Be}(1)\ll \kb T$ and $\kb T\gg m_e c^2$
the pairs behave like classical radiation with an internal energy density
\begin{equation}
u_{0,A,\rm pairs}\approx (7/4) u_{0,\gamma} {\rm e}_1 (1-{\rm e_2}),
\end{equation}
which gives a number density of pairs roughly equal to the number density of $m_e c^2$ photons.
When $\hbar\omega_{Be}(1)\ll \kb T$ and $\kb T\ll m_e c^2$,
then the pairs are classical and non-relativistic with an internal energy density
\begin{equation}
u_{0,B,\rm pairs}\approx \frac{2^{1/2}}{\pi^{3/2}} \frac{(m_e c^2)^4}{(\hbar c)^3} \left(\frac{\kb T}{m_e c^2}\right)^{3/2} {\rm e_1} (1-{\rm e_2}) .
\end{equation}
When $\hbar\omega_{Be}(1)\gg \kb T$ and $\kb T\gg m_e c^2$,
then photons and quantized relativistic pairs are in detailed balance with an internal energy density
\begin{equation}
u_{0,C,\rm pairs}\approx \frac{1}{12} \frac{\hbar c q b}{(\hbar c)^3} (\kb T)^2 {\rm e_1} {\rm e_2} .
\end{equation}
When $\hbar\omega_{Be}(1)\gg \kb T$ and $\kb T\ll m_e c^2$,
then the pairs are quantized and non-relativistic with an internal energy density
\begin{equation}
u_{0,D,\rm pairs}\approx \frac{(\hbar c q b)(m_e c^2)^2}{(2\pi^3)^{1/2}(\hbar c)^3} \left(\frac{\kb T}{m_e c^2}\right)^{1/2} {\rm e_1} {\rm e}_2 .
\end{equation}
Across these four regimes, a sufficient interpolation procedure is to simply sum all terms together
to obtain an internal energy density
\begin{equation}
u_{0,\rm pairs} \sim u_{0,A,\rm pairs} + u_{0,B,\rm pairs} + u_{0,C,\rm pairs} + u_{0,D,\rm pairs} ,
\end{equation}
pressure
\begin{equation}
p_{0,\rm pairs} \sim u_{0,A,\rm pairs}/3 + 2 u_{0,B,\rm pairs}/3 + u_{0,C,\rm pairs}/3 + 2 u_{0,D,\rm pairs}/3 ,
\end{equation}
and pair number density
\begin{equation}
n_{0,\rm pairs} \sim \frac{p_{0,\rm pairs}}{\kb T} .
\end{equation}
These expressions are only accurate to order unity
for $\kb T \gtrsim m_e c^2$ and otherwise less accurate,
but the QED effects tend to only be relevant at high temperatures.

\subsubsection{Opacity Effects on Electrons-Positrons}\label{opacitypairs}

The pair production processes $\gamma\gamma\to e^+ e^-$,
$\gamma e\to e e^+ e^-$, and $ee \to ee e^+ e^-$ are considered,
where the $ep\to ep e^+ e^-$ and $\gamma p\to p e^+ e^-$
have been shown to be less efficient for thermal plasmas considered
in this paper \citep{zdziarski82,sz86}.

Pairs act as radiation when annihilation into photons
dominates pair creation and pairs fill-in the region above the current layer
with a density scale-height of order $L_0$ as for photons.
In the limit that Kirchhoff's law applies,
the creation of photons occurs via the inverse reactions to $\gamma\gamma\to e^+ e^-$
and to $\gamma e\to e e^+ e^-$.  As seen below, these dominate
the inverse of $ee \to ee e^+ e^-$ except at $k_b T\gg m_ec ^2$,
such that any absorption opacity leads to photons that
can readily cross field lines and come into equilibrium
throughout the photon photosphere of size $L_0$.
Pairs also act as radiation (with the addition of Coulomb interactions)
when pairs flow down the current layer along straight field lines that open-up after a length $L_0$.
Then, both across and along the layer, the photon and pair density scale-heights are order $L_0$.
In the collisionless limit, the pairs are determined by their
prior history instead of equilibrium,
but this work only seeks to find the collisional layer structure.

The two-stream approximation is applied to pairs
whether they annihilate and travel above the layer or they
travel down field lines along the layer.
Order unity factors in equations~(\ref{gtau},\ref{htau})
would slightly change depending upon the allowed trajectories,
but such minor changes are ignored.
Kirchhoff's law is used as usual with angle and energy averaged rates given below,
where recall that $\<\sigma v\>/c = (1 + \delta_{12}) R_{12}/(c n_1 n_2)$ for
species $1$ and $2$ and number density rate $R_{12}$ (see., e.g., \citealt{weaver76}).
For simplicity, the underlying photon, electron, and pair distributions
are assumed to be thermal,
although the photon densities are modified
by the two-stream approximation to account for optical depth effects.
Because the (fully or marginally) optically thin energy density and pressure of pairs
are not required to be accurate in this work,
while the number density of pairs in the marginally optically thin limit should be somewhat accurate,
the number rates are treated directly while the energy rates are approximated.

Consider the $\gamma \gamma\to e^+ e^-$ process.
Following the notation and assumptions of \citet{weaver76},
let $C\equiv \exp(-\mu_\gamma/(\kb T))$ with photon chemical potential $\mu_\gamma$,
$\Phi\equiv m_e c^2/E_\gamma$ for the center-of-momentum photon energy $E_\gamma$,
$\Theta\equiv \kb T/(m_e c^2)$,
and $x\equiv 1/(\Theta\Phi)$
such that $x \kb T = E_\gamma$,
then
\begin{equation}
\sigma_{\gamma\gamma\to e^+ e^-} = \frac{3\sigma_T}{8}\Phi^2\left( (2+2\Phi^2-\Phi^4)\cosh^{-1}\left(\frac{1}{\Phi}\right) - \Phi'\right) ,
\end{equation}
for $\Phi<1$ and $\sigma_{\gamma\gamma\to e^+ e^-}\to 0$ for $\Phi>1$,
where $\Phi'\equiv (1+\Phi^2)(1-\Phi^2)^{1/2}$
and $\cosh^{-1}$ is the inverse and not reciprocal.
Then the number density rate for each pair unit is
\begin{align}
R_{0,\gamma\gamma\to e^+ e^-} &= n_\gamma^2 c \sum_{n=1,l=1}^{n=\infty,l=\infty} (\sqrt{nl}C^{n+1})^{-1} \times \\
&\  \int_{x=0}^{\infty} dx \left[x^4 \sigma_{\gamma\gamma\to e^+ e^-}[x] {\rm K}_1[2\sqrt{(nl)} x] \right] \nonumber ,
\end{align}
where ${\rm K}$ is the BesselK function \citep{weaver76},
and $\mu_\gamma\to 0$ such that $C=1$ is enforced
because the variation in photon number density is subsumed
into the prefactor number density of photons $n_\gamma$
(which would otherwise have been $n_{0,\gamma}$
and one would have to solve for $\mu_\gamma$
instead of using the two-stream approximation).
The energy density loss rate is
\begin{equation}
Q_{0,\gamma\gamma\to e^+ e^-} \approx E_{\gamma\gamma} R_{0,\gamma\gamma\to e^+ e^-} ,
\end{equation}
where the unit of pairs has thermal energy of roughly
where $E_{\gamma\gamma} \approx 2E_\gamma$
and $E_\gamma\sim u_{\gamma}/n_{\gamma}$.

The $\gamma e\to e e^+ e^-$ process has a fitted cross section
\begin{align}
\sigma_{\gamma e\to e e^+ e^-,A} &= 10^{-3}(y - 4)^2\times  \\
&\ [ 5.6 + 20.4(y - 4) - 10.9(y - 4)^2 \nonumber\\
&- 3.6(y - 4)^3 + 7.4(y - 4)^4 ] , \nonumber \\
\sigma_{\gamma e\to e e^+ e^-,B} &= 0.582814 - 0.29842y + 0.04354y^2 \\
&- 0.0012977y^3 , \nonumber \\
\sigma_{\gamma e\to e e^+ e^-,C} &= \frac{3.1247 - 1.3397y + 0.14612y^2}{1 + 0.4648y + 0.016683y^2} , \\
\sigma_{\gamma e\to e e^+ e^-,D} &= (84\ln(2y) - 218)/27 \\
&+ \frac{-1.333\ln^3(2y) + 3.863\ln^2(2y) - 11\ln(2y) + 27.9}{y} \nonumber ,
\end{align}
where $y=y_r=E_\gamma/(m_e c^2)$ is the variable photon energy per electron rest-mass energy,
such that
\begin{equation}
\sigma_{\gamma e\to e e^+ e^-} \approx \frac{3\sigma_T \alpha}{8\pi} \times\left[\begin{array}{ll}
0 & \mbox{if $y<4$} \\
\sigma_{\gamma e\to e e^+ e^-,A} & \mbox{if $4\le y < 4.6$}  \\
\sigma_{\gamma e\to e e^+ e^-,B} & \mbox{if $4.6 \le y < 6$}  \\
\sigma_{\gamma e\to e e^+ e^-,C} & \mbox{if $6 \le y < 14$}  \\
\sigma_{\gamma e\to e e^+ e^-,D} & \mbox{otherwise} \\
\end{array}\right],
\end{equation}
where $\alpha$ is the fine structure constant \citep{sg83}.
The underlying photon distribution is assumed to be Bose-Einstein,
such that the number density of photons is
\begin{equation}
n_{0,\gamma} = \frac{2(\kb T)^3}{(\hbar c)^3}\sum_{n=1}^{n=\infty} (n^3 C^n)^{-1} ,
\end{equation}
with distribution
\begin{equation}
f_\gamma = \frac{m_e c^2}{(\hbar c)^3 n_{0,\gamma}} \frac{(p_\gamma c)^2}{C{\rm e}^{(p_\gamma c)/(\kb T)}-1} ,
\end{equation}
for variable photon momentum $p_\gamma = E_\gamma/c$.
The number density rate for each pair unit is then
\begin{align}
R_{0,\gamma e\to e e^+ e^-} &= n_{e,\rm tot} n_\gamma \frac{c}{2{\rm K}_2[1/\Theta]} \times  \\
&\ \int_{y=0,y_r=4}^{\infty,\infty} dy dy_r \left[ y^{-2} f_\gamma y_r \sigma_{\gamma e\to e e^+ e^-}[y_r]{\rm e}^{-\frac{y/y_r + y_r/y}{2\Theta}}\right] \nonumber ,
\end{align}
\citep{sz86,zdziarski82,svensson84,svensson87}.
The energy density loss rate is
\begin{equation}
Q_{0,\gamma e\to e e^+ e^-} \approx E_{\gamma e} R_{0,\gamma e\to e e^+ e^-} ,
\end{equation}
where $E_{\gamma e}\sim E_\gamma + {\rm KE}_e$ and ${\rm KE}_e \sim u_e/n_e$.

The $e e\to e e e^+ e^-$ process has a number density rate for each pair unit of
\begin{equation}
R_{0,e e \to e e e^+ e^-} =  8.4\times 10^{-6} n_{e,\rm tot}^2 \sigma_T c \left(\ln(\Theta)\right)^3 ,
\end{equation}
for $\Theta>1$ and $R_{0,e e \to e e e^+ e^-}\to 0$ for $\Theta<1$
\citep{sz86,svensson84,svensson87}.
The energy density loss rate is
\begin{equation}
Q_{0,e e \to e e e^+ e^-} \approx E_{\rm e e} R_{0,e e \to e e e^+ e^-} ,
\end{equation}
where $E_{\rm e e}\approx 2{\rm KE}_e$.

The total optically thin energy and number density rates are
\begin{align}
R_{0,\rm pairs} &= R_{0,\gamma\gamma\to e^+ e^-} + R_{0,\gamma e\to e e^+ e^-} +  R_{0,e e \to e e e^+ e^-} , \\
Q_{0,\rm pairs} &= Q_{0,\gamma\gamma\to e^+ e^-} + Q_{0,\gamma e\to e e^+ e^-} +  Q_{0,e e \to e e e^+ e^-} .
\end{align}
The scattering optical depth is
\begin{equation}
\tau_{\rm pairs,\rm sca} \approx \sigma_{\rm es} n_{\rm pairs - \gamma,\rm eff} \Delta'_0 + \sigma_c (n_{\rm pairs} \Delta'_0 + n_e R'_j) ,
\end{equation}
where $n_{\rm pairs - \gamma,\rm eff}$ is the effective number density of photons
computed from Equation~(\ref{nabeff}) using $p_\gamma\sim (1/c)(u_{\gamma}/n_{\gamma})$,
and $\sigma_c$ is the Coulomb scattering cross section given later by Equation~(\ref{sigmac}).
The absorption optical depths are given by Kirchhoff's law as
\begin{align}
\tau_{n,\rm pairs,\rm abs} &\approx \frac{2R_{0,\rm pairs} \Delta'_0}{\vfs n_{0,\rm pairs}} ,\\
\tau_{u,\rm pairs,\rm abs} &\approx \frac{Q_{0,\rm pairs} \Delta'_0}{\vfs u_{0,\rm pairs}} ,
\end{align}
where the factor of $2R_{0,\rm pairs}$ appears because $R_{0,\rm pairs}$ corresponds
to the rate to produce a pair unit, while $n_{0,\rm pairs}$ is the total number of pairs
separately counting electrons and positrons.
$H_{\rm abs} = \Delta'_0$ as happens because there is only one depth involved for pairs and photons
involved in the absorption depths.
The effective free-stream velocity ($\vfs$) for pairs
traversing across field lines over the scale-height $\Delta'_0$
is determined by the distance per unit time traveled as photons and pairs.
The fractional distance traveled as photons is $L_\gamma \sim c/(R_{0,\gamma\gamma\to e^+ e^-} + R_{0,\gamma e\to e e^+ e^-})$
and the fractional distance traveled as pairs is $L_{\rm pairs}\sim v_e/R_{0,e^+ e^-\to \gamma \gamma}$.
The total fractional travel time is $T\sim 1/R_{0,e^+ e^-\to \gamma \gamma} + 1/(R_{0,\gamma\gamma\to e^+ e^-} + R_{0,\gamma e\to e e^+ e^-})$,
such that the effective average free-stream velocity is $\vfs\sim (L_\gamma+L_{\rm pairs})/T$.
Across the field lines $v_e\sim 0$ due to particle gyrations around magnetic field lines,
while parallel to the field lines, $v_e\sim c\sqrt{1-1/\gamma_e^2}$ is the electron-positron thermal speed.
In addition, along the length of the layer, the pairs can be advected at the speed $v_{\rm A}\sim c$.
For the temperatures and densities considered, a good approximation is found to be $v_{\rm fs}\sim v_e\sim c$,
where lower $v_e$ correspond to low temperatures where pairs are not dynamically important.
The total opacity is $\tau_{\rm pairs,\rm tot} = \tau_{\rm pairs,\rm sca} + \tau_{\rm pairs,\rm abs}$.
The two-stream approximation gives
\begin{align}
p_{\rm pairs} &= p_{0,\rm pairs} g[\tau_{u,\rm pairs}] ,\\
u_{\rm pairs} &= u_{0,\rm pairs} g[\tau_{u,\rm pairs}] ,\\
n_{\rm pairs} &= n_{0,\rm pairs} g[\tau_{n,\rm pairs}] ,\\
Q_{\rm pairs} &= \vfs u_{0,\rm pairs} h[\tau_{u,\rm pairs}] ,\\
R_{\rm pairs} &= \vfs n_{0,\rm pairs} h[\tau_{n,\rm pairs}] .
\end{align}

QED effects (e.g. 1-photon and modifications to 2-photon annihilation)
are only important when the pairs are optically thick to both absorption and scattering,
so the QED effects (see., e.g., \citealt{baring88b,baringh92})
on the optically thin rates need not be considered for the density of pairs.

\subsection{Neutrinos}\label{neutrinoemission}

At the highest densities ($\rho_b\gtrsim 10^{10}$g/cc)
and temperatures ($T\gtrsim 10^{10}$K) considered,
neutrino emission is the dominant source of cooling and pressure.
Kirchhoff's law is used as usual,
except because the true number density of neutrinos
is not required to be accurate in this work,
$\tau_{\nu,\rm abs}\approx \tau_{n,\nu,\rm abs}\approx \tau_{u,\nu,\rm abs}$
is set and only energy density loss rates are considered.

We follow \citet{km02} (see also \citealt{knp05}),
except their energy density loss rates assume that there exists
an optically thick thermalized photon and pair bath,
while in this work the photon and pair densities
are reduced by a factor $g[\tau_\gamma]$ and $g[\tau_{\rm pairs}]$, respectively.
When photons and pairs are involved in the reaction,
a $g[\tau]$ factor is applied for that given number density ($\propto T^3$)
as it enters the original integral,
so that these rates are consistent with the number density of photons and pairs.
Without this correction, then (for example) the pair annihilation rate
would be erroneously large at high temperatures in the regime
where photons and pairs are optically thin.
This assumes, as accurate in this work, that pairs dominate electrons
in number density when neutrinos are being produced.
The neutrino energy density loss rate for capture of non-degenerate pairs on nucleons is then
\begin{equation}
Q_{0,Ne\to} \approx 9.2\times 10^{33} T_{11}^6 (\rho_{10} X_{\rm nuc}) g[\tau_{\rm pairs}] ,
\end{equation}
and on degenerate pairs is
\begin{equation}
Q_{0,Ne\to} \approx 1.1\times 10^{31}\eta_e^9T_{11}^9 g[\tau_{\rm pairs}],
\end{equation}
where $Ne\to$ denotes a sum of processes $e^+ + n \to p + \bar{\nu}_e$ and $e^- + p \to n + \nu_e$
and $\to Ne$ is used to denote the sum of their inverse reactions.
The value of $\eta_e$ is used to linearly interpolate between regimes,
such that at $\eta_e=1$ only the degeneracy expression is used.
Neutrino pair production by annihilating pairs gives
\begin{equation}
Q_{0,e^+ + e^- \to \nu + \bar{\nu}} \approx 4.8\times 10^{33}T_{11}^9 g[\tau_{\rm pairs}]^2,
\end{equation}
and is negligible in the electron-degeneracy regime.
Neutrino pair production by non-degenerate free nucleon-nucleon bremsstrahlung gives
\begin{equation}
Q_{0,n+n \to n+n+ \nu + \bar{\nu}} \approx 1.5\times 10^{33}T_{11}^{5.5} (\rho_{13} X_{\rm nuc})^2 .
\end{equation}
The plasmon process gives
\begin{equation}
Q_{0,\tilde{\gamma} \to \nu + \bar{\nu}} \approx 1.5\times 10^{32}T_{11}^9 \gamma_p^6 {\rm e}^{-\gamma_p} (1+\gamma_p) \left(2+\frac{\gamma_p^2}{1+\gamma_p}\right) g[\tau_{\rm pairs}]g[\tau_\gamma],
\end{equation}
where $\tilde{\gamma}$ is a photon interacting with electrons
and $\gamma_p = 5.565\times 10^{-2}[(\pi^2+3\eta_e^2)/3]^{1/2}$.
Then the total optically thin neutrino energy density loss rate is
\begin{equation}
Q_{0,\nu} = Q_{0, Ne\to} + Q_{0, e^+ + e^- \to \nu + \bar{\nu}} + Q_{0, n+n \to n+n+ \nu + \bar{\nu}} + Q_{0, \tilde{\gamma} \to \nu + \bar{\nu}} .
\end{equation}

The neutrinos can be optically thick at sufficiently high densities/temperatures,
which is important to include because an artificially high neutrino cooling
rate would spuriously lead to compressible solutions for the collisional layer (see section~\ref{sec_collisionalrec}).
The electron type neutrino is treated most accurately among all the neutrino species
because it generally dominates the energy density loss rates.
Both anti-neutrino and neutrinos are treated using a single opacity
and all neutrino chemical potentials are assumed to be zero.
The scattering optical depth is
\begin{equation}\label{taunusca}
\tau_{\nu,\rm sca} = \sigma_{\nu,\rm bf} (n_{\nu-\rm bf,\rm eff}) R'_j + \sigma_{\nu e^\pm} (n_{\nu-e^\pm,\rm eff} \Delta'_0 + n_{e} R'_j) ,
\end{equation}
where $n_{\nu-\rm bf,\rm eff}$ is the effective number density of free baryons,
$n_{\rm bf} = X_{\rm nuc} n_b$ is the number density of free baryons,
and $n_{\nu-e^\pm,\rm eff}$ is the effective number density of electrons
as computed from Equation~(\ref{nabeff}).
Free nucleon scattering has
\begin{equation}
\sigma_{\nu,\rm bf} = 7.7\times 10^{-17} m_b (C_{s,p} Y_p + C_{s,n} Y_n) T_{11}^2 ,
\end{equation}
where $Y_p=Y_e$ and $Y_n=1-Y_p$,
$C_{s,p} = [4(C_{V}-1)^{2}+5\alpha_{a}^{2}]/24$ and
$C_{s,n}=(1+5\alpha_{a}^{2})/24$, with vector coupling
$C_{V} = 1/2+2\sin^{2}\theta_{W}$, $\alpha_{a} \approx 1.25$,
and Weinberg angle is $\sin^{2}\theta_{W} = 0.23$.
For electron-positron scattering
\begin{equation}
\sigma_{\nu e^\pm} = \sigma'_0 \left(1+\frac{\eta_e}{4}\right)\left[(C_V+C_A)^2 + \frac{1}{3}(C_V-C_A)^2\right]  \left(\frac{\kb T}{m_e c^2}\right)^2 ,
\end{equation}
where $C_A=1/2$ for electron neutrinos and $C_A=-1/2$ for electron anti-neutrinos,
$\sigma'_0 = (3\sigma_0/8) (2700\zeta[5])/(7\pi^4)\approx (3/2)\sigma_0$,
and $\sigma_0\approx 1.7\times 10^{-44}$.
Because both electron neutrinos are treated using a single opacity,
$C_A=1/2$ is set as applicable for the electron neutrino
because their emission rate is generally larger or equal
to the electron anti-neutrino emission rate when the neutrino chemical potential is zero.
The electron-neutrino scattering expression above assumes the neutrino energy is thermal,
which is inaccurate in the optically thin regime.  However, this only leads
to order unity corrections in our results for marginally optically thick regime.

The absorption optical depth is
\begin{align}\label{taunuabs}
\tau_{\nu,\rm abs} &= (\sigma_{\to Ne} + \sigma_{n+n+ \nu + \bar{\nu} \to n+n}) n_{\rm bf} R'_j \\
&+ (\sigma_{\nu + \bar{\nu}\to e^+ + e^-} + \sigma_{\nu + \bar{\nu}\to \tilde{\gamma} }) n_{\nu} \Delta'_0  \nonumber ,
\end{align}
where baryon reactions and photon-pair reactions have been collected together,
and from Kirchhoff's law $\sigma_{\to Ne} = Q_{0,Ne\to}/(n_{\rm bf} c u_{0,\nu})$,
$\sigma_{n+n+ \nu + \bar{\nu}\to n+n} = Q_{0,n+n \to n+n+ \nu + \bar{\nu} }/(n_{\rm bf} c u_{0,\nu})$,
$\sigma_{\nu + \bar{\nu} \to e^+ + e^- } = Q_{0,e^+ + e^- \to \nu + \bar{\nu}}/(n_{\nu} c u_{0,\nu})$,
$\sigma_{\nu + \bar{\nu}\to \tilde{\gamma}} = Q_{0,\tilde{\gamma} \to \nu + \bar{\nu}}/(n_{\nu} c u_{0,\nu})$,
where $u_{0,\nu} = (7/8) a_{\gamma} T^4$ that includes both neutrinos and anti-neutrinos.
Because the emission rate that is used merges both electron and pair capture processes,
a single $n_b$ is used instead of each $n_p$ and $n_n$ (for $Y_e=1/2$ this is accurate).
Notice that $n_\nu,n_{\rm bf}$ cancel out when obtaining the optical depth.

The total neutrino optical depth is $\tau_{\nu,\rm tot} = \tau_{\nu,\rm sca} + \tau_{\nu,\rm abs}$.
As for photons, the two-stream approximation is used to obtain the properties of the neutrinos
for general optical depths.
The $\mu$ and $\tau$ neutrinos have been neglected up to this point
because their optically thin emission rates are smaller than the electron types.
However, in the optically thick regime all neutrino species have the same densities and pressure
when assuming zero neutrino chemical potentials.
To approximately capture this simple thermalization effect,
note that the absorption and scattering opacities in the optically thick limit
are similar to within an order of magnitude among neutrino species.
So, the optical depth factors $g[\tau_\nu]$ and $h[\tau_\nu]$ are used to
interpolate to the optically thick regime in order to include the $\mu$ and $\tau$ neutrinos.
The neutrino energy density loss rate is
\begin{equation}
Q_\nu = \frac{F_\nu}{H_{\nu,\rm abs}} ,
\end{equation}
where $F_\nu = c u_{0,\nu} h[\tau_{\nu}] (1+2h[\tau_{\nu}])$,
$H_{\nu,\rm abs} \equiv \tau_{\nu,\rm abs} / (d\tau_{\nu,\rm abs}/ds)$,
$d\tau_{\nu,\rm abs}/ds = n_{\rm bf} (\sigma_{\to Ne} + \sigma_{n+n+ \nu + \bar{\nu} \to n+n}) + n_{\nu} (\sigma_{\nu + \bar{\nu} \to e^+ + e^-}+\sigma_{\nu + \bar{\nu} \to \tilde{\gamma}}) \equiv Q_{0,\nu}/(c u_{0,\nu}) $.
The $(1+2h[\tau_{\nu}])$ factor in $F_\nu$ approximately accounts for $\mu$ and $\tau$
neutrinos in the optically thick regime by modifying
the value of $u_{0,\nu}$ to be three times larger as required.
The pressure, internal energy density, and number density of neutrinos are
\begin{align}
n_\nu &= n_{0,\nu} g[\tau_\nu] (1+2g[\tau_\nu]),\\
u_\nu &= u_{0,\nu} g[\tau_\nu] (1+2g[\tau_\nu]),\\
p_\nu &= (1/3) u_\nu ,
\end{align}
where $n_{0,\nu} = (45\zeta[3]/(2\pi^4)) (a_{\gamma}/\kb) T^3 \approx (1/3)(a_{\gamma}/\kb) T^3$.
The $(1+2g[\tau_\nu])$ factor approximately accounts
for $\mu$ and $\tau$ neutrinos in the optically thick regime.
Future work can consider all neutrino species in detail
and consider QED effects on the neutrinos that may be important when $\tau_{\nu}\gtrsim 1$.

\section{Collisional and Collisionless Reconnection}
\label{sec_collisionalcollisionlessreconnection}

The goal of the following sections is to determine the thickness of
the collisional current layer.  The key point is that collisional
reconnection dominates unless dissipation (occurring on the scale of
the layer thickness) is dominated by collisionless effects that occur
on the scale of the plasma skin depth.

The resistivity ($\eta$) is computed (section~\ref{classicalresistivity})
in order to determine the thickness of the collisional current layer
(sections~\ref{sec_reconnectionmodels},\ref{sec_collisionalrec}).
These calculations assume the background jet's values of $b^2$ and $\rho_b$,
the current sheet length $L_0$,
and sheet separation length $\Delta_0$.
Then, collisionless reconnection is discussed (section~\ref{sec_collisionless}).

\subsection{Classical Resistivity}\label{classicalresistivity}

Consider the collisional resistivity on current-carrying
electrons and positrons (with number density $n_{e'}\approx n_{e,\rm tot}$)
as due to:
photon drag,
proton collisions,
or electron-positron collisions\footnote{
For the regimes of interest in this paper,
the generalized Ohm's law shows that the
current rise time for pairs ($t_{\rm rise}\sim t_p^2/t_\delta$,
the square of the electron plasma time over the light crossing time of the current layer)
is always shorter than the pair annihilation time
($t_a\sim (n_{e,\rm tot}\sigma_{\rm es} c)^{-1}$) such that pairs
always contribute to the current.
Protons are assumed to contribute negligibly to the current density,
which is accurate for the regimes considered.}~\footnote{
The collisional resistivity computed here assumes all particles are non-degenerate,
but the degeneracy of nucleons and electron-positrons is computed
and the fully degenerate electron-proton resistivity code by \citet{pot99}
was used to check how electron-proton collisions are affected.
Our work's results end up not depending upon electron-proton or pair-proton collisions,
so the degeneracy effects (such as discussed in \citealt{rossi08} for
electron-proton collisions) can be neglected in this study.
Computing radiative drag and pair drag on the degenerate
current-carrying electrons and positrons is left for future work.}.

The resistivity is presumed to be associated with comoving 4-current $j = e n_{e'} \gamma_d v_d$
with comoving relative electron drift 4-speed $\gamma_d v_d$ and electron charge $e$.
Then, the resistivity $\eta' = (\nu_{ec} m_e)/(n_{e'} e^2)$
can be determined by using
the electron drift momentum $p_e=\gamma_d v_d m_e$,
comoving electric field strength $|E_{\rm co}|$,
effective collisional frequency $\nu_{ec}$,
interaction force balance $e|E_{\rm co}| \approx p_e \nu_{ec}$,
and Ohm's law $e j = e |E_{\rm co}|/\eta'$.
The magnetic diffusivity is then given by $\eta = \eta' c^2/(4\pi)$,
which can also be written as $\eta = d^2_{e'} \nu_{ec}$,
where $d_{e'} = c/\omega_{pe'}$ is the skin depth of current-carrying electrons and positrons,
$\omega_{pe',\rm non-rel} = \sqrt{4\pi n_{e'} e^2/m_e}$
is the current-carrying electron-positron non-relativistic plasma frequency,
$\omega_{pe'} = \omega_{pe',\rm non-rel} G^{30}_{10}\left(\mu^2/4|^2_{-1/2,1,3/2}\right)/(2 {\rm K}_2(\mu))$
is the associated relativistic plasma frequency,
$\mu=m_e c^2/(\kb T_e)$,
$G()$ is the MeijerG function,
and ${\rm K}_2()$ is the modified Bessel function of the second kind \citep{bergman01}.

In addition to collisions between charged particles,
current-carrying electrons and positrons with drift momentum $p_e=\gamma_d m_e v_d$
experience a photon radiative drag force of
$F_e = -(4/3)(\gamma_d v_d/c) u_{\gamma} \sigma_{\rm es}$,
where $F_e = e |E_{\rm co}|$.
This gives an effective collisional frequency of $\nu_{ec}\approx F_e/p_e$.
From the magnetic diffusivity of $\eta = d^2_{e'} \nu_{ec}$
one obtains the resistivity due to photon drag of
\begin{equation}
\eta_\gamma \approx (4/3)d^2_{e'} (u_{\gamma}\sigma_{\rm es} c) /(m_e c^2) ,
\end{equation}
\citep{gu08}.

The Coulomb resistivity for current-carrying electrons and positrons interacting
with protons, electrons, or positrons is determined
by the mean free path given by $\lambda_{\rm mfp} \approx 1/((n_{\rm pairs} + n_p) \sigma)$,
where in general $n_{\rm pairs} + n_p \ge n_p = n_{e}$
so that pairs contribute an extra opacity that sets a lower-limit on the mean free path.
The use of $n_{\rm pairs} + n_p$ estimates the fact that electron-positron pair bath
and the ion bath contribute (to order unity) the same to the per-particle Coulomb collision
on the current-carrying electrons and positrons.
For an effective collisional rate as given above ($\nu_{ec}$),
the magnetic diffusivity is then $\eta = ((n_{\rm pairs} + n_p)/n_{e'})(v_d \sigma m_e c^2)/(e^2 4\pi)$
if each scatter is effective.
If $n_{\rm pairs}\to 0$ or $n_{e'}\sim n_{\rm pairs}\gg n_p$,
then $\eta = (v_d \sigma m_e c^2)/(e^2 4\pi)$,
which is independent of the density of current-carrying electrons and positrons.
In order to determine the resistivity, the cross section ($\sigma$) must be determined.

Coulomb collisions have naive cross section of $\sigma_0 = \pi\lambda_c^2$,
where $\lambda_c \approx \sqrt{5/(16\pi)} e^2/KE$ is the length scale over which Coulomb forces are important,
where the kinetic energy of electrons in the center-of-momentum frame is given by
$KE\approx \sqrt{(m_e \gamma_{\rm th} v_{\rm th})^2 c^2 + m_e^2 c^4} - m_e c^2$,
where $v_e \sim v_{\rm th,e}\equiv v_{\rm th}\approx c\sqrt{\Theta_{e'}(2+\Theta_{e'})}/(1+\Theta_{e'})$
is the thermal speed of electrons,
$\gamma_{\rm th} = 1/\sqrt{1-(v_{\rm th}/c)^2}$,
and $\Theta_{e'}\approx (u_e/n_e)/(m_e c^2)$.
There are additional relativistic corrections to $\lambda_c$
of order unity not considered here (see, e.g., \citealt{McKinley:1948:CSR}).
Also note that there are QED corrections that suppress the electron-electron cross section,
but there is only weak suppression of electron-proton interactions
that then dominate the Coulomb cross section \citep{sm87,sj07}.
For weakly coupled plasmas the Debye screening
gives a corrected Coulomb cross section of
\begin{equation}\label{sigmac}
\sigma_c \approx \sigma_0 \ln\Lambda ,
\end{equation}
where $\ln\Lambda \approx \ln(\lambda_D/\lambda_c)$ is the Coulomb logarithm
and $\lambda_D \approx v_{\rm th}/\omega_{pe'}$ is the Debye length.

The classical resistivity for current-carrying electrons and positrons
interacting with protons, electrons, and positrons that accounts for Debye screening
gives the Spitzer resistivity of $\eta_s \approx d^2_{e'} \nu_c$.
While the relativistic expression is used,
in the non-relativistic limit this becomes simply
\begin{equation}
\eta_s \approx \left(\frac{5\sqrt{3}}{32\pi}\right) c r_e (\theta_e)^{-3/2} \ln\Lambda ,
\end{equation}
where $\theta_e=(\kb T)/(m_e c^2)$ is the dimensionless electron temperature,
and $r_e = e^2/(m_e c^2)$ is the classical electron radius.

The total resistivity is then taken to be
\begin{equation}
\eta = \eta_\gamma + \eta_s ,
\end{equation}
which is a simple sum because the resistivities act on the same
current-carrying electrons and positrons.
In most of parameter space for the GRB jets considered,
photon drag dominates the Spitzer resistivity.
Note that neutrino drag is generally negligible.

\subsection{Reconnection Models}
\label{sec_reconnectionmodels}

\begin{figure}
  \begin{center}

      \includegraphics[width=2.5in]{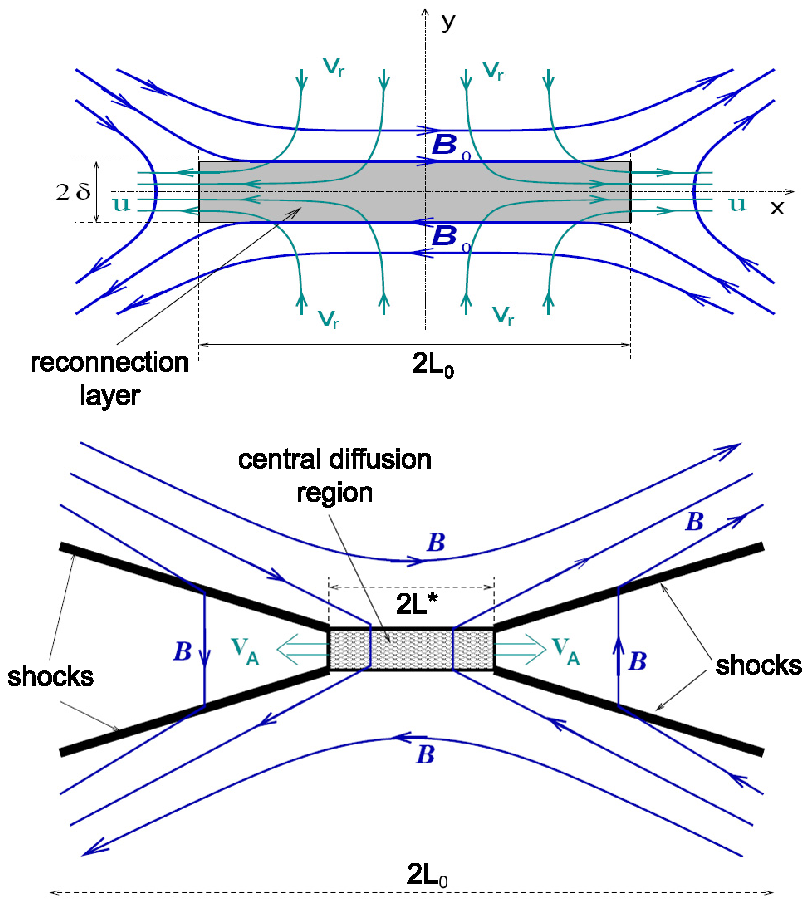}
  \end{center}
  \caption{
  Structure of Sweet-Parker and Petschek reconnection layers of length $2L_0$:
  Top panel shows Sweet-Parker reconnection with narrow ejection channel of thickness $2\delta_{\rm SP}$.
  Bottom panel shows Petschek reconnection with broad ejection channel between two shocks.
  The reconnection velocity is denoted by $v_r$,
  and the ejection velocity by $u$ ($u\sim v_{\rm A}$, the~\alf~speed).
  In Petschek reconnection, most of the energy dissipation takes place in the slow-mode shocks,
  while in Sweet-Parker reconnection dissipation occurs within the central dissipation region.
  Both velocities are shown as green lines.
  The magnetic field, $B$ and $B_0$, is shown as blue lines.
  Sweet-Parker-like reconnection occurs in collisional plasmas and is slow,
  while Petschek-like reconnection occurs in collisionless plasmas and is fast.
  }
  \label{fig_sweetpark_petschek}
\end{figure}

Given the length ($L_0$) of the current layer (as determined in section~\ref{sec_layerformationandlength}),
one can derive two classes of solutions for magnetic dissipation:
the slow Sweet-Parker type reconnection \citep{sweet58,parker57,parker63}
and the fast Petschek-type reconnection \citep{petschek64}.
Which of these two modes of reconnection is ultimately realized will be determined
by the plasma collisionality as discussed in section~\ref{sec_condition}.

As shown in Figure~\ref{fig_sweetpark_petschek},
the Sweet-Parker regime corresponds to an elongated reconnection layer.
Plasma is forced into a confined space before being ejected
out along a geometrically narrow channel,
which causes the reconnection rate to be slow\footnote{Many
effects are neglected, such as
secondary instabilities \citep{dka03,dka05},
2D tearing instability \citep{lsc07,jh09}
(but a temperature-dependent resistivity,
non-uniform flow along the layer,
and pressure anisotropy
can stabilize the tearing mode; \citealt{biskamp86,priestforbes00}),
interchange instability \citep{us77},
resistive kink mode \citep{sk79},
an anisotropic Spitzer resistivity,
realistic thermal conduction,
and viscosity.}.

The Petschek model uses the fact that topological changes in the electromagnetic
field do not have to take place in the same region where most of dissipation occurs.
The Petschek model has a very small inner Sweet-Parker-like layer
(called the central diffusion region) at the X-point where topological changes occur,
but the current layer generates slow-mode
shocks that dissipate most of the electromagnetic energy
and deflect the incoming plasma into two relatively wide exhaust channels.
The ejected plasma therefore has a large geometrical exit channel
allowing faster ingoing plasma and so faster reconnection.
Petschek reconnection occurs with an asymptotic reconnection speed
that typically relaxes to $v_r/v_{\rm A} \sim 1/\ln{S}$
(including in the relativistic regime; \citealt{lyubarsky05}).
For a Lundquist number $S\gg 1$,
the Petschek reconnection rate is $v_r/v_{\rm A}\approx \pi/(4\ln(S (v_r/v_{\rm A})^2))$,
giving $v_r\sim 0.018c$ for $S=10^{22}$ for a highly magnetized jet \citep{lyubarsky05}.

\subsection{Collisional Reconnection}\label{sec_collisionalrec}

Collisional reconnection is assumed to be approximately
like the Sweet-Parker model of a stationary dissipating current sheet.
\citet{lyubarsky05} found that the relativistic Sweet-Parker reconnection rate,
layer thickness, and ejection velocity all behave as expected from non-relativistic Sweet-Parker theory.
For Sweet-Parker layer extent $2L_0$, the Sweet-Parker layer has thickness
\begin{equation}
\delta_{\rm SP} \sim \sqrt{\frac{L_0 \eta}{v_{\rm A}}} = L_0\, S^{-1/2} ,
\end{equation}
which is associated with a reconnection velocity of
\begin{equation}
v_{r,\rm sp} \sim \frac{\eta}{\delta_{\rm SP}} = v_{\rm A}\, S^{-1/2} ,
\end{equation}
and an ejection velocity of $u \sim v_{\rm A}$.
where recall that $S=L_0 v_{\rm A}/\eta$ is the Lundquist number.
The reconnection timescale is $\tau_{sp} \sim \tau_A \sqrt{S}$,
where $\tau_A = L_0/v_{\rm A}$ is the fluid transit flow time along the length of the layer.
For very large $S$, the SP solution may be unstable to plasmoid formation (see., e.g., \citealt{uls10}).

\citet{um11} found that one can readily extend the non-relativistic Sweet-Parker
theory to include radiative cooling.
The results in their section 2 apply for arbitrary optical depths.
They obtained such results by using the energy equation rather than the incompressible assumption
as done in \citet{lyubarsky05}.
An interesting point is that the effects of cooling decouple from rest of equations,
which allows one to obtain the Sweet-Parker type result from pressure equilibrium as usual
and then independently obtain the constraint from radiative cooling conditions that determine
a baryon compression ratio, $A=n_c/n_0$, i.e., the
ratio of plasma density in the center of the layer, $n_c$, to the plasma density
in the upstream region, $n_0$.

The solution that \citet{um11} obtained for general optical depths ultimately involves
two equations: pressure equilibrium across the layer and radiative energy balance across the layer.
While a global jet+layers structure is not sought in the present paper,
a global solution would show that radiation pressure, like the magnetic pressure,
would have its own structure across the jet.
This justifies balancing the total gas+radiation pressure of the complex
against the electromagnetic pressure of the jet.

To obtain the radiative Sweet-Parker solution,
first one imposes the pressure balance condition
\begin{equation}\label{forcebalance}
p_{\rm EM} \sim p_g(T) ,
\end{equation}
where $p_{\rm EM}=b^2/(8\pi)$ is the electromagnetic pressure.
This equation corresponds to a loss of magnetic pressure
within the layer that is recovered by a balanced thermal gas pressure.

Second, the energy equation determines the effects of radiative transport \citep{um11}.
The result is that one sets the compression ratio
of the rest-mass density within the current layer of size
$\delta_{\rm SP'}$ (given below) to be
\begin{equation}\label{Aum}
A = \frac{Q_g(T)}{Q_{\rm SP}} ,
\end{equation}
where recall that $Q_g(T)$ is the gas energy volume loss rate, and
\begin{equation}
Q_{\rm SP} = b^2/(4\pi L_0/v_{\rm A}) ,
\end{equation}
where $v_{\rm A}$  is the usual upstream~\alf~velocity \citep{um11}.
The quantity $Q_{\rm SP}$ is the energy density dissipation rate
if the layer corresponds to the standard thin Sweet-Parker solution
with thickness $\delta_{\rm sp}$.
Note that $A$ does not depend upon the resistivity.
Equation~(\ref{Aum}) applies only in the strong cooling regime, i.e., if $A \gg 1$.
If $A<1$, then one sets $A=1$ corresponding to weak cooling limit
giving back the original non-radiative incompressible Sweet-Parker solution.
This $A$ factor is applied to all baryon densities ($n_b$)
and baryon-associated electron densities ($n_e$),
except when used in opacity integrals because total baryonic mass across the jet is conserved
regardless of localized compression in each thin current sheet.
For example, there are no changes to the opacity integrals over baryonic-associated mass
when performed for equations~(\ref{tauradsca},~\ref{tauradabs},~\ref{taunusca},~\ref{taunuabs})\footnote{
Because the free nucleon fraction depends upon density and not column density,
for a pre-collapse baryon density $\rho_{\rm b}$ one should integrate to get
$\tau\propto A X_{\rm nuc}[A \rho_{\rm b}] \delta + X_{\rm nuc}[\rho_{\rm b}] (R_j-\delta)$.
However, because generally $R_j/\delta \gg A$ is found, this is a negligible effect
and $\tau\propto X_{\rm nuc}[\rho_{\rm b}] R_j$ is set as usual.}.

Once the energy density loss rate ($Q_g$)
is comparable to (or larger than) the energy dissipation rate ($Q_{\rm SP}$),
the layer must compress and undergo more rapid reconnection
in order to balance the radiative losses.
The reconnection velocity and current layer thickness are,
respectively, given by
\begin{align}
v_{r,\rm sp'} &\sim v_{r,sp} A^{1/2} , \\
\delta_{\rm SP'} &\sim \delta_{\rm SP} A^{-1/2} ,
\end{align}
\citep{um11}.
This is only an approximation of a so-far undeveloped
fully relativistic radiation reconnection theory.
However, if $A=1$, then the radiation only contributes
to the pressure and internal energy and the original
non-relativistic Sweet-Parker solution should be accurate
even in the relativistic regime \citep{lyubarsky05}.
In most of parameter space considered in the present study,
$A\sim 1$ is found to hold.
This implies that the usual Sweet-Parker solution is generally valid to order unity.

The time-rate of change of reconnected magnetic flux is
\begin{equation}\label{dAdt}
\left|\frac{\partial P_z/dz}{c \partial t}\right| = \left|-E_z\right| = \left|v_r B_y/c\right| ,
\end{equation}
where $B_y\sim |b|$ (the comoving electromagnetic field strength in the jet).
In the non-relativistic incompressible limit, this gives a time-rate of change of magnetic flux of
$\partial P_z/(dz c \partial t) \sim  |v_{\rm A} S^{-1/2} |b| /c| \ll v_{\rm A} |b|/c$,
which implies a slow reconnection rate compared to~\alf~timescales.
The reconnection timescale is given by the time needed to reconnect
a finite amount of flux of order $|b| \Delta_0 dz$
present in the volume between multiple current sheets.
This gives $\tau_r \sim (|b| \Delta_0 dz)/(\partial P_z/\partial t) \sim (\Delta_0/v_{\rm A}) \sqrt{S}$,
which is much longer than it takes an~\alf~wave to cross $\Delta_0$.

In the relativistic regime the inflow speed may reach $v_r\sim v_{\rm A}\sim c$
and significant Lorentz contraction might occur
when there is a significant loss of energy through radiation
or if the Lundquist number $S$ is smaller than
the comoving magnetization $b^2/(8\pi\rho_b c^2)$ \citep{lu03}.
However, when including both the energy and momentum equations \citep{lyubarsky05},
this is found not to occur in the Sweet-Parker regime.
This validates the assumptions used in this work.
Relativistic reconnection is an active area of research
\citep{watanabe06,kbl07,hz07,zh07,zh08,zhk09a,tenbarge2010},
so aspects of this work may require modification.

\subsection{Collisionless Reconnection}
\label{sec_collisionless}

Sweet-Parker reconnection is slow because plasma has to flow through a narrow channel.
Microphysics (e.g. anomalous resistivity or the Hall effect)
alone can not enhance the reconnection rate
if the current layer is preserved in the Sweet-Parker configuration.
For example, as the current layer thickness $\delta \to d_i$, Hall effects dominate the resistivity.
The Sweet-Parker analysis yields a reconnection speed of $v_r \sim (d_i/L_0) v_A$,
which still gives $v_r\ll v_A$ simply because of the assumed Sweet-Parker geometry.
However, if a transition to collisionless reconnection occurs as discussed in section~\ref{sec_condition},
then the Sweet-Parker geometry can be disrupted into a Petschek-like geometry
leading to fast reconnection.

One candidate for fast collisionless reconnection is the Hall effect.
In the Generalized Ohm's law the Hall term is $j\times B/(n_{e'} e c)$
operating on the ion skin depth where electrons and ions decouple.
The Hall effect involves a two-fluid laminar
reconnection configuration with $v_r \le 0.1 v_{\rm A}$
\citep{mandt94,shay98,birn01,bhat01,cassak05,yamada06,daughton06}.

Another candidate is an anomalous resistivity,
which can be due to plasma microinstabilities
\citep{Parker:1963:SFP,ce77,tu77,Parker:1979:CMF,Syrovatskii:1981:PSR,Biskamp:1986:MRC,Taylor:1986:RMR,Parker:1988:NSX,scholer89,Masuda:1994:LTH,kulsrud98,erkaev01,kulsrud01,bs01,mlk05,Melrose:1986:ISL,Hasegawa:1975:PIN,Begelman:1988:TCI}.
Anomalous resistivity can be triggered, e.g.,
when the drift velocity ($v_d$) exceeds some critical velocity ($v_c$),
such that $v_d = j/(e n_{e'}) > v_c \sim v_{\rm th}$.
This process can drive plasma instabilities developing microturbulence
where scattering of electrons by waves enhances resistivity.
As the layer's thickness $\delta$ decreases down to the critical thickness
$\delta_c = c B/(4\pi j_e)$ where $j_e = e n_{e'} \gamma_{\rm th} v_{\rm th}$,
anomalous resistivity turns on (this scale is usually comparable to $d_i$).
Anomalous resistivity allows for higher resistivity than Spitzer,
but it also enables the Petschek-like geometry \citep[see, e.g.,][]{kulsrud01,bs01,yamada06,zweibel09}.

Overall, collisionless Petschek-type reconnection operates on the ion skin depth
\begin{equation}
d_{i} = c/\omega_{pi} ,
\end{equation}
where $\omega_{pi}$ is the ion plasma frequency.
The ejection velocity is the same as in the Sweet-Parker case with velocity order $v_{\rm A}$.


\label{lastpage}
\end{document}